%% file: DESY-06-116.tex
\documentclass[zpreprint,zbstdefault]{zeus_paper}
\usepackage[english]{babel}
\input zeus_def.tex
\input  DESY-06-116-cit.tex
\begin{document}
%------------------------------------------------------------------------------
%       Title sheet
%------------------------------------------------------------------------------
\include{DESY-06-116-tit}
\include{DESY-06-116-aut}
%------------------------------------------------------------------------------
%       Text
%------------------------------------------------------------------------------
\include{DESY-06-116-txt}

%------------------------------------------------------------------------------
%       Bibliography
%------------------------------------------------------------------------------
\include{DESY-06-116-ref}
%------------------------------------------------------------------------------
%       Tables
%------------------------------------------------------------------------------
\include{DESY-06-116-tab}

%------------------------------------------------------------------------------
%       Figures
%------------------------------------------------------------------------------
\include{DESY-06-116-fig}

%------------------------------------------------------------------------------
\end{document}

%% file: zeus_def.tex
\newcommand{\ZcoosysA}{%
The ZEUS coordinate system is a right-handed Cartesian system, with the $Z$
axis pointing in the proton beam direction, referred to as the ``forward
direction'', and the $X$ axis pointing left towards the center of HERA.
The coordinate origin is at the nominal interaction point.\xspace}

\newcommand{\ZcoosysfnA}{\footnote{\ZcoosysA}}

\chardef\usc=95
\chardef\til=126
\catcode`\@=11 % @ signs are now treated as letters
\DeclareRobustCommand\xdotspace{\futurelet\@let@token\@xdotspace}
\def\@xdotspace{%
  \ifx\@let@token.\else
  \ifx\@let@token\bgroup.\else
  \ifx\@let@token\egroup.\else
  \ifx\@let@token\/.\else
  \ifx\@let@token\ .\else
  \ifx\@let@token~.\else
  \ifx\@let@token!.\else
  \ifx\@let@token,.\else
  \ifx\@let@token:.\else
  \ifx\@let@token;.\else
  \ifx\@let@token?.\else
  \ifx\@let@token/.\else
  \ifx\@let@token'.\else
  \ifx\@let@token).\else
  \ifx\@let@token-.\else
  \ifx\@let@token\@xobeysp.\else
  \ifx\@let@token\space.\else
  \ifx\@let@token\@sptoken.\else
   .\space
   \fi\fi\fi\fi\fi\fi\fi\fi\fi\fi\fi\fi\fi\fi\fi\fi\fi\fi}
\catcode`\@=12 % @ signs are no longer letters

\newcommand{\stru}[2]{%
   \relax\ifmmode\hbox{\vrule height#1 depth#2 width0pt}%
   \else\vrule height#1 depth#2 width0pt\fi}
\newcommand{\Ronum}[1]{\uppercase\expandafter{\romannumeral#1}}
\newcommand{\ronum}[1]{\expandafter{\romannumeral#1}}
\DeclareRobustCommand{\LaTeXZ}{%
  \LaTeX\kern-.05em4\kern-.1em
  {\raisebox{-0.2ex}{$\scriptstyle\text{ZEUS}$}}\xspace}
\newcommand{\eq}[1]{(\ref{eq-#1})}

\DeclareMathAlphabet{\mathbf}{OT1}{cmr}{bx}{sl}
\newcommand{\eVdist}{\kern-0.06667em}

\newcommand{\Gev}{{\text{Ge}\eVdist\text{V\/}}}

\newcommand{\gev}{{\,\text{Ge}\eVdist\text{V\/}}}

\newcommand{\pbi}{\,\text{pb}^{-1}}

\newcommand{\cm}{\,\text{cm}}

\newcommand{\ns}{\,\text{ns}}

\newcommand{\mrad}{\,\text{mrad}}

\newcommand{\Tesla}{\,\text{T}}

\newcommand{\slashfrac}[2]{%
  \raisebox{0.5ex}{\ensuremath #1}\kern-0.12em/\kern-0.08em
  \raisebox{-.8ex}{\ensuremath #2}}
\newcommand{\sqr}[3]{%
    {\vcenter{\hrule height.#3ex\hbox{\vrule width.#2ex height#1ex
     \kern#1ex\vrule width.#3ex}\hrule height.#2ex}}}

\catcode`\@=11 % @ signs are now treated as letters
\newcommand{\parenbar}{\mathpalette\p@renb@r}
\def\p@renb@r#1#2{\vbox{%
  \ifx#1\scriptscriptstyle \dimen@.7em\dimen@ii.2em\else
  \ifx#1\scriptstyle \dimen@.8em\dimen@ii.25em\else
  \dimen@1em\dimen@ii.4em\fi\fi \offinterlineskip
  \ialign{\hfill##\hfill\cr
    \vbox{\hrule width\dimen@ii}\cr
    \noalign{\vskip-.3ex}%
    \hbox to\dimen@{$\mathchar300\hfil\mathchar301$}\cr
    \noalign{\vskip-.3ex}%
    $#1#2$\cr}}}
\catcode`\@=12 % @ signs are no longer letters

\newcommand{\IP}{{\rm I$\kern-0.01667em$P}\xspace}

\mathchardef\qsm=63
\mathchardef\pls=43
\mathchardef\mns=512
\mathchardef\plm=518
\mathchardef\eql=61
\mathchardef\smallleft=300
\mathchardef\smallright=301
\mathchardef\les=316
\mathchardef\gre=318
\mathchardef\leq=532
\mathchardef\grq=533
\catcode`\@=11 % @ signs are now treated as letters
\newcounter{pict@width}
\newcounter{pict@height}
\newlength{\pict@scale}
\newcommand{\psfigadd}[4]{%
\setcounter{pict@width}{1*\ratio{#2+\pict@scale/2}{\pict@scale}}
\setcounter{pict@height}{1*\ratio{#3+\pict@scale/2}{\pict@scale}}
\setlength{\unitlength}{\pict@scale}
\hbox to #2{\hspace{-\fill}\begin{picture}(\thepict@width,\thepict@height)
\put(0,0){\psfig{figure=#1,width=#2,height=#3,clip=}}
\SetScale{0.283466457}
\SetWidth{1.763889}
{#4}
\end{picture}}
}
\newcounter{pict@widthfst}
\newcounter{pict@widthscd}
\newcounter{pict@widthtot}
\newcommand{\psfigaddtwo}[7]{%
\setcounter{pict@widthfst}{1*\ratio{#2+\pict@scale/2}{\pict@scale}}
\setcounter{pict@widthscd}{1*\ratio{#2+#4+\pict@scale/2}{\pict@scale}}
\setcounter{pict@widthtot}{1*\ratio{#2+#4+#6+\pict@scale/2}{\pict@scale}}
\setcounter{pict@height}{1*\ratio{#3+\pict@scale/2}{\pict@scale}}
\setlength{\unitlength}{\pict@scale}
\hbox{\hspace{-\fill}\begin{picture}(\thepict@widthtot,\thepict@height)
\put(0,0){\psfig{figure=#1,width=#2,height=#3,clip=}}
\put(\thepict@widthscd,0){\psfig{figure=#5,width=#6,height=#3,clip=}}
\SetScale{0.283466457}
\SetWidth{1.763889}
{#7}
\end{picture}}
}
\newcommand{\psfigror}[4]{%
\setcounter{pict@width}{1*\ratio{#2+\pict@scale/2}{\pict@scale}}
\setcounter{pict@height}{1*\ratio{#3+\pict@scale/2}{\pict@scale}}
\setlength{\unitlength}{\pict@scale}
\hbox{\begin{picture}(\thepict@width,\thepict@height)
\put(0,\thepict@height){\psfig{figure=#1,width=#3,height=#2,clip=,angle=270}}
\SetScale{0.283466457}
\SetWidth{1.763889}
{#4}
\end{picture}}
}
\newcommand{\psfigrol}[4]{%
\setcounter{pict@width}{1*\ratio{#2+\pict@scale/2}{\pict@scale}}
\setcounter{pict@height}{1*\ratio{#3+\pict@scale/2}{\pict@scale}}
\setlength{\unitlength}{\pict@scale}
\hbox{\begin{picture}(\thepict@width,\thepict@height)
\put(0,0){\psfig{figure=#1,width=#3,height=#2,clip=,angle=90}}
\SetScale{0.283466457}
\SetWidth{1.763889}
{#4}
\end{picture}}
}
\catcode`\@=12 % @ signs are no longer letters
\newlength\listtextwidth

\catcode`\@=11 % @ signs are now treated as letters
\newlength{\@tabfninsert}
\newlength{\@tabfnwidth}
\newcommand{\tabfootnote}[2]{%
  \setlength{\@tabfninsert}{0.8em}
  \setlength{\@tabfnwidth}{\textwidth}
  \addtolength{\@tabfnwidth}{-\@tabfninsert}
  \addtolength{\@tabfnwidth}{-0.4em}
  \noindent\makebox[\@tabfninsert][r]{\footnotesize$^{#1}$\hfil}\hfill%
  \parbox[t]{\@tabfnwidth}{\footnotesize #2\hfill}}
\catcode`\@=12 % @ signs are no longer letters

%% file: DESY-06-116-cit.tex
\def\citeCTD{{\cite{%
nim:a279:290,*npps:b32:181,*nim:a338:254%
}}\xspace}
\def\citeCAL{{\cite{%
nim:a309:77,*nim:a309:101,*nim:a321:356,*nim:a336:23%
}}\xspace}

%% file: DESY-06-116-tit.tex
\prepnum {DESY 06-116}
\title{
Measurement of neutral current cross sections at high Bjorken-$x$ with the 
ZEUS detector at HERA
}                                                       
   
\author{ZEUS Collaboration}

\date{July, 2006}

\abstract{
A new method is employed to measure the neutral current cross section up 
to Bjorken-$x$ values of one with the ZEUS detector at HERA using an 
integrated luminosity of $65.1$~$\pbi$ for $e^+p$ collisions and
$16.7$~$\pbi$ for $e^-p$ collisions at $\sqrt{s}=318$~$\gev$
and $38.6$~$\pbi$ for $e^+p$ collisions at $\sqrt{s}=300$~$\gev$. 
Cross sections have been extracted for 
$Q^2 \ge 648$~$\gev^{2}$ and are compared to predictions using different parton 
density functions. For the highest $x$ bins, the data have a tendency to lie 
above the expectations using recent parton density function parametrizations.  
%The data will provide new constraints on the form of
%the parton density functions at the highest values of $x$.
}

\makezeustitle

%% file: DESY-06-116-aut.tex
\pagenumbering{Roman}                                                                              
                                    % this "%"s are for cosmetics only                             
%\begin{document}                                                                                   
                                                   %                                               
\begin{center}                                                                                     
{                      \Large  The ZEUS Collaboration              }                               
\end{center}                                                                                       
  S.~Chekanov,                                                                                     
  M.~Derrick,                                                                                      
  S.~Magill,                                                                                       
  S.~Miglioranzi$^{   1}$,                                                                         
  B.~Musgrave,                                                                                     
  D.~Nicholass$^{   1}$,                                                                           
  \mbox{J.~Repond},                                                                                
  R.~Yoshida\\                                                                                     
 {\it Argonne National Laboratory, Argonne, Illinois 60439-4815}, USA~$^{n}$                       
\par \filbreak                                                                                     
  M.C.K.~Mattingly \\                                                                              
 {\it Andrews University, Berrien Springs, Michigan 49104-0380}, USA                               
\par \filbreak                                                                                     
  N.~Pavel~$^{\dagger}$, A.G.~Yag\"ues Molina \\                                                   
  {\it Institut f\"ur Physik der Humboldt-Universit\"at zu Berlin,                                 
           Berlin, Germany}
% ...
\newpage                                                                                           
$^{\    1}$ also affiliated with University College London, UK \\                                  
% ...
$^{\dagger}$ deceased \\                                                                           
%                                                                                                  
% \par         % if index listing & table fit to 1 page, put gap here                              
\newpage   % alternatively: go to newpage, if page is too small                                    
                                                           %                                       
% \institute_references_start    % do not touch or move this line !                                
                                                           %                                       
\begin{tabular}[h]{rp{14cm}}                                                                       
$^{a}$ &  supported by the Natural Sciences and Engineering Research Council of Canada (NSERC) \\  
$^{b}$ &  supported by the German Federal Ministry for Education and Research (BMBF), under        
          contract numbers HZ1GUA 2, HZ1GUB 0, HZ1PDA 5, HZ1VFA 5\\                                
$^{c}$ &  supported in part by the MINERVA Gesellschaft f\"ur Forschung GmbH, the Israel Science   
          Foundation (grant no. 293/02-11.2) and the U.S.-Israel Binational Science Foundation \\  
$^{d}$ &  supported by the German-Israeli Foundation and the Israel Science Foundation\\           
$^{e}$ &  supported by the Italian National Institute for Nuclear Physics (INFN) \\                
$^{f}$ &  supported by the Japanese Ministry of Education, Culture, Sports, Science and Technology 
          (MEXT) and its grants for Scientific Research\\                                          
$^{g}$ &  supported by the Korean Ministry of Education and Korea Science and Engineering          
          Foundation\\                                                                             
$^{h}$ &  supported by the Netherlands Foundation for Research on Matter (FOM)\\                   
$^{i}$ &  supported by the Polish State Committee for Scientific Research, grant no.               
          620/E-77/SPB/DESY/P-03/DZ 117/2003-2005 and grant no. 1P03B07427/2004-2006\\             
$^{j}$ &  partially supported by the German Federal Ministry for Education and Research (BMBF)\\   
$^{k}$ &  supported by RF Presidential grant N 1685.2003.2 for the leading scientific schools and  
          by the Russian Ministry of Education and Science through its grant for Scientific        
          Research on High Energy Physics\\                                                        
$^{l}$ &  supported by the Spanish Ministry of Education and Science through funds provided by     
          CICYT\\                                                                                  
$^{m}$ &  supported by the Particle Physics and Astronomy Research Council, UK\\                   
$^{n}$ &  supported by the US Department of Energy\\                                               
$^{o}$ &  supported by the US National Science Foundation\\                                        
$^{p}$ &  supported by the Polish Ministry of Scientific Research and Information Technology,      
          grant no. 112/E-356/SPUB/DESY/P-03/DZ 116/2003-2005 and 1 P03B 065 27\\                  
$^{q}$ &  supported by FNRS and its associated funds (IISN and FRIA) and by an Inter-University    
          Attraction Poles Programme subsidised by the Belgian Federal Science Policy Office\\     
$^{r}$ &  supported by the Malaysian Ministry of Science, Technology and                           
Innovation/Akademi Sains Malaysia grant SAGA 66-02-03-0048\\                                       
\end{tabular}                                                                                      
                                                           %                                       
% \institute_references_end     % do not touch or move this line !                                 
                                                           %                                       
%\end{document}                                                                                     

%% file: DESY-06-116-txt.tex
\pagenumbering{arabic} 
\pagestyle{plain}
% ----------------------------------------------------------------------------
%      motivation and standard model cross section
% ----------------------------------------------------------------------------
\section{Introduction}
Only limited information is available on structure functions at high 
Bjorken-$x$ in the deep inelastic scattering (DIS) regime. This is 
largely due to limitations in beam energies, in measurement techniques 
and the small cross section at 
high $x$.
In this paper, a new method is described and used to measure the 
neutral current (NC) cross section in electron-proton 
scattering up to $x=1$ with data from the ZEUS detector at HERA. 
%A new method is described and employed to measure the neutral current (NC) 
%cross section up to Bjorken $x$ values equal to one.

At HERA, proton beams of 920~GeV (820~GeV prior to 1998), collide with either 
electron or positron beams of 27.5~GeV. The 
electron\footnote{In the following, we use the term 
electron to represent both electrons and positrons unless specifically noted 
otherwise.} 
interacts with the proton via the exchange of a gauge boson. 
%The exchanged boson can be a neutral particle (photon or Z$^0$), leading to a
%NC interaction.
The description of DIS is usually given in terms 
of three Lorentz-invariant quantities, $Q^2$, $x$ and $y$, which are related 
by $Q^2=sxy$, where the masses of the electron and proton are neglected, 
$s$ is the square of the center-of-mass energy, $Q^2$ is the negative of the
square of the transferred four-momentum,
$x$ is the Bjorken variable~\cite{bjorken:1969ja} 
and $y$ is the inelasticity.
The NC electron-proton differential scattering cross section 
is typically written in terms of the proton structure functions as
\begin{equation}
  \frac{d^2\sigma_{\rm Born}^{\rm SM}(e^\pm p)}{dx \,dQ^2} =  
  \frac{2 \pi \alpha^2}{x Q^4}
  \left[
    Y_+ F_2 \left(x, Q^2 \right) \mp Y_- xF_3 \left(x, Q^2 \right) - 
    y^2 F_L \left(x, Q^2 \right)
  \right] \ ,
  \label{eq-Born}
\end{equation}
where $Y_{\pm} \equiv 1 \pm (1-y)^2$ and $\alpha$ denotes the fine-structure
constant. At leading order (LO) in QCD, the longitudinal structure
function, $F_L$, is zero and the structure functions $F_2$ and $xF_3$ can be
expressed as products of electroweak couplings and parton density functions
(PDFs).

The form of the PDFs
is typically parametrized as 
$q(x) = Ax^{-\lambda}f(x)(1-x)^{\eta}$, where $f(x)$ interpolates between
the low-$x$ and high-$x$ domains.  Such a form allows for an accurate
description of the data at low $x$~\cite{pl:b531:216,CTEQ6M,pr:d70:052001,
epl:c30:1,pr:d68:014002}.  
For $x\geq 0.3$, the PDFs are found to decrease very quickly.  
However, a direct confrontation with data has not been possible to date for
$x \rightarrow 1$.
The highest measured points in the DIS regime are for 
$x=0.75$~\cite{pl:b223:485}. Data at higher $x$ 
exist~\cite{pl:b282:475,pr:d67:092001} but these are 
in the resonance production region and cannot be easily
interpreted in terms of parton distributions. 
The highest $x$ value for HERA structure function data reported to date is 
$x=0.65$~\cite{pr:d70:052001,epl:c30:1}. The differences 
between different PDF sets increase rapidly as $x$ increases,
% as seen in  Fig.??? 
even though they use similar data and have common parametrization
for $x \rightarrow 1$. The uncertainties for $x>0.75$ are large and 
hard to quantify.
 
%The high $x$ cross section at $Q^2=1000$~GeV$^2$, evaluated for several
%recent PDF sets,  are compared to that calculated using CTEQ6M in
%Fig.~\ref{fig-diff-pdf}.  The uncertainty band associated with CTEQ6M 
%is also shown.  As is clear from the figure, the uncertainty at high 
%$x$ is large. It is also likely understimated since the different PDFs
%use similar parametrizations.

%Although the different PDFs yield cross sections within the
%uncertainty band of CTEQ6M, the uncertainty is likely underestimated since
%they all use a similar parametrizations.

This paper presents a reanalysis of previously published ZEUS 
data~\cite{epj:c21:443,epj:c28:175,pr:d70:052001}
with a new reconstruction technique designed to extract cross sections 
extending up to $x=1$ at high $Q^{2}$. The data
correspond to an integrated luminosity of
$38.6$~$\pbi$ for $e^+p$ collisions at $\sqrt{s}=300$~$\gev$ recorded in 96-97, 
$16.7$~$\pbi$ for $e^-p$ collisions  at $\sqrt{s}=318$~$\gev$ recorded in 98-99 and
$65.1$~$\pbi$ for $e^+p$ collisions at $\sqrt{s}=318$~$\gev$  recorded in 99-00.

%%%%%%%%%%%%%%%%%%%%%%%%%%%%%%%%%%%%%%%%%%%%%%%%%%%%%%%%%%%%%%%%%%%%
\section{The ZEUS experiment at HERA}
%%%%%%%%%%%%%%%%%%%%%%%%%%%%%%%%%%%%%%%%%%%%%%%%%%%%%%%%%%%%%%%%%%%
\label{sec-zeus}
ZEUS is a multipurpose detector described elsewhere~\cite{zeus:1993:bluebook}. 
A schematic depiction of the ZEUS detector is given in Fig.~\ref{skzeus}.
A brief outline of the components that are
most relevant for this analysis is given below.

The high-resolution uranium--scintillator calorimeter (CAL)~\citeCAL consists 
of three parts: the forward (FCAL), the barrel (BCAL) and the rear (RCAL)
calorimeters. Each part is divided into modules and further 
subdivided into towers; each tower is
longitudinally segmented into one electromagnetic section (EMC) and either one
(in RCAL) or two (in BCAL and FCAL) hadronic sections (HAC). 
The smallest subdivision of the calorimeter is called a cell. 
The CAL energy resolutions,
measured under test-beam conditions, are $\sigma(E)/E=0.18/\sqrt{E}$ for
electrons and $\sigma(E)/E=0.35/\sqrt{E}$ for hadrons, with $E$ in GeV. The
timing resolution of the CAL is $\sim 1 \ns$ for energy deposits larger than
$4.5$~$\gev$.

Charged particles are tracked in the central tracking detector (CTD)~\citeCTD,
which operates in a magnetic field of $1.43\Tesla$ provided by a thin
superconducting solenoid. The CTD consists of 72~cylindrical drift-chamber
layers, organized in nine superlayers covering the polar-angle\ZcoosysfnA\
region \mbox{$15^\circ<\theta<164^\circ$}. The transverse-momentum resolution
for full-length tracks is $\sigma(p_T)/p_T=0.0058 \,
p_T\oplus0.0065\oplus0.0014/p_T$, with $p_T$ in $\gev$.

The luminosity is measured using the Bethe-Heitler reaction $ep
\rightarrow e\gamma p$ \cite{desy-92-066,*zfp:c63:391,*acpp:b32:2025}.
The resulting small-angle photons were measured by the luminosity
monitor, a lead-scintillator calorimeter placed in the HERA tunnel 107
m from the interaction point in the electron beam direction.

%%%%%%%%%%%%%%%%%%%%%%%%%%%%%%%%%%%%%%%%%%%%%%%%%%%%%%%%%%%%%%%%%%%%
\section{New reconstruction method}
%%%%%%%%%%%%%%%%%%%%%%%%%%%%%%%%%%%%%%%%%%%%%%%%%%%%%%%%%%%%%%%%%%%
\label{sec-method}

Figure~\ref{skzeus} also shows a schematic depiction of 
a high-$Q^2$ NC event in the ZEUS 
detector: a scattered electron and a jet are outlined in the CAL, while the 
proton remnant largely disappears down the forward beam pipe.
The electron is typically scattered at a large angle and is easily recognized
in the detector. 
%Such events can therefore be measured with high acceptance
%independently of Bjorken $x$.
%For large $x$, the jet from the scattered quark is boosted in the forward (proton) direction
%and $\theta_{\rm {jet}}$ decreases. 
%When $x$ is too high, a part of the jet is lost in the beam pipe.
% For $x$ not too high, 
Such events have been analyzed by a variety of techniques in the past, 
such as the double-angle method, all of which limited the maximum value of $x$
which could be measured. In the new techniques presented here, 
the hadronic system can be used to measure $x$ 
by reconstructing the energy and angle of the jet 
produced by the scattered quark. Above some $x$ value that depends
on $Q^2$, the jet is at a small angle and not well reconstructed. 
An integrated cross section
above an $x$ cut value is then measured.

The scattered electron was identified and reconstructed by combining 
calorimeter and CTD information \cite{epj:c11:427}. The algorithm 
starts by identifying CAL clusters topologically consistent
with an electromagnetic shower. If the electron candidate was in the range
$23^\circ < \theta_e < 156^\circ$, 
a well reconstructed matched track was required.
The scattered-electron energy, $E'_e$, was determined from the calorimeter 
energy deposit and was corrected for the energy lost in inactive material 
in front of the CAL. The electron energy resolution was $5\%$ for 
$E'_e > 20$~$\gev$.
The electron angle $\theta_e$ was determined 
using the matched track, when available, and the position of the calorimeter 
cluster and the event vertex if the electron was outside the CTD acceptance.
The electron angular resolution was 2~mrad for $\theta_e < 23^\circ$,
3~mrad for $23^\circ < \theta_e < 156^\circ$ and 
5~mrad for $\theta_e > 156^\circ$~\cite{pr:d70:052001}.

Jets were reconstructed from the remaining clusters with the longitudinally 
invariant $k_{T}$ 
cluster algorithm \cite{np:b406:187} in the inclusive mode~\cite{pr:d48:3160}.
Each cluster energy was corrected for energy loss in dead material,
and clusters identified as backsplash~\cite{epj:c11:427} 
from the FCAL into the BCAL or RCAL were rejected.
 The jet variables were defined according to the
Snowmass convention \cite{proc:snowmass:1990:134}:
\begin{alignat*}{2}
 E_{T, \rm {jet}} = \sum_i E_{T,i}\; ,\;\;\;
 \eta_{\rm {jet}} = \frac{\sum_i E_{T,i} \eta_i}{E_{T,\rm {jet}}},  
 \\ 
 \theta_{\rm {jet}} = 2\tan^{-1}(e^{-\eta_{\rm {jet}}})\; , \;\;\;
 E_{\rm {jet}} = \sum_i E_{i}, \nonumber
\end{alignat*}
where $E_{i}$, $E_{T,i}$ and $\eta_{i}$ are the energy, transverse energy 
and pseudorapidity of the CAL clusters. 
The jet energy and angular resolutions were  
$\sigma_{E_{\rm {jet}}}/E_{\rm {jet}} = 55\%/\sqrt{E_{\rm {jet}}} \oplus 2\%$ and
$\sigma_{\theta_{\rm {jet}}}/\theta_{\rm {jet}} = 1.6\% \oplus 1.9\%/\sqrt{\theta_{\rm {jet}}}$~\cite{epj:c16:253}.

 %These events are nevertheless found with high acceptance and the 
%integrated cross section above an $x$ cut is measured. 
%The $x$ value for which jets are well contained increases as $Q^2$ increases.
%At the $Q^2$ values considered in this analysis, 
%the scattered electron is at large angles and 
%well contained in the detector.
%
%The new method employed in this analysis combines electron and 
%jet information to allow a measurement of the cross section up to $x=1$. 
Events were first sorted into $Q^2$ bins using information 
from the electron only: 
\begin{equation*}
  Q^2=2 E_e E'_e (1+\cos\theta_e), 
  \label{eq-q2}
\end{equation*}  
where $E_e$ is the electron beam energy. 
The electron was well reconstructed in the whole kinematic region, yielding
a relative resolution in $Q^2$ for all $x$ of about $5\%$.
The jet information was then used to calculate $x$ 
for events with a well reconstructed jet:
\begin{equation*}
 x = \frac{E_{\rm {jet}}(1+\cos\theta_{\rm {jet}})}
      {2E_{p}(1-\frac{E_{\rm {jet}}(1-\cos\theta_{\rm {jet}})}{2E_{e}})}, 
  \label{eq-x}
\end{equation*}
where $E_{p}$ is the proton beam energy. 
The relative resolution in $x$ varied from $15\%$ to $4\%$ as 
$x$ increased from 0.06 to 0.7 in events where a jet could be reconstructed. 
At high $x$, $\theta_{\rm {jet}}$ is small and 
$x \approx E_{\rm {jet}}/E_{p}$,
where $E_{\rm {jet}}$ has good resolution.
The events with a reconstructed jet were sorted into $x$ bins to 
allow a measurement of the double differential cross-section 
$d^2\sigma_{\rm Born}/dxdQ^2$.  
Events with no jet reconstructed within the fiducial volume were
assumed to come from high $x$ and were collected in a bin with 
$x_{\rm {edge}}<x<1$. 
%Since these bins are generally
%large and the form of the PDF is not well known in this region, 
An integrated cross section was calculated from
$\int_{x_{\rm {edge}}}^{1} (d^2\sigma_{\rm Born}/dxdQ^2) dx$.  
Due to their poorer $x$ resolution, events with more than one 
jet were discarded.
The correction to the cross section for multi-jet events was taken from the
Monte Carlo simulation described below, and ranged from $9\%$ at $x=0.1$ to
$1\%$ at $x=0.6$.  The systematic uncertainty associated with this cut 
is discussed in Section~\ref{subsec-corsys}.  More details of this 
reconstruction technique are given elsewhere~\cite{thesis:yujin:2006}.

%%%%%%%%%%%%%%%%%%%%%%%%%%%%%%%%%%%%%%%%%%%%%%%%%%%%%%
\section{Monte Carlo simulations}     
%%%%%%%%%%%%%%%%%%%%%%%%%%%%%%%%%%%%%%%%%%%%%%%%%%%%%%
\label{sec-MC}
Monte Carlo (MC) simulations were used to evaluate the efficiency for selecting
events, for determining the accuracy of the kinematic reconstruction, and for
estimating
the background rate. The
statistical uncertainties from the MC samples were negligible in comparison to
those of the data.

 Standard Model (SM) NC DIS events were simulated with 
{\sc DjangoH} version 1.1~\cite{proc:hera:1991:1419,*spi:www:djangoh11}
 which includes an interface to the
{\sc Heracles} 4.6.1~\cite{cpc:69:155} program.
{\sc Heracles} includes the corrections for the initial- and final-state
electroweak radiation, vertex and propagator corrections,
and two-boson exchange.
The hadronic final state was simulated using the {\sc Meps} model of 
{\sc Lepto} 6.5~\cite{cpc:101:108}, which includes order-$\alpha_S$ matrix 
elements (ME) with a lower and upper cutoff on the soft and collinear 
divergences. Both the ME cut-offs and the parton evolutions are
treated by parton showers based on the DGLAP evolution equations.
The fragmentation of the scattered partons into observable hadrons is 
performed with the Lund string hadronization model by 
{\sc Jetset}~\cite{cpc:82:74}. 
The CTEQ6D PDF set~\cite{CTEQ6M}
was used.

The ZEUS detector response was simulated using a program based on {\sc Geant}
3.13~\cite{tech:cern-dd-ee-84-1}. The generated events were passed through the
detector simulation, subjected to the same trigger requirements as the data and
processed by the same reconstruction programs.

The vertex distribution in data is a crucial input to the MC simulation for the
correct evaluation of the event-selection efficiency. Therefore, the $Z$-vertex
distribution used in the MC simulation was determined from a sample of NC DIS
events in which the event-selection efficiency was independent of the $Z$ 
position of the event vertex~\cite{pr:d70:052001}.

%%%%%%%%%%%%%%%%%%%%%%%%%%%%%%%%%%%%%%%%%%%%%%%%%%%%%%
\section{Event selection}     
%%%%%%%%%%%%%%%%%%%%%%%%%%%%%%%%%%%%%%%%%%%%%%%%%%%%%%
\label{Sect:EvtSel}

\subsection{Trigger}
\label{sec-Trigger}

ZEUS operates a three-level trigger 
system~\cite{pr:d70:052001,epj:c32:1}. 
At the first-level, only coarse calorimeter and tracking information 
are available.
Events were selected using criteria based on energy deposits 
in the CAL consistent with an isolated electron. 
In addition, events with high $E_{T}$ in coincidence with a CTD track 
were accepted. 
At the second level, a cut on $\delta > 29$~GeV was imposed to select
 NC events, with $\delta$ defined as:
\begin{equation*}
  \delta \equiv \sum\limits_{i} (E-p_Z)_{i} = \sum\limits_{i} 
   ( E_i - E_i \cos
  \theta_{i} ) \; ,
%  \label{eq-Delta}
\end{equation*} 
  where the sum runs over all calorimeter energy deposits $E_i$ with
  polar angles $\theta_i$.
%was used to select NC DIS 
Timing information from the calorimeter was used 
to reject events inconsistent with the bunch-crossing time. 
At the third level, events were fully reconstructed and selected according
to requirements similar to, but looser than, the
offline cuts described below.

The main uncertainty in the trigger efficiency comes from the first level.
The efficiency in data and MC simulation agreed to within $\sim 0.5\%$ and the
overall efficiency was above $95\%$.

\subsection{Offline selection}
 \label{sec-selec:offline}
The following criteria were imposed to select NC DIS events offline:

\begin{itemize}
    \item an electron with $E'_{e} > 25$~GeV was required. 
      An isolation cut was imposed by requiring that less 
      than 4~GeV be deposited in calorimeter cells not associated with 
      the scattered electron in an $\eta$-$\phi$ cone of radius 
      $R_{\rm cone} = 0.8$ centered on the candidate cluster. 
      For those electrons in the CTD acceptance, a matched track was required  
      which passed within 10~cm of the cluster center.
      The matched track was required to  
      traverse at least four of the nine superlayers of the CTD. 
      The momentum of the track, $p_{\rm {trk}}$, had to be at 
      least 10~GeV. 
      For electrons outside 
      the forward tracking acceptance of the CTD, the tracking
      requirement in the electron selection was replaced by a cut on the
      transverse momentum of the electron, $p_{T,e} > 30$~GeV; 
\item a fiducial-volume cut was applied to the electron to guarantee that
      the experimental acceptance was well understood. 
      It excluded the transition regions
      between the FCAL and the BCAL~\cite{thesis:xliu:2003}. 
      It also excluded the regions within 1.5~cm of the  module gaps in the 
      BCAL.
      As the kinematic region considered in this analysis is at high $Q^2$, 
      events with electrons in the RCAL were discarded. 
\end{itemize}

The following cuts were used to select an essentially background 
free and well reconstructed event sample:

\begin{itemize}
\item either 0 or 1 valid jets.  Valid jets were required to have 
     $E_{T,\rm {jet}}>$~10~GeV and $\theta_{\rm {jet}}>$~0.12~rad;

\item a reconstructed vertex with $-50 < Z_{\rm vtx} < 50$~cm, a range
  consistent with $ep$ interactions;

\item $\delta > 40$~GeV to suppress photoproduction events, in which 
  the scattered electron escaped through the beam hole in the RCAL.
  This cut value was $\delta>47$~GeV for events in the highest $x$ bins.
  This cut also rejected events with large initial-state QED radiation.  
  In addition, $\delta<65$~GeV was required to remove 
 ``overlay'' events in which a 
  normal DIS event coincided with
  additional energy deposits in the RCAL from some other reaction. 
  This requirement had 
  a negligible effect on the efficiency for selecting NC DIS events.

\item $y_e < 0.95$ to further reduce background from photoproduction events, 
  where $y_e$ was defined as

\begin{equation*}
  y_e = 1 - \frac{E'_e}{2E_e}(1 - \cos\theta_e);
  \label{eq-ye}
\end{equation*}
 
\item  $P_T / \sqrt{E_T}< 4\sqrt{\gev}$ to remove cosmic rays and 
  beam-related backgrounds.  The variables $P_T$ and $E_T$ were defined by: 
\begin{eqnarray*}
  P_T^2 & = & P_X^2 + P_Y^2 = ( \sum\limits_{i} E_i \sin \theta_i \cos
    \phi_i )^2+ ( \sum\limits_{i} E_i \sin \theta_i \sin \phi_i)^2, \\ 
  E_T & = & \sum\limits_{i} E_i \sin \theta_i,
\end{eqnarray*}
  where the sums run over all calorimeter energy deposits, $E_i$, with 
  polar and azimuthal angles $\theta_i$ and $\phi_i$ with respect to the 
  event vertex, respectively;

\item $\geq5$ HAC cells with energy above $110$~MeV to remove elastic
  Compton scattering events ($e p \rightarrow e \gamma p$) and further 
  reduce the size of the QED radiative corrections. 
  The contribution from deeply virtual 
  Compton scattering was found to be negligible;

\item $y_{\rm {JB}} < 1.3 \cdot Q^2_{\rm {edge}}/(s \cdot x_{\rm {edge}})$ 
   to limit event migration from small $x$ to large $x$ for zero-jet events. 
   The variable $y_{\rm {JB}}$ was calculated with the 
  Jacquet-Blondel method~\cite{proc:epfacility:1979:391}. 
  The quantities $x_{\rm {edge}}$
  and $Q^2_{\rm {edge}}$ are the lower $x$ and upper $Q^2$ edges of the 
  bins defined for 
  the cross-section measurement (see Section~\ref{sec-Binning}).
\end{itemize}

After these selections, 10298 events remained in the 99-00 $e^+p$ data, 2664
in the 98-99 $e^-p$ data and 5935 in the 96-97 $e^+p$ data in the bins
used to extract the cross sections. The numbers of events in 
the zero-jet bins were 1292, 293 and 493, respectively.

Monte Carlo distributions are compared with those from data in 
Figs.~\ref{kincp}-\ref{0jcp} for several variables.
The MC distributions are normalized to the measured luminosity.  
Only the comparison to 99-00 $e^+p$ data is shown; 
the comparisons of 98-99 $e^-p$ and 96-97 $e^{+}p$ data with MC 
distributions show similar features. 
The first set of plots, Fig.~\ref{kincp}, shows
general properties for the full sample of events.  
Good agreement between data and MC simulation is observed, with
no indication of residual backgrounds. 
The small disagreement observed in Fig.~2c has been verified to have
negligible impact on the results presented in this paper.
Figure~\ref{ecp} shows distributions related to the  scattered electron. 
Figure~\ref{jcp} presents a series of control plots for jet quantities.  
The MC reproduces the data distribution for the number of reconstructed 
jets to high accuracy.  
This is important since the MC is used to correct for the inefficiency 
resulting from the requirement of zero or one jet in the event. 
The remaining distributions in this figure are for the jet quantities 
in one jet events.  
Figure~\ref{0jcp} shows distributions for the class of events with zero jets.
Overall, 13\% more data events for 99-00 $e^+p$, 
2\% more data events for 98-99 $e^-p$ and 
5\% more data events for 96-97 $e^+p$ 
are observed for zero-jet events than expected in the simulation.  
An offset in the $\delta$ distribution is seen, with the MC distribution
slightly lower than the data.
This offset can however be 
explained by shifting the electron energy scale by $1\%$, which 
is within its estimated uncertainty.

%%%%%%%%%%%%%%%%%%%%%%%%%%%%%%%%%%%%%%%%%%%%%%%%%%%%%%%%%%%
\section{Analysis}
\label{Sect:Analysis}
%%%%%%%%%%%%%%%%%%%%%%%%%%%%%%%%%%%%%%%%%%%%%%%%%%%%%%%%%%%
\subsection{Binning, acceptance and cross-section determination}
\label{sec-Binning}
The bins in the ($x,Q^2$) kinematic space used in this analysis are shown in Fig.~\ref{9600bin}.  
The bin widths in $Q^2$ were chosen to 
correspond to three times the resolution of the reconstructed $Q^2$.
The minimum value of $Q^2$ corresponds roughly to the acceptance of the BCAL.
The lower $x$ edge of the bin for zero-jet events, $x_{\rm {edge}}$, was determined
from the condition $\theta_{\rm {jet}}>0.12$~rad. For the bins where a jet was 
reconstructed, the bin widths in $x$ were chosen to correspond to three 
times the resolution of the reconstructed $x$.
%The bin structures are shown in Fig~\ref{9600bin}. 
  
The MC simulation was used to study the $x$ distribution of the zero-jet 
events. 
Figure~\ref{xmig} shows the true $x$ distribution for the 99-00 $e^+p$ 
MC events in different $Q^2$ bins.
Similar distributions are observed in the 98-99 $e^-p$ and 96-97 $e^{+}p$ 
MC. As can be seen in this figure, 
the zero-jet events originate predominantly from the interval $x_{\rm {edge}}<x<1$. 
We note that these distributions depend on the particular PDF chosen 
and that there are uncertainties at large $x$.

The efficiency, defined as the number of events generated and reconstructed
in a bin after all selection cuts divided by the number of
events that were generated in that bin, was typically 40\%. In some
low-$Q^2$ bins, dominated by events in which the electron is scattered into
the RCAL or the B/RCAL transition region and removed by the fiducial cut, 
the efficiency was lower. 
The purity, defined as the number of events
generated and reconstructed in a bin after all selection cuts divided by the
total number of events reconstructed in that bin, was typically 50\%. 
The efficiency and purity in the $(x,Q^{2})$ bins for the 99-00 $e^+p$ simulation 
are shown in Fig.~\ref{ep}. 
The 96-97 $e^+p$ and 98-99 $e^{-}p$ simulations yielded similar values.
The efficiency and purity in zero-jet bins are comparable to those in
the mid-$x$ bins.

The double-differential cross section was determined as
\begin{equation*}
  \frac{d^2 \sigma_{\rm Born}(x, Q^2)}{dx dQ^2} = \frac{N_{\rm data}(\Delta x, \Delta Q^2)}
  {N_{\rm MC}(\Delta x, \Delta Q^2)} (1+\delta(Q^2))\frac{d^2
  \sigma_{\rm Born}^{\rm SM}}{dx dQ^2} ,
%  \label{eq-xsect}
\end{equation*}
and the integrated cross section was determined as
\begin{equation*}
  \int_{x_{\rm {edge}}}^{1} \frac{d^2 \sigma_{\rm Born}(x, Q^2)}{dx dQ^2} dx= 
   \frac{N_{\rm data}(\Delta x, \Delta Q^2)}{N_{\rm MC}(\Delta x, \Delta Q^2)} (1+\delta(Q^2))
  \int_{x_{\rm {edge}}}^{1} \frac{d^2\sigma_{\rm Born}^{\rm SM}}{dx dQ^2}dx ,
%  \label{eq-xsecti}
\end{equation*}

where $N_{\rm data}(\Delta x, \Delta Q^2)$ is the number of data events
in a bin $(\Delta x, \Delta Q^2)$ and $N_{\rm MC}(\Delta x, \Delta Q^2)$ 
is the number of signal MC events normalized to the luminosity of the data. 
The SM prediction, $d^2 \sigma^{\rm SM}_{\rm Born}(x, Q^2)/dx dQ^2$, 
was evaluated according to Eq.~\eq{Born} with the same PDF and 
electroweak (EW) parameters as used in the MC simulation.  This procedure 
implicitly takes the acceptance, bin-centering and leading-order 
radiative corrections from the MC simulation. The variation of the cross 
sections resulting from different choices of PDF in the MC are described in 
the next section.  The values of $(x,Q^2)$ at which the cross sections 
are quoted
are given in Tables~\ref{tab:9697cros}-\ref{tab:9900crosl}.

Monte Carlo studies indicated that the radiative corrections have little dependence
on $x$ for the kinematic reconstruction method used here. 
The correction for higher-order radiative effects, $\delta(Q^2)$,
calculated from HECTOR~\cite{cpc:94:128} varied from $3\%$ at low $Q^2$ 
to $0\%$ at high $Q^2$.

% ----------------------------------------------------------------------------
%       Systematic errors
% ----------------------------------------------------------------------------
\subsection{Systematic uncertainties}
\label{sec-SysErr}

Systematic uncertainties associated with the MC simulations were estimated by
re-calculating the cross section after modifying the simulation to account for
known uncertainties. Cut values were varied where this method
was not applicable.

\subsubsection{Uncorrelated systematic uncertainties}

The following systematic uncertainties are either small or exhibit 
no bin-to-bin correlations:
\begin{itemize}
\item electron energy resolution in the MC simulation. The effect on the cross 
  sections was evaluated by changing the resolution by $\pm 1\%$ in the MC. 
  This resulted in $\pm 1\%$ effects over almost the full kinematic range. 
  The effect increased to $\pm 2\%$ for the double-differential cross 
  section in several low $Q^{2}$ bins and for several integrated cross section 
   bins;
\item electron angle. Uncertainties in the electron scattering-angle 
  determination are known to be at most $1 \mrad$~\cite{epj:c28:175}. 
  The resulting systematic 
  effects on the cross-section measurement were at most $2\%$;
\item electron-isolation requirement. Variation of the
  electron-isolation energy by $\pm2 \gev$ caused negligible effects 
in the measured
  cross section in the low-$Q^2$ region and $2.5\%$ in the high-$Q^2$ region;
\item FCAL alignment. The FCAL jet position was varied by 
  $\pm 0.5 \cm$ in both $X$ and $Y$ directions. The resulting
  changes in the cross sections were negligible;
\item reconstructed-vertex uncertainty. The cut on the reconstructed 
  $Z$ vertex was changed by $\pm 2$~cm;
  The uncertainties in the cross sections associated with
  this variation were negligible over the full kinematic range;
\item background uncertainty.  The estimated background from all sources was 
less than
  $1\%$ and gave negligible uncertainty.

\end{itemize}

\subsubsection{Correlated systematic uncertainties}
\label{subsec-corsys}
The significant correlated systematic uncertainties are listed below and labeled for
further reference.  They were determined to result from the following sources:
\begin{itemize}
\item $\{\delta_1\}$ electron-energy scale. The systematic uncertainty 
  resulting from uncertainty in the
  electron energy scale was checked by changing the energy scale by $\pm1\%$.
  This resulted in typically
  $2\%$ systematic variations in the cross sections;
  \item $\{\delta_2\}$ jet-energy scale. The uncertainty in the cross sections 
  arising from the measurement of the jet energy was checked by changing 
  the energy scale by $\pm1\%$. The effect in the highest-$x$ bins was 
  negligible
  over the full $Q^{2}$ region. The uncertainty in the double-differential 
  cross-section bins was $0-7\%$ for $0.1 < x < 0.7$;

\item $\{\delta_3\}$ FCAL first inner ring (FIR) EMC energy scale. 
  The effect of the FIR EMC energy scale uncertainty on the cross section was 
  checked by 
  changing the energy scale by $\pm5\%$, which gave $0-3.5\%$ 
  uncertainty as $x$ increased from $0.1$ to $0.9$;

\item $\{\delta_4\}$ different PDFs. The uncertainty in the extracted cross 
   section resulting from uncertainties in the shape of the PDFs at high $x$ was checked
   by comparing the cross sections calculated from different sets: CTEQ4D, 
   CTEQ6D, MRST99, ZEUS-S and ZEUS-JETS. The effect was
   less than $1\%$ at low $x$ and increased to $5\%$ at high $x$;
 
\item $\{\delta_5\}$ simulation of the hadronic final state and jet-selection 
   procedure. The invariant $k_{T}$ jet algorithm was replaced with a cone 
   algorithm~\cite{proc:snowmass:1990:134} with cone radius $0.7$, and cross 
  sections were re-evaluated. The uncertainty was found to be $\pm 1.6\%$ 
  in the highest $x$ bins and $\pm2.5\%$ in the lower $x$ bins. In addition,
  the analysis was redone under the following conditions: including multijet
  events for the events with $x<x_{\rm{edge}}$; varying the jet $E_T$ and
  $\theta_{\rm jet}$ cuts for the jet selection; and varying the $y_{\rm {JB}}$ 
  cut. These checks produced small differences consistent with expected
  statistical variations and were not included in the systematic uncertainty;

\item $\{\delta_6\}$ higher order radiative corrections and a possible 
  dependence on $x$. The uncertainty 
  was estimated to be about $2\%$ at low $x$, increasing 
  to $12\%$ at $x > 0.8$;

\item $\{\delta_7\}$  uncertainties on the luminosity. 
  The uncertainties for the 96-97 $e^+p$ sample, 
   98-99 $e^-p$ sample and 99-00 $e^+p$ sample are
  1.6\%, 1.8\% and 2.25\%, respectively.
\end{itemize}

%%%%%%%%%%%%%%%%%%%%%%%%%%%%%%%%%%%%%%%%%%%%%%%%%%%%%%%%%%%%%%%%%%%%
\subsection{Results}
%%%%%%%%%%%%%%%%%%%%%%%%%%%%%%%%%%%%%%%%%%%%%%%%%%%%%%%%%%%%%%%%%%%
\label{sec-cross}

The measured Born level cross sections for 96-97 $e^+p$, 98-99 $e^-p$ 
and 99-00 $e^+p$ and their systematic uncertainties are shown in 
Tables~\ref{tab:9697cros}-\ref{tab:9900crosl}. The statistical uncertainties
on the cross sections correspond to the central $68\%$ probability interval 
evaluated using a Bayesian approach with
flat prior and a Poisson likelihood. 
For bins with zero measured events, a $68\%$ probability limit,
calculated including the uncorrelated systematic uncertainty, is given. 
The cross sections 
are shown in Figs.~\ref{cros96}, \ref{cros98} and \ref{cros00} 
and compared to SM expectations using
the CTEQ6M PDFs~\cite{CTEQ6M}. 
The double differential cross sections are
represented by solid points, and generally agree well with the expectations.  
The cross section in
the highest-$x$ bins is plotted as
$$\frac{1}{1-x_{\rm {edge}}}\int_{x_{\rm {edge}}}^{1}\frac{d^2\sigma_{\rm Born}}{dxdQ^2}dx 
\;\; .$$
In these bins, the expected cross section is drawn as a horizontal line, 
while the measured cross section is displayed as the open symbol
at the center of the bin.  
The error bars represent the quadratic sum of the correlated systematic
uncertainty and the combined statistical and uncorrelated
systematic uncertainty determined from the Bayesian probability analysis.

The ratios of the measured cross sections to the SM expectation using 
the CTEQ6M PDFs for 96-97 $e^+p$ , 98-99 $e^-p$ and 99-00 $e^+p$
are shown in Figs.~\ref{ratio96}, \ref{ratio98} and \ref{ratio00} 
respectively.  
The ratios of the expectations using the ZEUS-S PDF~\cite{ZEUS-S}
to that using CTEQ6M and 
ZEUS-JETS PDF~\cite{ZEUS-JET} to that using CTEQ6M are also shown.  
The uncertainty for the CTEQ6M fit is displayed in the figure as a 
shaded band.
The measured double-differential cross sections generally agree well
with all three sets of expectations. 
For the highest $x$ bins, which extend to previously unmeasured
kinematic ranges, the data have a tendency to lie above the expectations.  

The data presented here, specifically the zero-jet data at high $x$, extend
the kinematic coverage for DIS.
These results are expected to have a significant impact on the
valence-quark distributions at high $x$, where little data are available to
date. It should however be noted that there is overlap with the data
presented in previous ZEUS publications, and these new results should
therefore not be used simultaneously with the previously published ZEUS
data~\cite{epj:c21:443,epj:c28:175,pr:d70:052001} in fits to extract 
model parameters. In the kinematic region of overlap of this technique
 with the previous ZEUS technique, the extracted cross sections are 
in excellent agreement. 

%%%%%%%%%%%%%%%%%%%%%%%%%%%%%%%%%%%%%%%%%%%%%%%%%%%%%%%%%%%
\section{Summary}
\label{Sect:Summary}
%%%%%%%%%%%%%%%%%%%%%%%%%%%%%%%%%%%%%%%%%%%%%%%%%%%%%%%%%%%
This paper has presented a reanalysis of previously published ZEUS data with 
a new technique designed for the reconstruction of large $x$ events, which  
allows for the extraction of the cross section up to $x=1$. 
In the previously measured kinematic region, the data and simulation
based on the CTEQ6M PDF are in good agreement.
The Standard Model predictions tend to underestimate the 
data at the highest values of $x$.

%%%%%%%%%%%%%%%%%%%%%%%%%%%%%%%%%%%%%%%%%%%%%%%%%%%%%%%%%
\section{Acknowledgments}
%%%%%%%%%%%%%%%%%%%%%%%%%%%%%%%%%%%%%%%%%%%%%%%%%%%%%%%%%
We are grateful to the DESY directorate for their strong support and
encouragement. We thank the HERA machine group whose outstanding efforts
allowed the measurements presented here. We also
wish to express our thanks 
for the support of the DESY computing and network services. The 
design, construction and installation of the ZEUS detector has 
been made possible by the efforts of many people not listed as authors.

%% file: DESY-06-116-ref.tex
{
\def\bibname{\Large\bf References}
\def\refname{\Large\bf References}
\pagestyle{plain}
\ifzeusbst
  \bibliographystyle{./BiBTeX/bst/l4z_default}
\fi
\ifzdrftbst
  \bibliographystyle{./BiBTeX/bst/l4z_draft}
\fi
\ifzbstepj
  \bibliographystyle{./BiBTeX/bst/l4z_epj}
\fi
\ifzbstnp
  \bibliographystyle{./BiBTeX/bst/l4z_np}
\fi
\ifzbstpl
  \bibliographystyle{./BiBTeX/bst/l4z_pl}
\fi
{\raggedright
\bibliography{./BiBTeX/user/syn.bib,%
              ./BiBTeX/bib/l4z_articles.bib,%
              ./BiBTeX/bib/l4z_books.bib,%
              ./BiBTeX/bib/l4z_conferences.bib,%
              ./BiBTeX/bib/l4z_h1.bib,%
              ./BiBTeX/bib/l4z_misc.bib,%
              ./BiBTeX/bib/l4z_old.bib,%
              ./BiBTeX/bib/l4z_preprints.bib,%
              ./BiBTeX/bib/l4z_replaced.bib,%
              ./BiBTeX/bib/l4z_temporary.bib,%
              ./BiBTeX/bib/l4z_zeus.bib}}
}
\vfill\eject

%% file: DESY-06-116-tab.tex
%------------------------------------------------------------------------------
\input{DESY-06-116-tab-1}
\clearpage
\input{DESY-06-116-tab-2}

\clearpage
\input{DESY-06-116-tab-3}
\clearpage
\input{DESY-06-116-tab-4}

\clearpage
\input{DESY-06-116-tab-5}
\clearpage
\input{DESY-06-116-tab-6}

\clearpage

%% file: DESY-06-116-tab-1.tex
\begin{center}
\footnotesize
\renewcommand{\arraystretch}{1.2}
\tablehead{
\multicolumn{13}{l}{
{\normalsize}}\\
\hline
{$Q^2$} & 
\multicolumn{1}{c|}{$x$} & 
${d^2\sigma}/{dxdQ^2}$  &
$N$& 
$\delta_s$& 
$\delta_t$& 
$\delta_u$& 
$\delta_1$ &
$\delta_2$ &
$\delta_3$ &
$\delta_4$ &
$\delta_5$ &
$\delta_6$ \\
{($\Gev^2$)} & 
$$ &
{($\rm pb/GeV^{2}$)} &
$$ &
(\%) &
(\%) &
(\%) &
(\%) &
(\%) &
(\%) &
(\%) &
(\%) &
(\%) \\
\hline \hline}
\tablelasttail{
\hline
\multicolumn{11}{r}{} \\}
\normalsize
\bottomcaption{
    The cross-section table for 96-97 $e^+p$ NC scattering.
    The first two columns of the table contain the $Q^2$ and $x$ values 
    at which the cross section is quoted, the third contains the measured
    cross section ${d^2\sigma}/{dxdQ^2}$ corrected to the electroweak 
    Born level or the upper limit in case of zero observed 
    events, the fourth contains the number of events reconstructed in 
    the bin, $N$, the fifth contains the 
    statistical uncertainty, $\delta_s$, and the sixth contains the 
     total systematic uncertainty, $\delta_t$. 
    The right part of the table lists the total uncorrelated
    systematic uncertainty, $\delta_u$, followed by the bin-to-bin correlated
    systematic uncertainties $\delta_1$--\,$\delta_6$ defined in
    the text.
    The upper (lower) numbers refer to the variation of 
    the cross section, whereas the signs of
    the numbers reflect the direction of change in the cross
    sections. Note that the normalization uncertainty, $\delta_7$ is
    not listed.
    } 
\label{tab:9697cros}
\begin{supertabular}{|c|l|c|c|c|c||c|c|c|c|c|c|c|}
$ 648$ & $0.09$ & $2.66$ & $ 114 $ & $^{+10}_{-8.5}$ & $ ^{+7.9}_{-7.8} $ & $^{+2.0}_{-1.6} $ & $^{+6.0}_{-6.1} $ & $^{-0.3}_{+1.2} $ & $^{-0.2}_{+0.1} $ & $^{+1.4}_{-1.4}$ & $^{+2.4}_{-2.4}$ & $^{+3.3}_{-3.3}$\\
$ $ & $0.15$ & $1.08$ & $ 40 $ & $^{+19}_{-13}$ & $ ^{+8.3}_{-8.7} $ & $^{+2.1}_{-2.2} $ & $^{+6.4}_{-6.7} $ & $^{-1.3}_{+0.4} $ & $^{+1.6}_{-2.1} $ & $^{+0.9}_{-0.9}$ & $^{+2.4}_{-2.4}$ & $^{+3.3}_{-3.3}$\\
$ $ & $0.21$ & $7.97\cdot 10^{-1}$ & $ 33 $ & $^{+21}_{-15}$ & $ ^{+11}_{-8.9} $ & $^{+3.6}_{-2.0} $ & $^{+9.2}_{-8.1} $ & $^{-0.4}_{+2.5} $ & $^{+2.7}_{-0.2} $ & $^{+0.5}_{-0.5}$ & $^{+2.4}_{-2.4}$ & $^{+1.0}_{-1.0}$\\
$ 761$ & $0.06$ & $3.06$ & $ 299 $ & $^{+6.1}_{-5.4}$ & $ ^{+5.0}_{-4.8} $ & $^{+1.3}_{-0.7} $ & $^{+1.0}_{-0.6} $ & $^{-0.4}_{-0.1} $ & $^{-0.4}_{+0.2} $ & $^{+1.7}_{-1.7}$ & $^{+2.4}_{-2.4}$ & $^{+3.3}_{-3.3}$\\
$ $ & $0.11$ & $1.72$ & $ 187 $ & $^{+7.9}_{-6.8}$ & $ ^{+5.3}_{-5.6} $ & $^{+0.2}_{-1.0} $ & $^{+2.4}_{-3.0} $ & $^{-0.9}_{+1.5} $ & $^{-0.0}_{-0.3} $ & $^{+1.3}_{-1.3}$ & $^{+2.4}_{-2.4}$ & $^{+3.3}_{-3.3}$\\
$ $ & $0.16$ & $8.11\cdot 10^{-1}$ & $ 92 $ & $^{+12}_{-9.4}$ & $ ^{+6.2}_{-8.2} $ & $^{+0.8}_{-2.1} $ & $^{+4.2}_{-6.1} $ & $^{-1.4}_{-0.3} $ & $^{+0.3}_{-2.0} $ & $^{+0.8}_{-0.8}$ & $^{+2.4}_{-2.4}$ & $^{+3.3}_{-3.3}$\\
$ $ & $0.23$ & $6.09\cdot 10^{-1}$ & $ 88 $ & $^{+12}_{-9.6}$ & $ ^{+6.7}_{-4.5} $ & $^{+0.9}_{-1.5} $ & $^{+4.8}_{-2.8} $ & $^{-0.6}_{+2.6} $ & $^{+1.9}_{+0.5} $ & $^{+0.6}_{-0.6}$ & $^{+2.4}_{-2.4}$ & $^{+1.0}_{-1.0}$\\
$ 891$ & $0.07$ & $1.89$ & $ 314 $ & $^{+6.0}_{-5.4}$ & $ ^{+4.8}_{-4.7} $ & $^{+1.0}_{-0.3} $ & $^{-0.8}_{+0.8} $ & $^{-0.3}_{+0.3} $ & $^{-0.3}_{+0.3} $ & $^{+1.6}_{-1.6}$ & $^{+2.4}_{-2.4}$ & $^{+3.3}_{-3.3}$\\
$ $ & $0.12$ & $9.96\cdot 10^{-1}$ & $ 214 $ & $^{+7.3}_{-6.4}$ & $ ^{+4.8}_{-4.6} $ & $^{+1.2}_{-0.6} $ & $^{+1.1}_{+0.2} $ & $^{-0.9}_{+0.2} $ & $^{+0.0}_{-0.3} $ & $^{+1.2}_{-1.2}$ & $^{+2.4}_{-2.4}$ & $^{+3.3}_{-3.3}$\\
$ $ & $0.18$ & $5.96\cdot 10^{-1}$ & $ 145 $ & $^{+9.1}_{-7.6}$ & $ ^{+4.8}_{-4.7} $ & $^{+1.4}_{-0.8} $ & $^{+0.4}_{+0.4} $ & $^{-1.1}_{+0.9} $ & $^{+0.9}_{-1.1} $ & $^{+0.7}_{-0.7}$ & $^{+2.4}_{-2.4}$ & $^{+3.3}_{-3.3}$\\
$ $ & $0.25$ & $2.92\cdot 10^{-1}$ & $ 91 $ & $^{+12}_{-9.4}$ & $ ^{+4.0}_{-5.3} $ & $^{+0.7}_{-1.1} $ & $^{+0.5}_{-3.0} $ & $^{-1.9}_{+1.5} $ & $^{+1.6}_{-2.0} $ & $^{+0.8}_{-0.8}$ & $^{+2.4}_{-2.4}$ & $^{+1.0}_{-1.0}$\\
$1045$ & $0.08$ & $1.16$ & $ 251 $ & $^{+6.8}_{-5.9}$ & $ ^{+4.8}_{-4.8} $ & $^{+0.6}_{-0.4} $ & $^{-1.2}_{+1.0} $ & $^{-0.5}_{+0.2} $ & $^{-0.2}_{+0.5} $ & $^{+1.6}_{-1.6}$ & $^{+2.4}_{-2.4}$ & $^{+3.3}_{-3.3}$\\
$ $ & $0.13$ & $6.10\cdot 10^{-1}$ & $ 182 $ & $^{+8.0}_{-6.9}$ & $ ^{+4.6}_{-4.8} $ & $^{+0.7}_{-1.1} $ & $^{-0.8}_{+0.2} $ & $^{-1.0}_{+0.8} $ & $^{-0.2}_{-0.3} $ & $^{+1.1}_{-1.1}$ & $^{+2.4}_{-2.4}$ & $^{+3.3}_{-3.3}$\\
$ $ & $0.19$ & $3.79\cdot 10^{-1}$ & $ 143 $ & $^{+9.1}_{-7.7}$ & $ ^{+5.4}_{-4.9} $ & $^{+1.3}_{-0.8} $ & $^{-1.7}_{+2.1} $ & $^{+0.2}_{+0.7} $ & $^{+1.9}_{-1.2} $ & $^{+0.6}_{-0.6}$ & $^{+2.4}_{-2.4}$ & $^{+3.3}_{-3.3}$\\
$ $ & $0.27$ & $1.94\cdot 10^{-1}$ & $ 94 $ & $^{+11}_{-9.3}$ & $ ^{+4.4}_{-4.4} $ & $^{+2.0}_{-0.5} $ & $^{-0.3}_{+1.1} $ & $^{-2.9}_{+1.8} $ & $^{+1.0}_{-1.0} $ & $^{+0.9}_{-0.9}$ & $^{+2.4}_{-2.4}$ & $^{+1.0}_{-1.0}$\\
\hline
\end{supertabular}
\end{center}

\begin{center}
\footnotesize
\renewcommand{\arraystretch}{1.2}
\tablehead{
\multicolumn{13}{l}{
{\normalsize {\bf Table \thetable\ } (continued):}}\\
\hline
{$Q^2$} & 
\multicolumn{1}{c|}{$x$} & 
${d^2\sigma}/{dxdQ^2}$  &
$N$& 
$\delta_s$& 
$\delta_t$& 
$\delta_u$& 
$\delta_1$ &
$\delta_2$ &
$\delta_3$ &
$\delta_4$ &
$\delta_5$ &
$\delta_6$ \\
{($\Gev^2$)} & 
$$ &
{($\rm pb/GeV^{2}$)} &
$$ &
(\%) &
(\%) &
(\%) &
(\%) &
(\%) &
(\%) &
(\%) &
(\%) &
(\%) \\
\hline \hline}
\tabletail{
\hline
}
\tablelasttail{\hline}
\normalsize
\begin{supertabular}{|c|l|c|c|c|c||c|c|c|c|c|c|c|}
$1224$ & $0.09$ & $7.98\cdot 10^{-1}$ & $ 233 $ & $^{+7.0}_{-6.2}$ & $ ^{+4.8}_{-5.0} $ & $^{+0.0}_{-0.7} $ & $^{-1.6}_{+1.3} $ & $^{-0.2}_{-0.0} $ & $^{-0.5}_{+0.3} $ & $^{+1.6}_{-1.6}$ & $^{+2.4}_{-2.4}$ & $^{+3.3}_{-3.3}$\\
$ $ & $0.14$ & $4.26\cdot 10^{-1}$ & $ 169 $ & $^{+8.3}_{-7.2}$ & $ ^{+5.1}_{-4.6} $ & $^{+1.0}_{-0.4} $ & $^{-0.4}_{+2.2} $ & $^{-1.1}_{+0.9} $ & $^{+0.1}_{-0.4} $ & $^{+0.9}_{-0.9}$ & $^{+2.4}_{-2.4}$ & $^{+3.3}_{-3.3}$\\
$ $ & $0.21$ & $2.04\cdot 10^{-1}$ & $ 91 $ & $^{+12}_{-9.4}$ & $ ^{+3.9}_{-3.9} $ & $^{+1.8}_{-0.4} $ & $^{+1.0}_{+0.2} $ & $^{-0.4}_{-0.6} $ & $^{+1.1}_{-2.2} $ & $^{+0.6}_{-0.6}$ & $^{+2.4}_{-2.4}$ & $^{+1.0}_{-1.0}$\\
$ $ & $0.30$ & $1.25\cdot 10^{-1}$ & $ 78 $ & $^{+13}_{-10}$ & $ ^{+4.7}_{-4.1} $ & $^{+0.6}_{-1.7} $ & $^{-1.7}_{+0.6} $ & $^{-0.7}_{+2.7} $ & $^{+1.8}_{-0.1} $ & $^{+1.1}_{-1.1}$ & $^{+2.4}_{-2.4}$ & $^{+1.0}_{-1.0}$\\
$1431$ & $0.06$ & $7.68\cdot 10^{-1}$ & $ 170 $ & $^{+8.3}_{-7.1}$ & $ ^{+5.0}_{-5.0} $ & $^{+1.0}_{-0.5} $ & $^{-1.1}_{+1.4} $ & $^{-0.8}_{-0.3} $ & $^{-0.2}_{-0.4} $ & $^{+1.8}_{-1.8}$ & $^{+2.4}_{-2.4}$ & $^{+3.3}_{-3.3}$\\
$ $ & $0.10$ & $3.83\cdot 10^{-1}$ & $ 126 $ & $^{+9.7}_{-8.1}$ & $ ^{+5.1}_{-4.8} $ & $^{+1.0}_{-1.2} $ & $^{-0.7}_{+1.6} $ & $^{+1.0}_{+0.4} $ & $^{-0.1}_{+0.5} $ & $^{+1.5}_{-1.5}$ & $^{+2.4}_{-2.4}$ & $^{+3.3}_{-3.3}$\\
$ $ & $0.16$ & $2.27\cdot 10^{-1}$ & $ 112 $ & $^{+10}_{-8.6}$ & $ ^{+4.6}_{-5.0} $ & $^{+0.5}_{-0.3} $ & $^{-1.0}_{+0.5} $ & $^{-1.8}_{+0.8} $ & $^{-0.2}_{-0.6} $ & $^{+0.8}_{-0.8}$ & $^{+2.4}_{-2.4}$ & $^{+3.3}_{-3.3}$\\
$ $ & $0.23$ & $1.19\cdot 10^{-1}$ & $ 70 $ & $^{+13}_{-11}$ & $ ^{+4.3}_{-3.7} $ & $^{+1.9}_{-0.9} $ & $^{-1.0}_{+2.0} $ & $^{-0.0}_{-0.2} $ & $^{+1.1}_{-1.3} $ & $^{+0.7}_{-0.7}$ & $^{+2.4}_{-2.4}$ & $^{+1.0}_{-1.0}$\\
$ $ & $0.32$ & $7.17\cdot 10^{-2}$ & $ 59 $ & $^{+15}_{-11}$ & $ ^{+4.6}_{-4.1} $ & $^{+0.8}_{-0.4} $ & $^{-0.3}_{+0.6} $ & $^{-2.0}_{+2.3} $ & $^{+1.8}_{-1.1} $ & $^{+1.4}_{-1.4}$ & $^{+2.4}_{-2.4}$ & $^{+1.0}_{-1.0}$\\
$1672$ & $0.07$ & $4.21\cdot 10^{-1}$ & $ 126 $ & $^{+9.7}_{-8.1}$ & $ ^{+5.3}_{-5.0} $ & $^{+1.6}_{-0.5} $ & $^{-1.6}_{+1.9} $ & $^{-0.1}_{+0.1} $ & $^{-0.1}_{-0.1} $ & $^{+1.8}_{-1.8}$ & $^{+2.4}_{-2.4}$ & $^{+3.3}_{-3.3}$\\
$ $ & $0.11$ & $3.25\cdot 10^{-1}$ & $ 144 $ & $^{+9.1}_{-7.7}$ & $ ^{+5.0}_{-5.0} $ & $^{+0.6}_{-0.7} $ & $^{-2.0}_{+2.0} $ & $^{-0.3}_{+0.3} $ & $^{-0.3}_{+0.1} $ & $^{+1.3}_{-1.3}$ & $^{+2.4}_{-2.4}$ & $^{+3.3}_{-3.3}$\\
$ $ & $0.17$ & $1.35\cdot 10^{-1}$ & $ 80 $ & $^{+13}_{-10}$ & $ ^{+4.5}_{-4.9} $ & $^{+0.4}_{-1.4} $ & $^{-1.3}_{-0.2} $ & $^{-0.8}_{+0.5} $ & $^{+0.2}_{-0.1} $ & $^{+0.7}_{-0.7}$ & $^{+2.4}_{-2.4}$ & $^{+3.3}_{-3.3}$\\
$ $ & $0.25$ & $7.96\cdot 10^{-2}$ & $ 55 $ & $^{+15}_{-12}$ & $ ^{+3.3}_{-4.8} $ & $^{+0.3}_{-2.4} $ & $^{-2.0}_{-0.1} $ & $^{-1.5}_{+0.6} $ & $^{+0.6}_{-1.2} $ & $^{+0.8}_{-0.8}$ & $^{+2.4}_{-2.4}$ & $^{+1.0}_{-1.0}$\\
$ $ & $0.35$ & $4.27\cdot 10^{-2}$ & $ 45 $ & $^{+17}_{-13}$ & $ ^{+6.4}_{-4.8} $ & $^{+4.1}_{-1.0} $ & $^{-1.7}_{+1.5} $ & $^{-2.1}_{+2.7} $ & $^{+1.5}_{-1.5} $ & $^{+1.8}_{-1.8}$ & $^{+2.4}_{-2.4}$ & $^{+1.0}_{-1.0}$\\
$1951$ & $0.08$ & $3.32\cdot 10^{-1}$ & $ 130 $ & $^{+9.6}_{-8.1}$ & $ ^{+4.9}_{-5.4} $ & $^{+0.5}_{-1.5} $ & $^{-2.2}_{+1.2} $ & $^{-0.5}_{+0.6} $ & $^{-0.1}_{-0.1} $ & $^{+1.8}_{-1.8}$ & $^{+2.4}_{-2.4}$ & $^{+3.3}_{-3.3}$\\
$ $ & $0.13$ & $1.66\cdot 10^{-1}$ & $ 86 $ & $^{+12}_{-9.7}$ & $ ^{+4.7}_{-4.8} $ & $^{+0.9}_{-0.9} $ & $^{-1.2}_{+1.1} $ & $^{-0.6}_{-0.6} $ & $^{-0.3}_{+0.1} $ & $^{+1.2}_{-1.2}$ & $^{+2.4}_{-2.4}$ & $^{+3.3}_{-3.3}$\\
$ $ & $0.19$ & $8.47\cdot 10^{-2}$ & $ 66 $ & $^{+14}_{-11}$ & $ ^{+5.3}_{-4.9} $ & $^{+0.9}_{-1.4} $ & $^{-1.6}_{+2.6} $ & $^{-0.7}_{+0.9} $ & $^{+0.1}_{-0.2} $ & $^{+0.7}_{-0.7}$ & $^{+2.4}_{-2.4}$ & $^{+3.3}_{-3.3}$\\
$ $ & $0.27$ & $4.53\cdot 10^{-2}$ & $ 41 $ & $^{+18}_{-13}$ & $ ^{+4.2}_{-3.4} $ & $^{+1.6}_{-0.4} $ & $^{-0.3}_{+1.9} $ & $^{-0.6}_{+0.8} $ & $^{+0.9}_{-0.6} $ & $^{+1.0}_{-1.0}$ & $^{+2.4}_{-2.4}$ & $^{+1.0}_{-1.0}$\\
$ $ & $0.38$ & $2.69\cdot 10^{-2}$ & $ 35 $ & $^{+20}_{-14}$ & $ ^{+4.8}_{-4.8} $ & $^{+0.3}_{-1.4} $ & $^{-1.3}_{-0.3} $ & $^{-1.8}_{+2.4} $ & $^{+1.8}_{-1.4} $ & $^{+2.2}_{-2.2}$ & $^{+2.4}_{-2.4}$ & $^{+1.0}_{-1.0}$\\
$2273$ & $0.09$ & $1.88\cdot 10^{-1}$ & $ 88 $ & $^{+12}_{-9.6}$ & $ ^{+4.8}_{-5.4} $ & $^{+0.3}_{-1.5} $ & $^{-2.0}_{+1.1} $ & $^{-0.9}_{+0.4} $ & $^{-0.4}_{-0.0} $ & $^{+1.7}_{-1.7}$ & $^{+2.4}_{-2.4}$ & $^{+3.3}_{-3.3}$\\
$ $ & $0.14$ & $8.43\cdot 10^{-2}$ & $ 60 $ & $^{+15}_{-11}$ & $ ^{+5.0}_{-5.0} $ & $^{+0.4}_{-0.7} $ & $^{-2.1}_{+2.1} $ & $^{+0.8}_{-0.5} $ & $^{-0.2}_{+0.2} $ & $^{+1.0}_{-1.0}$ & $^{+2.4}_{-2.4}$ & $^{+3.3}_{-3.3}$\\
$ $ & $0.21$ & $6.76\cdot 10^{-2}$ & $ 59 $ & $^{+15}_{-11}$ & $ ^{+3.5}_{-3.9} $ & $^{+0.4}_{-1.0} $ & $^{-0.9}_{+0.1} $ & $^{-1.9}_{+1.4} $ & $^{+0.1}_{-0.4} $ & $^{+0.7}_{-0.7}$ & $^{+2.4}_{-2.4}$ & $^{+1.0}_{-1.0}$\\
$ $ & $0.30$ & $3.05\cdot 10^{-2}$ & $ 37 $ & $^{+19}_{-14}$ & $ ^{+4.0}_{-3.8} $ & $^{+1.4}_{-0.7} $ & $^{-0.3}_{+0.9} $ & $^{-1.3}_{+1.3} $ & $^{+0.6}_{-1.1} $ & $^{+1.2}_{-1.2}$ & $^{+2.4}_{-2.4}$ & $^{+1.0}_{-1.0}$\\
$ $ & $0.41$ & $1.56\cdot 10^{-2}$ & $ 25 $ & $^{+24}_{-16}$ & $ ^{+6.9}_{-5.2} $ & $^{+2.2}_{-0.7} $ & $^{+1.4}_{+1.0} $ & $^{-1.8}_{+2.6} $ & $^{+3.1}_{-0.5} $ & $^{+2.7}_{-2.7}$ & $^{+2.4}_{-2.4}$ & $^{+2.7}_{-2.7}$\\
$2644$ & $0.06$ & $1.78\cdot 10^{-1}$ & $ 71 $ & $^{+13}_{-11}$ & $ ^{+5.4}_{-5.5} $ & $^{+0.5}_{-1.3} $ & $^{-2.1}_{+2.5} $ & $^{-0.5}_{+0.0} $ & $^{-0.1}_{-0.6} $ & $^{+2.0}_{-2.0}$ & $^{+2.4}_{-2.4}$ & $^{+3.3}_{-3.3}$\\
$ $ & $0.11$ & $9.53\cdot 10^{-2}$ & $ 53 $ & $^{+16}_{-12}$ & $ ^{+4.7}_{-4.8} $ & $^{+0.6}_{-0.7} $ & $^{-1.2}_{+0.4} $ & $^{+0.0}_{+0.7} $ & $^{-0.2}_{+0.1} $ & $^{+1.6}_{-1.6}$ & $^{+2.4}_{-2.4}$ & $^{+3.3}_{-3.3}$\\
$ $ & $0.16$ & $6.08\cdot 10^{-2}$ & $ 53 $ & $^{+16}_{-12}$ & $ ^{+4.9}_{-4.8} $ & $^{+0.8}_{-0.5} $ & $^{-1.5}_{+1.8} $ & $^{-0.6}_{+0.5} $ & $^{-0.4}_{-0.1} $ & $^{+0.9}_{-0.9}$ & $^{+2.4}_{-2.4}$ & $^{+3.3}_{-3.3}$\\
$ $ & $0.23$ & $3.70\cdot 10^{-2}$ & $ 41 $ & $^{+18}_{-13}$ & $ ^{+3.4}_{-3.7} $ & $^{+0.7}_{-0.7} $ & $^{-0.6}_{+0.6} $ & $^{-1.6}_{+0.7} $ & $^{+0.3}_{-0.4} $ & $^{+0.7}_{-0.7}$ & $^{+2.4}_{-2.4}$ & $^{+1.0}_{-1.0}$\\
$ $ & $0.33$ & $1.54\cdot 10^{-2}$ & $ 24 $ & $^{+25}_{-17}$ & $ ^{+5.6}_{-4.9} $ & $^{+3.9}_{-1.9} $ & $^{-2.3}_{+1.1} $ & $^{-0.5}_{+1.2} $ & $^{+1.4}_{-1.7} $ & $^{+1.6}_{-1.6}$ & $^{+2.4}_{-2.4}$ & $^{+1.0}_{-1.0}$\\
$ $ & $0.45$ & $7.21\cdot 10^{-3}$ & $ 13 $ & $^{+36}_{-21}$ & $ ^{+5.9}_{-7.8} $ & $^{+0.8}_{-3.0} $ & $^{-2.5}_{+0.6} $ & $^{-4.4}_{+2.5} $ & $^{-0.3}_{+1.2} $ & $^{+3.3}_{-3.3}$ & $^{+2.4}_{-2.4}$ & $^{+2.7}_{-2.7}$\\
$3073$ & $0.07$ & $9.55\cdot 10^{-2}$ & $ 51 $ & $^{+16}_{-12}$ & $ ^{+5.5}_{-5.2} $ & $^{+1.7}_{-1.6} $ & $^{-1.1}_{+2.1} $ & $^{-0.1}_{-0.1} $ & $^{+0.1}_{-0.5} $ & $^{+2.0}_{-2.0}$ & $^{+2.4}_{-2.4}$ & $^{+3.3}_{-3.3}$\\
$ $ & $0.12$ & $6.15\cdot 10^{-2}$ & $ 42 $ & $^{+18}_{-13}$ & $ ^{+5.1}_{-5.3} $ & $^{+1.9}_{-1.4} $ & $^{-2.0}_{+1.0} $ & $^{-1.0}_{+0.6} $ & $^{-0.6}_{+0.0} $ & $^{+1.4}_{-1.4}$ & $^{+2.4}_{-2.4}$ & $^{+3.3}_{-3.3}$\\
$ $ & $0.18$ & $3.67\cdot 10^{-2}$ & $ 40 $ & $^{+19}_{-13}$ & $ ^{+4.9}_{-5.2} $ & $^{+0.6}_{-0.9} $ & $^{-2.4}_{+2.0} $ & $^{-0.4}_{-0.8} $ & $^{-0.2}_{-0.1} $ & $^{+0.8}_{-0.8}$ & $^{+2.4}_{-2.4}$ & $^{+3.3}_{-3.3}$\\
$ $ & $0.26$ & $2.30\cdot 10^{-2}$ & $ 31 $ & $^{+21}_{-15}$ & $ ^{+4.0}_{-3.9} $ & $^{+0.6}_{-0.6} $ & $^{-1.2}_{+0.6} $ & $^{-1.8}_{+2.3} $ & $^{+0.3}_{+0.2} $ & $^{+0.9}_{-0.9}$ & $^{+2.4}_{-2.4}$ & $^{+1.0}_{-1.0}$\\
$ $ & $0.36$ & $1.25\cdot 10^{-2}$ & $ 24 $ & $^{+25}_{-17}$ & $ ^{+4.5}_{-4.3} $ & $^{+0.7}_{-0.8} $ & $^{-1.1}_{+0.6} $ & $^{-1.4}_{+2.1} $ & $^{+1.0}_{-0.9} $ & $^{+2.1}_{-2.1}$ & $^{+2.4}_{-2.4}$ & $^{+1.0}_{-1.0}$\\
$ $ & $0.49$ & $5.38\cdot 10^{-3}$ & $ 13 $ & $^{+36}_{-21}$ & $ ^{+6.5}_{-7.2} $ & $^{+0.7}_{-1.3} $ & $^{+0.6}_{-0.5} $ & $^{-4.0}_{+3.2} $ & $^{-0.2}_{-1.9} $ & $^{+3.8}_{-3.8}$ & $^{+2.4}_{-2.4}$ & $^{+2.7}_{-2.7}$\\
$3568$ & $0.09$ & $5.61\cdot 10^{-2}$ & $ 39 $ & $^{+19}_{-14}$ & $ ^{+5.8}_{-5.1} $ & $^{+2.3}_{-0.7} $ & $^{-1.7}_{+2.2} $ & $^{+0.4}_{+0.7} $ & $^{+0.3}_{-0.8} $ & $^{+1.9}_{-1.9}$ & $^{+2.4}_{-2.4}$ & $^{+3.3}_{-3.3}$\\
$ $ & $0.14$ & $2.97\cdot 10^{-2}$ & $ 27 $ & $^{+23}_{-16}$ & $ ^{+4.9}_{-6.1} $ & $^{+0.6}_{-3.5} $ & $^{-1.2}_{+1.6} $ & $^{-1.8}_{+0.7} $ & $^{-0.6}_{+0.7} $ & $^{+1.1}_{-1.1}$ & $^{+2.4}_{-2.4}$ & $^{+3.3}_{-3.3}$\\
$ $ & $0.21$ & $1.67\cdot 10^{-2}$ & $ 22 $ & $^{+26}_{-17}$ & $ ^{+3.8}_{-3.4} $ & $^{+0.7}_{-0.7} $ & $^{-1.1}_{+1.8} $ & $^{-0.1}_{+0.7} $ & $^{-0.1}_{-0.1} $ & $^{+0.7}_{-0.7}$ & $^{+2.4}_{-2.4}$ & $^{+1.0}_{-1.0}$\\
$ $ & $0.29$ & $1.04\cdot 10^{-2}$ & $ 19 $ & $^{+29}_{-18}$ & $ ^{+4.2}_{-3.4} $ & $^{+1.1}_{-0.4} $ & $^{+0.6}_{+1.9} $ & $^{-0.9}_{+1.3} $ & $^{+0.1}_{-0.3} $ & $^{+1.2}_{-1.2}$ & $^{+2.4}_{-2.4}$ & $^{+1.0}_{-1.0}$\\
$ $ & $0.40$ & $6.43\cdot 10^{-3}$ & $ 16 $ & $^{+32}_{-19}$ & $ ^{+6.6}_{-5.7} $ & $^{+1.3}_{-0.1} $ & $^{-1.2}_{+4.1} $ & $^{-2.7}_{+1.6} $ & $^{+0.8}_{-1.0} $ & $^{+2.6}_{-2.6}$ & $^{+2.4}_{-2.4}$ & $^{+2.7}_{-2.7}$\\
$ $ & $0.53$ & $1.37\cdot 10^{-3}$ & $ 4 $ & $^{+79}_{-29}$ & $ ^{+7.8}_{-7.5} $ & $^{+1.5}_{-0.6} $ & $^{-0.2}_{+1.6} $ & $^{-4.6}_{+4.5} $ & $^{+0.8}_{-0.3} $ & $^{+4.3}_{-4.3}$ & $^{+2.4}_{-2.4}$ & $^{+2.7}_{-2.7}$\\
$4145$ & $0.10$ & $3.18\cdot 10^{-2}$ & $ 26 $ & $^{+24}_{-16}$ & $ ^{+5.2}_{-6.5} $ & $^{+0.4}_{-3.5} $ & $^{-2.6}_{+2.2} $ & $^{-1.0}_{+0.2} $ & $^{+0.2}_{-0.5} $ & $^{+1.7}_{-1.7}$ & $^{+2.4}_{-2.4}$ & $^{+3.3}_{-3.3}$\\
$ $ & $0.16$ & $2.61\cdot 10^{-2}$ & $ 32 $ & $^{+21}_{-15}$ & $ ^{+5.6}_{-5.2} $ & $^{+3.1}_{-1.0} $ & $^{-2.5}_{+1.5} $ & $^{-0.3}_{+0.5} $ & $^{-0.4}_{+0.3} $ & $^{+0.9}_{-0.9}$ & $^{+2.4}_{-2.4}$ & $^{+3.3}_{-3.3}$\\
$ $ & $0.23$ & $7.72\cdot 10^{-3}$ & $ 14 $ & $^{+35}_{-20}$ & $ ^{+7.6}_{-4.4} $ & $^{+6.8}_{-1.1} $ & $^{-2.5}_{+0.7} $ & $^{-1.3}_{+1.3} $ & $^{-0.2}_{-0.4} $ & $^{+0.7}_{-0.7}$ & $^{+2.4}_{-2.4}$ & $^{+1.0}_{-1.0}$\\
$ $ & $0.33$ & $8.50\cdot 10^{-3}$ & $ 20 $ & $^{+28}_{-19}$ & $ ^{+4.3}_{-4.6} $ & $^{+0.7}_{-1.1} $ & $^{-2.3}_{+1.8} $ & $^{-1.7}_{+1.5} $ & $^{+0.4}_{+0.3} $ & $^{+1.6}_{-1.6}$ & $^{+2.4}_{-2.4}$ & $^{+1.0}_{-1.0}$\\
$ $ & $0.44$ & $3.03\cdot 10^{-3}$ & $ 9 $ & $^{+46}_{-23}$ & $ ^{+5.3}_{-13} $ & $^{+0.0}_{-11} $ & $^{-1.6}_{-0.7} $ & $^{-4.0}_{+1.0} $ & $^{+0.5}_{-2.1} $ & $^{+3.3}_{-3.3}$ & $^{+2.4}_{-2.4}$ & $^{+2.7}_{-2.7}$\\
$ $ & $0.58$ & $2.96\cdot 10^{-4}$ & $ 1 $ & $^{+220}_{-30}$ & $ ^{+11}_{-7.4} $ & $^{+1.3}_{-2.4} $ & $^{+0.0}_{-2.0} $ & $^{-2.6}_{+7.7} $ & $^{-0.9}_{+4.4} $ & $^{+4.7}_{-4.7}$ & $^{+2.4}_{-2.4}$ & $^{+2.7}_{-2.7}$\\
$4806$ & $0.12$ & $1.76\cdot 10^{-2}$ & $ 18 $ & $^{+30}_{-19}$ & $ ^{+4.8}_{-7.0} $ & $^{+0.6}_{-5.0} $ & $^{-1.6}_{+1.3} $ & $^{+0.4}_{-0.4} $ & $^{+0.3}_{-0.3} $ & $^{+1.5}_{-1.5}$ & $^{+2.4}_{-2.4}$ & $^{+3.3}_{-3.3}$\\
$ $ & $0.18$ & $1.36\cdot 10^{-2}$ & $ 20 $ & $^{+28}_{-18}$ & $ ^{+5.2}_{-4.8} $ & $^{+1.1}_{-0.6} $ & $^{-0.4}_{+1.2} $ & $^{-1.6}_{+2.2} $ & $^{-0.9}_{+0.3} $ & $^{+0.7}_{-0.7}$ & $^{+2.4}_{-2.4}$ & $^{+3.3}_{-3.3}$\\
$ $ & $0.26$ & $5.44\cdot 10^{-3}$ & $ 12 $ & $^{+38}_{-21}$ & $ ^{+3.7}_{-3.6} $ & $^{+0.6}_{-0.8} $ & $^{-0.8}_{+1.6} $ & $^{-1.4}_{+0.8} $ & $^{+0.1}_{+0.0} $ & $^{+0.9}_{-0.9}$ & $^{+2.4}_{-2.4}$ & $^{+1.0}_{-1.0}$\\
$ $ & $0.36$ & $3.48\cdot 10^{-3}$ & $ 10 $ & $^{+43}_{-23}$ & $ ^{+5.1}_{-4.5} $ & $^{+1.6}_{-0.5} $ & $^{-1.2}_{+2.8} $ & $^{-2.1}_{+1.3} $ & $^{-0.2}_{-0.7} $ & $^{+2.1}_{-2.1}$ & $^{+2.4}_{-2.4}$ & $^{+1.0}_{-1.0}$\\
$ $ & $0.49$ & $1.12\cdot 10^{-3}$ & $ 4 $ & $^{+79}_{-29}$ & $ ^{+7.6}_{-6.5} $ & $^{+1.8}_{-0.6} $ & $^{-1.6}_{+2.8} $ & $^{-2.8}_{+3.9} $ & $^{+1.0}_{-0.5} $ & $^{+3.9}_{-3.9}$ & $^{+2.4}_{-2.4}$ & $^{+2.7}_{-2.7}$\\
$ $ & $0.63$ & $9.18\cdot 10^{-4}$ & $ 4 $ & $^{+79}_{-29}$ & $ ^{+12}_{-8.1} $ & $^{+1.0}_{-2.0} $ & $^{-2.3}_{-0.3} $ & $^{-4.8}_{+9.6} $ & $^{-1.0}_{+4.8} $ & $^{+4.9}_{-4.9}$ & $^{+2.4}_{-2.4}$ & $^{+0.1}_{-0.1}$\\
$5561$ & $0.09$ & $2.26\cdot 10^{-2}$ & $ 15 $ & $^{+33}_{-20}$ & $ ^{+6.9}_{-9.6} $ & $^{+1.8}_{-6.5} $ & $^{-4.1}_{+3.5} $ & $^{+3.1}_{-3.1} $ & $^{-0.1}_{-0.5} $ & $^{+1.9}_{-1.9}$ & $^{+2.4}_{-2.4}$ & $^{+3.3}_{-3.3}$\\
$ $ & $0.14$ & $1.42\cdot 10^{-2}$ & $ 21 $ & $^{+27}_{-17}$ & $ ^{+5.2}_{-5.2} $ & $^{+0.7}_{-1.4} $ & $^{-2.2}_{+2.5} $ & $^{-0.3}_{+0.6} $ & $^{-0.1}_{-0.4} $ & $^{+1.1}_{-1.1}$ & $^{+2.4}_{-2.4}$ & $^{+3.3}_{-3.3}$\\
$ $ & $0.21$ & $3.41\cdot 10^{-3}$ & $ 6 $ & $^{+60}_{-26}$ & $ ^{+3.5}_{-4.2} $ & $^{+0.6}_{-0.9} $ & $^{-2.0}_{+1.5} $ & $^{-1.4}_{-0.3} $ & $^{-0.7}_{+0.3} $ & $^{+0.7}_{-0.7}$ & $^{+2.4}_{-2.4}$ & $^{+1.0}_{-1.0}$\\
$ $ & $0.30$ & $4.92\cdot 10^{-3}$ & $ 14 $ & $^{+35}_{-20}$ & $ ^{+4.3}_{-4.4} $ & $^{+0.4}_{-0.7} $ & $^{-2.4}_{+1.1} $ & $^{-1.7}_{+2.5} $ & $^{+0.1}_{-0.3} $ & $^{+1.2}_{-1.2}$ & $^{+2.4}_{-2.4}$ & $^{+1.0}_{-1.0}$\\
$ $ & $0.41$ & $2.56\cdot 10^{-3}$ & $ 9 $ & $^{+46}_{-23}$ & $ ^{+5.0}_{-5.8} $ & $^{+0.7}_{-1.5} $ & $^{-1.8}_{+0.0} $ & $^{-2.1}_{+1.0} $ & $^{+0.4}_{-0.2} $ & $^{+2.8}_{-2.8}$ & $^{+2.4}_{-2.4}$ & $^{+2.7}_{-2.7}$\\
$ $ & $0.54$ & $9.13\cdot 10^{-4}$ & $ 4 $ & $^{+79}_{-29}$ & $ ^{+8.0}_{-8.0} $ & $^{+0.4}_{-2.2} $ & $^{-1.8}_{-0.1} $ & $^{-4.2}_{+5.1} $ & $^{-0.3}_{-1.1} $ & $^{+4.6}_{-4.6}$ & $^{+2.4}_{-2.4}$ & $^{+2.7}_{-2.7}$\\
$ $ & $0.69$ & $6.11\cdot 10^{-4}$ & $ 3 $ & $^{+96}_{-30}$ & $ ^{+13}_{-9.9} $ & $^{+5.8}_{-2.1} $ & $^{+0.3}_{+1.8} $ & $^{-7.3}_{+9.2} $ & $^{-3.1}_{+3.1} $ & $^{+4.6}_{-4.6}$ & $^{+2.4}_{-2.4}$ & $^{+0.1}_{-0.1}$\\
$6966$ & $0.11$ & $6.88\cdot 10^{-3}$ & $ 13 $ & $^{+36}_{-21}$ & $ ^{+6.3}_{-9.9} $ & $^{+0.4}_{-6.6} $ & $^{-4.2}_{+2.3} $ & $^{+3.5}_{-3.6} $ & $^{-0.1}_{-1.5} $ & $^{+1.8}_{-1.8}$ & $^{+2.4}_{-2.4}$ & $^{+3.3}_{-3.3}$\\
$ $ & $0.17$ & $6.24\cdot 10^{-3}$ & $ 25 $ & $^{+24}_{-16}$ & $ ^{+4.7}_{-7.8} $ & $^{+0.0}_{-5.3} $ & $^{-2.9}_{+1.7} $ & $^{-2.0}_{+0.3} $ & $^{-0.6}_{-0.2} $ & $^{+0.8}_{-0.8}$ & $^{+2.4}_{-2.4}$ & $^{+3.3}_{-3.3}$\\
$ $ & $0.25$ & $2.85\cdot 10^{-3}$ & $ 15 $ & $^{+33}_{-20}$ & $ ^{+3.7}_{-3.7} $ & $^{+0.6}_{-0.8} $ & $^{-1.6}_{+1.8} $ & $^{-0.4}_{+0.8} $ & $^{-0.5}_{+0.0} $ & $^{+0.7}_{-0.7}$ & $^{+2.4}_{-2.4}$ & $^{+1.0}_{-1.0}$\\
$ $ & $0.34$ & $2.52\cdot 10^{-3}$ & $ 22 $ & $^{+26}_{-17}$ & $ ^{+4.3}_{-4.3} $ & $^{+0.7}_{-0.8} $ & $^{-1.3}_{+1.9} $ & $^{-1.9}_{+1.3} $ & $^{+0.1}_{-0.1} $ & $^{+1.8}_{-1.8}$ & $^{+2.4}_{-2.4}$ & $^{+1.0}_{-1.0}$\\
$ $ & $0.47$ & $6.97\cdot 10^{-4}$ & $ 7 $ & $^{+54}_{-25}$ & $ ^{+7.1}_{-6.4} $ & $^{+1.1}_{-0.4} $ & $^{-0.7}_{+1.3} $ & $^{-3.3}_{+4.2} $ & $^{+0.3}_{-0.5} $ & $^{+3.7}_{-3.7}$ & $^{+2.4}_{-2.4}$ & $^{+2.7}_{-2.7}$\\
$ $ & $0.61$ & $6.75\cdot 10^{-4}$ & $ 8 $ & $^{+49}_{-24}$ & $ ^{+9.4}_{-9.9} $ & $^{+1.3}_{-0.6} $ & $^{-1.1}_{+3.0} $ & $^{-7.9}_{+6.6} $ & $^{-0.5}_{-0.1} $ & $^{+5.0}_{-5.0}$ & $^{+2.4}_{-2.4}$ & $^{+0.1}_{-0.1}$\\
$ $ & $0.78$ & $ < 4.42\cdot 10^{-5}$ & $ 0 $ & $ $ & $  $ & $  $ & $  $ & $  $ & $  $ & $  $ & $  $ & $  $\\
$9055$ & $0.16$ & $5.35\cdot 10^{-3}$ & $ 16 $ & $^{+32}_{-19}$ & $ ^{+6.5}_{-12} $ & $^{+0.4}_{-9.9} $ & $^{-4.6}_{+3.8} $ & $^{+2.6}_{-3.4} $ & $^{+0.6}_{-1.4} $ & $^{+1.0}_{-1.0}$ & $^{+2.4}_{-2.4}$ & $^{+3.3}_{-3.3}$\\
$ $ & $0.23$ & $2.96\cdot 10^{-3}$ & $ 17 $ & $^{+31}_{-19}$ & $ ^{+3.8}_{-4.9} $ & $^{+1.3}_{-1.7} $ & $^{-2.6}_{+1.2} $ & $^{-1.9}_{+1.2} $ & $^{-1.1}_{+0.3} $ & $^{+0.4}_{-0.4}$ & $^{+2.4}_{-2.4}$ & $^{+1.0}_{-1.0}$\\
$ $ & $0.33$ & $1.23\cdot 10^{-3}$ & $ 12 $ & $^{+38}_{-21}$ & $ ^{+4.0}_{-4.5} $ & $^{+0.6}_{-0.7} $ & $^{-2.2}_{+1.6} $ & $^{-1.7}_{+1.4} $ & $^{-0.3}_{-0.1} $ & $^{+1.5}_{-1.5}$ & $^{+2.4}_{-2.4}$ & $^{+1.0}_{-1.0}$\\
$ $ & $0.45$ & $4.31\cdot 10^{-4}$ & $ 5 $ & $^{+67}_{-27}$ & $ ^{+6.3}_{-6.9} $ & $^{+0.4}_{-0.7} $ & $^{-1.3}_{+0.4} $ & $^{-4.2}_{+3.4} $ & $^{+0.1}_{-0.3} $ & $^{+3.4}_{-3.4}$ & $^{+2.4}_{-2.4}$ & $^{+2.7}_{-2.7}$\\
$ $ & $0.59$ & $2.12\cdot 10^{-4}$ & $ 3 $ & $^{+96}_{-30}$ & $ ^{+9.8}_{-7.0} $ & $^{+0.6}_{-0.7} $ & $^{-1.4}_{-0.4} $ & $^{-2.3}_{+7.3} $ & $^{+0.1}_{+0.3} $ & $^{+5.1}_{-5.1}$ & $^{+2.4}_{-2.4}$ & $^{+2.7}_{-2.7}$\\
$ $ & $0.73$ & $6.86\cdot 10^{-5}$ & $ 1 $ & $^{+220}_{-30}$ & $ ^{+10}_{-25} $ & $^{+1.1}_{-6.2} $ & $^{-2.3}_{-2.5} $ & $^{-23}_{+8.3} $ & $^{-0.9}_{+2.2} $ & $^{+4.6}_{-4.6}$ & $^{+2.4}_{-2.4}$ & $^{+0.1}_{-0.1}$\\
$ $ & $0.90$ & $ < 8.12\cdot 10^{-6}$ & $ 0 $ & $ $ & $  $ & $  $ & $  $ & $  $ & $  $ & $  $ & $  $ & $  $\\
$14807$ & $0.76$ & $1.03\cdot 10^{-5}$ & $ 1 $ & $^{+220}_{-30}$ & $ ^{+15}_{-16} $ & $^{+2.1}_{-2.6} $ & $^{-2.7}_{+1.6} $ & $^{-15}_{+14} $ & $^{-1.5}_{+0.7} $ & $^{+2.5}_{-2.5}$ & $^{+2.4}_{-2.4}$ & $^{+0.1}_{-0.1}$\\
$ $ & $0.92$ & $ < 1.28\cdot 10^{-6}$ & $ 0 $ & $ $ & $  $ & $  $ & $  $ & $  $ & $  $ & $  $ & $  $ & $  $\\
\hline
\end{supertabular}
\end{center}

%% file: DESY-06-116-tab-2.tex
\begin{center}
\footnotesize
\renewcommand{\arraystretch}{1.2}
\tablehead{
\hline
{$Q^2$} & 
\multicolumn{1}{c|}{$x_{\rm edge}$} & 
$\int_{x_{\rm edge}}^{1} {d^2\sigma}/{dxdQ^2}$  &
$N$& 
$\delta_{s}$& 
$\delta_t$ & 
$\delta_u$ &
$\delta_1$ &
$\delta_2$ &
$\delta_3$ &
$\delta_4$ &
$\delta_5$ &
$\delta_6$ \\
{($\Gev^2$)} & 
$$ &
{($\rm pb/GeV^{2}$)} &
$$ &
(\%) &
(\%) &
(\%) &
(\%) &
(\%) &
(\%) &
(\%) &
(\%) &
(\%) \\
\hline \hline}
\tablelasttail{
\hline}
\normalsize
\bottomcaption{
    The integral cross section table for 96-97 $e^+p$ NC scattering.
    The first two columns of the table contain the $Q^2$ and $x_{\rm edge}$ values 
    for the bin, the third contains the measured
    cross section $\int_{x_{\rm edge}}^{1} {d^2\sigma}/{dxdQ^2}$ corrected to the electroweak 
    Born level or the upper limit in case of zero observed 
    events, the fourth contains the number of events reconstructed in 
    the bin, $N$, the fifth contains the 
    statistical uncertainty, $\delta_s$, and the sixth contains the 
     total systematic uncertainty, $\delta_t$.
    The right part of the table lists the total uncorrelated
    systematic uncertainty, $\delta_u$, followed by the bin-to-bin correlated
    systematic uncertainties $\delta_1$--\,$\delta_6$ defined in
    the text.
    The upper (lower) numbers refer to the variation of 
    the cross section, whereas the signs of
    the numbers reflect the direction of change in the cross
    sections. Note that the normalization uncertainty, $\delta_7$ is
    not listed.
    } 
\label{tab:9697crosl}
\begin{supertabular}{|c|l|c|c|c|c||c|c|c|c|c|c|c|}
$ 648$ & $0.25$ & $8.49\cdot 10^{-2}$ & $ 34 $ & $^{+20}_{-14}$ & $ ^{+8.5}_{-6.7} $ & $^{+4.9}_{-2.7} $ & $^{+6.1}_{-5.2} $ & $^{+0.6}_{+0.6} $ & $^{-0.8}_{+1.1} $ & $^{+2.0}_{-2.0}$ & $^{+1.6}_{-1.6}$ & $^{+0.1}_{-0.1}$\\
$ 761$ & $0.27$ & $4.66\cdot 10^{-2}$ & $ 74 $ & $^{+13}_{-10}$ & $ ^{+5.1}_{-3.6} $ & $^{+1.8}_{-1.1} $ & $^{+3.0}_{-0.6} $ & $^{+0.9}_{+0.9} $ & $^{-0.7}_{+0.7} $ & $^{+2.5}_{-2.5}$ & $^{+1.6}_{-1.6}$ & $^{+0.1}_{-0.1}$\\
$ 891$ & $0.30$ & $3.19\cdot 10^{-2}$ & $ 109 $ & $^{+11}_{-8.7}$ & $ ^{+6.5}_{-3.7} $ & $^{+4.6}_{-0.1} $ & $^{+2.2}_{+0.5} $ & $^{+1.1}_{+1.1} $ & $^{-1.0}_{+1.1} $ & $^{+2.7}_{-2.7}$ & $^{+1.6}_{-1.6}$ & $^{+0.1}_{-0.1}$\\
$1045$ & $0.32$ & $2.08\cdot 10^{-2}$ & $ 101 $ & $^{+11}_{-9.1}$ & $ ^{+6.9}_{-5.8} $ & $^{+3.8}_{-0.6} $ & $^{-4.4}_{+4.2} $ & $^{+0.9}_{+0.9} $ & $^{-1.0}_{+0.7} $ & $^{+2.9}_{-2.9}$ & $^{+1.6}_{-1.6}$ & $^{+0.1}_{-0.1}$\\
$1224$ & $0.35$ & $1.03\cdot 10^{-2}$ & $ 65 $ & $^{+14}_{-11}$ & $ ^{+6.0}_{-3.9} $ & $^{+3.0}_{-0.2} $ & $^{+0.6}_{+3.2} $ & $^{+0.7}_{+0.7} $ & $^{-0.4}_{+0.7} $ & $^{+3.1}_{-3.1}$ & $^{+1.6}_{-1.6}$ & $^{+0.1}_{-0.1}$\\
$1431$ & $0.38$ & $5.65\cdot 10^{-3}$ & $ 42 $ & $^{+18}_{-13}$ & $ ^{+7.2}_{-6.0} $ & $^{+2.3}_{-0.1} $ & $^{-4.3}_{+5.3} $ & $^{+0.8}_{+0.8} $ & $^{-1.1}_{+0.7} $ & $^{+3.4}_{-3.4}$ & $^{+1.6}_{-1.6}$ & $^{+0.1}_{-0.1}$\\
$1672$ & $0.41$ & $3.66\cdot 10^{-3}$ & $ 29 $ & $^{+22}_{-15}$ & $ ^{+5.8}_{-4.6} $ & $^{+2.2}_{-0.2} $ & $^{-1.3}_{+2.8} $ & $^{+0.5}_{+0.5} $ & $^{-0.7}_{+1.5} $ & $^{+3.7}_{-3.7}$ & $^{+1.6}_{-1.6}$ & $^{+0.1}_{-0.1}$\\
$1951$ & $0.44$ & $2.07\cdot 10^{-3}$ & $ 21 $ & $^{+27}_{-17}$ & $ ^{+6.8}_{-5.5} $ & $^{+2.3}_{-0.6} $ & $^{-2.5}_{+4.1} $ & $^{+0.6}_{+0.6} $ & $^{-1.1}_{+1.0} $ & $^{+4.1}_{-4.1}$ & $^{+1.6}_{-1.6}$ & $^{+0.1}_{-0.1}$\\
$2273$ & $0.48$ & $1.22\cdot 10^{-3}$ & $ 13 $ & $^{+36}_{-21}$ & $ ^{+7.3}_{-8.3} $ & $^{+4.1}_{-1.1} $ & $^{-5.7}_{+2.9} $ & $^{+0.4}_{+0.4} $ & $^{-2.9}_{+0.9} $ & $^{+4.6}_{-4.6}$ & $^{+1.6}_{-1.6}$ & $^{+0.1}_{-0.1}$\\
$2644$ & $0.52$ & $2.87\cdot 10^{-4}$ & $ 4 $ & $^{+79}_{-29}$ & $ ^{+8.7}_{-6.8} $ & $^{+2.2}_{-3.0} $ & $^{-2.1}_{+6.5} $ & $^{+0.5}_{+0.5} $ & $^{-2.3}_{+0.9} $ & $^{+4.7}_{-4.7}$ & $^{+1.6}_{-1.6}$ & $^{+0.1}_{-0.1}$\\
$3073$ & $0.56$ & $ < 8.32\cdot 10^{-5}$ & $ 0 $ &$ $ & $  $ & $  $ & $  $ & $  $ & $  $ & $  $ & $  $ & $  $\\
$3568$ & $0.60$ & $6.04\cdot 10^{-5}$ & $ 1 $ & $^{+220}_{-30}$ & $ ^{+16}_{-15} $ & $^{+4.5}_{-3.1} $ & $^{-6.1}_{+6.1} $ & $^{+1.4}_{+1.4} $ & $^{-2.1}_{+3.7} $ & $^{+4.7}_{-4.7}$ & $^{+1.6}_{-1.6}$ & $^{+13}_{-13}$\\
$4145$ & $0.65$ & $ < 5.30\cdot 10^{-5}$ & $ 0 $ & $ $ & $  $ & $  $ & $  $ & $  $ & $  $ & $  $ & $  $ & $  $\\
$4806$ & $0.70$ & $ < 5.79\cdot 10^{-5}$ & $ 0 $ & $ $ & $  $ & $  $ & $  $ & $  $ & $  $ & $  $ & $  $ & $  $\\
$5561$ & $0.76$ & $ < 3.87\cdot 10^{-5}$ & $ 0 $ & $ $ & $  $ & $  $ & $  $ & $  $ & $  $ & $  $ & $  $ & $  $\\
$6966$ & $0.89$ & $ < 1.61\cdot 10^{-6}$ & $ 0 $ & $ $ & $  $ & $  $ & $  $ & $  $ & $  $ & $  $ & $  $ & $  $\\
\hline
\end{supertabular}
\end{center}

%% file: DESY-06-116-tab-3.tex
\begin{center}
\footnotesize
\renewcommand{\arraystretch}{1.2}
\tablehead{
\multicolumn{13}{l}{
{\normalsize }}\\
\hline
{$Q^2$} & 
\multicolumn{1}{c|}{$x$} & 
${d^2\sigma}/{dxdQ^2}$  &
$N$& 
$\delta_s$& 
$\delta_t$& 
$\delta_u$& 
$\delta_1$ &
$\delta_2$ &
$\delta_3$ &
$\delta_4$ &
$\delta_5$ &
$\delta_6$ \\
{($\Gev^2$)} & 
$$ &
{($\rm pb/GeV^{2}$)} &
$$ &
(\%) &
(\%) &
(\%) &
(\%) &
(\%) &
(\%) &
(\%) &
(\%) &
(\%) \\
\hline \hline}
\tablelasttail{
\hline
\multicolumn{11}{r}{} \\}
\normalsize
\bottomcaption{
    The cross section table for 98-99 $e^-p$ NC scattering.
    The first two columns of the table contain the $Q^2$ and $x$ values 
    at which the cross section is quoted, the third contains the measured
    cross section ${d^2\sigma}/{dxdQ^2}$ corrected to the electroweak 
    Born level or the upper limit in case of zero observed 
    events, the fourth contains the number of events reconstructed in 
    the bin, $N$, the fifth contains the 
    statistical uncertainty, $\delta_s$, and the sixth contains the 
     total systematic uncertainty, $\delta_t$.
    The right part of the table lists the total uncorrelated
    systematic uncertainty, $\delta_u$, followed by the bin-to-bin correlated
    systematic uncertainties $\delta_1$--\,$\delta_6$ defined in
    the text.
    The upper (lower) numbers refer to the variation of 
    the cross section, whereas the signs of
    the numbers reflect the direction of change in the cross
    sections. Note that the normalization uncertainty, $\delta_7$ is
    not listed.
    } 
\label{tab:9899cros}
\begin{supertabular}{|c|l|c|c|c|c||c|c|c|c|c|c|c|}
$ 648$ & $0.08$ & $2.76$ & $ 51 $ & $^{+16}_{-12}$ & $ ^{+8.3}_{-9.0} $ & $^{+0.7}_{-0.4} $ & $^{-7.0}_{+5.6} $ & $^{-1.1}_{+1.1} $ & $^{-3.0}_{+3.6} $ & $^{+0.1}_{-0.1} $ & $^{+2.4}_{-2.4} $ & $^{+3.4}_{-3.4}$\\
$ $ & $0.13$ & $1.78$ & $ 23 $ & $^{+25}_{-17}$ & $ ^{+9.1}_{-12} $ & $^{+3.6}_{-3.3} $ & $^{-8.4}_{+6.0} $ & $^{-1.3}_{+0.1} $ & $^{-6.1}_{+3.3} $ & $^{+0.0}_{-0.0} $ & $^{+2.4}_{-2.4} $ & $^{+3.4}_{-3.4}$\\
$ $ & $0.19$ & $8.63\cdot 10^{-1}$ & $ 14 $ & $^{+35}_{-20}$ & $ ^{+11}_{-15} $ & $^{+1.0}_{-9.2} $ & $^{-9.2}_{+8.8} $ & $^{-1.4}_{+1.3} $ & $^{-5.4}_{+4.3} $ & $^{+0.0}_{+0.0} $ & $^{+2.4}_{-2.4} $ & $^{+3.4}_{-3.4}$\\
$ 761$ & $0.09$ & $1.71$ & $ 92 $ & $^{+12}_{-9.4}$ & $ ^{+6.1}_{-5.8} $ & $^{+1.0}_{-1.4} $ & $^{-2.5}_{+3.6} $ & $^{-0.9}_{+0.8} $ & $^{-1.6}_{+1.2} $ & $^{+0.1}_{-0.1} $ & $^{+2.4}_{-2.4} $ & $^{+3.3}_{-3.3}$\\
$ $ & $0.14$ & $8.70\cdot 10^{-1}$ & $ 44 $ & $^{+17}_{-13}$ & $ ^{+6.5}_{-6.5} $ & $^{+1.1}_{-1.1} $ & $^{-4.3}_{+4.2} $ & $^{-0.4}_{+0.3} $ & $^{-0.7}_{+1.4} $ & $^{+0.1}_{-0.1} $ & $^{+2.4}_{-2.4} $ & $^{+3.3}_{-3.3}$\\
$ $ & $0.21$ & $6.10\cdot 10^{-1}$ & $ 38 $ & $^{+19}_{-14}$ & $ ^{+6.8}_{-6.7} $ & $^{+1.3}_{-1.1} $ & $^{-4.0}_{+2.8} $ & $^{-1.4}_{+2.8} $ & $^{-2.0}_{+2.8} $ & $^{+0.0}_{-0.0} $ & $^{+2.4}_{-2.4} $ & $^{+3.3}_{-3.3}$\\
$ 891$ & $0.10$ & $1.36$ & $ 137 $ & $^{+9.3}_{-7.9}$ & $ ^{+4.8}_{-4.7} $ & $^{+0.6}_{-0.2} $ & $^{+0.0}_{-0.4} $ & $^{-0.6}_{+1.2} $ & $^{-0.2}_{-0.1} $ & $^{+0.1}_{-0.1} $ & $^{+2.4}_{-2.4} $ & $^{+3.2}_{-3.2}$\\
$ $ & $0.15$ & $6.16\cdot 10^{-1}$ & $ 61 $ & $^{+15}_{-11}$ & $ ^{+5.0}_{-4.8} $ & $^{+0.5}_{-0.2} $ & $^{-0.2}_{+0.6} $ & $^{-1.4}_{+0.2} $ & $^{+0.0}_{+1.7} $ & $^{+0.1}_{-0.1} $ & $^{+2.4}_{-2.4} $ & $^{+3.2}_{-3.2}$\\
$ $ & $0.22$ & $5.20\cdot 10^{-1}$ & $ 68 $ & $^{+14}_{-11}$ & $ ^{+5.5}_{-5.3} $ & $^{+2.2}_{-2.2} $ & $^{-1.3}_{+1.7} $ & $^{+0.4}_{+1.2} $ & $^{-0.2}_{-0.2} $ & $^{+0.0}_{+0.0} $ & $^{+2.4}_{-2.4} $ & $^{+3.2}_{-3.2}$\\
$1045$ & $0.07$ & $1.56$ & $ 148 $ & $^{+9.0}_{-7.6}$ & $ ^{+4.9}_{-5.2} $ & $^{+0.1}_{-0.6} $ & $^{+1.9}_{-2.4} $ & $^{+0.2}_{+0.2} $ & $^{+0.8}_{-1.1} $ & $^{+0.0}_{-0.0} $ & $^{+2.4}_{-2.4} $ & $^{+2.9}_{-2.9}$\\
$ $ & $0.11$ & $6.53\cdot 10^{-1}$ & $ 84 $ & $^{+12}_{-9.7}$ & $ ^{+4.6}_{-4.6} $ & $^{+0.3}_{-0.4} $ & $^{+0.2}_{-0.2} $ & $^{-1.3}_{+1.3} $ & $^{+0.4}_{-0.6} $ & $^{+0.1}_{-0.1} $ & $^{+2.4}_{-2.4} $ & $^{+2.9}_{-2.9}$\\
$ $ & $0.17$ & $4.61\cdot 10^{-1}$ & $ 67 $ & $^{+14}_{-11}$ & $ ^{+4.6}_{-4.8} $ & $^{+1.0}_{-1.4} $ & $^{+1.0}_{-0.2} $ & $^{-0.1}_{-0.2} $ & $^{-0.6}_{-1.2} $ & $^{+0.1}_{-0.1} $ & $^{+2.4}_{-2.4} $ & $^{+2.9}_{-2.9}$\\
$ $ & $0.24$ & $2.02\cdot 10^{-1}$ & $ 43 $ & $^{+18}_{-13}$ & $ ^{+4.8}_{-5.4} $ & $^{+1.1}_{-0.9} $ & $^{+0.4}_{-2.5} $ & $^{-1.5}_{+1.4} $ & $^{-0.4}_{+0.1} $ & $^{+0.1}_{-0.1} $ & $^{+2.4}_{-2.4} $ & $^{+2.9}_{-2.9}$\\
$1224$ & $0.07$ & $7.47\cdot 10^{-1}$ & $ 87 $ & $^{+12}_{-9.6}$ & $ ^{+4.7}_{-4.9} $ & $^{+0.3}_{-1.1} $ & $^{+1.4}_{-1.7} $ & $^{+0.0}_{-0.4} $ & $^{+0.6}_{-0.4} $ & $^{+0.0}_{-0.0} $ & $^{+2.4}_{-2.4} $ & $^{+2.9}_{-2.9}$\\
$ $ & $0.12$ & $4.90\cdot 10^{-1}$ & $ 80 $ & $^{+13}_{-10}$ & $ ^{+4.7}_{-4.7} $ & $^{+0.3}_{-0.3} $ & $^{+0.9}_{-1.5} $ & $^{-0.9}_{+1.3} $ & $^{+0.5}_{-0.1} $ & $^{+0.1}_{-0.1} $ & $^{+2.4}_{-2.4} $ & $^{+2.9}_{-2.9}$\\
$ $ & $0.18$ & $3.83\cdot 10^{-1}$ & $ 70 $ & $^{+13}_{-11}$ & $ ^{+4.8}_{-4.6} $ & $^{+1.5}_{-0.2} $ & $^{-0.5}_{-1.0} $ & $^{-0.9}_{+0.3} $ & $^{+1.1}_{+0.2} $ & $^{+0.1}_{-0.1} $ & $^{+2.4}_{-2.4} $ & $^{+2.9}_{-2.9}$\\
$ $ & $0.26$ & $1.63\cdot 10^{-1}$ & $ 44 $ & $^{+17}_{-13}$ & $ ^{+5.1}_{-4.9} $ & $^{+1.0}_{-1.7} $ & $^{+0.5}_{+0.0} $ & $^{-0.7}_{+1.8} $ & $^{+1.6}_{-1.0} $ & $^{+0.1}_{-0.1} $ & $^{+2.4}_{-2.4} $ & $^{+2.9}_{-2.9}$\\
\hline
\end{supertabular}
\end{center}

\begin{center}
\footnotesize
\renewcommand{\arraystretch}{1.2}
\tablehead{
\multicolumn{13}{l}{
{\normalsize {\bf Table \thetable\ } (continued):}}\\
\hline
{$Q^2$} & 
\multicolumn{1}{c|}{$x$} & 
${d^2\sigma}/{dxdQ^2}$  &
$N$& 
$\delta_s$& 
$\delta_t$& 
$\delta_u$& 
$\delta_1$ &
$\delta_2$ &
$\delta_3$ &
$\delta_4$ &
$\delta_5$ &
$\delta_6$ \\
{($\Gev^2$)} & 
$$ &
{($\rm pb/GeV^{2}$)} &
$$ &
(\%) &
(\%) &
(\%) &
(\%) &
(\%) &
(\%) &
(\%) &
(\%) &
(\%) \\
\hline \hline}
\tabletail{
\hline
}
\tablelasttail{\hline}
\normalsize
\begin{supertabular}{|c|l|c|c|c|c||c|c|c|c|c|c|c|}
$1431$ & $0.09$ & $5.65\cdot 10^{-1}$ & $ 79 $ & $^{+13}_{-10}$ & $ ^{+4.8}_{-4.8} $ & $^{+0.3}_{-0.7} $ & $^{+1.7}_{-1.4} $ & $^{+0.6}_{+0.3} $ & $^{+0.7}_{-1.3} $ & $^{+0.1}_{-0.1} $ & $^{+2.4}_{-2.4} $ & $^{+2.8}_{-2.8}$\\
$ $ & $0.14$ & $2.69\cdot 10^{-1}$ & $ 55 $ & $^{+15}_{-12}$ & $ ^{+4.6}_{-4.9} $ & $^{+1.0}_{-0.5} $ & $^{+0.9}_{-1.9} $ & $^{-1.1}_{+0.9} $ & $^{+0.4}_{+0.0} $ & $^{+0.1}_{-0.1} $ & $^{+2.4}_{-2.4} $ & $^{+2.8}_{-2.8}$\\
$ $ & $0.20$ & $1.92\cdot 10^{-1}$ & $ 41 $ & $^{+18}_{-13}$ & $ ^{+5.0}_{-5.3} $ & $^{+1.1}_{-0.7} $ & $^{+2.1}_{-1.7} $ & $^{-1.3}_{+0.7} $ & $^{+0.8}_{-2.0} $ & $^{+0.0}_{-0.0} $ & $^{+2.4}_{-2.4} $ & $^{+2.8}_{-2.8}$\\
$ $ & $0.29$ & $1.29\cdot 10^{-1}$ & $ 41 $ & $^{+18}_{-13}$ & $ ^{+4.8}_{-4.7} $ & $^{+1.0}_{-0.9} $ & $^{+0.9}_{-1.5} $ & $^{-0.6}_{+1.6} $ & $^{-0.1}_{-0.1} $ & $^{+0.2}_{-0.2} $ & $^{+2.4}_{-2.4} $ & $^{+2.8}_{-2.8}$\\
$1672$ & $0.10$ & $3.25\cdot 10^{-1}$ & $ 57 $ & $^{+15}_{-11}$ & $ ^{+4.3}_{-4.2} $ & $^{+0.3}_{-0.5} $ & $^{+1.2}_{-0.7} $ & $^{-0.3}_{-0.2} $ & $^{+0.4}_{-0.2} $ & $^{+0.1}_{-0.1} $ & $^{+2.4}_{-2.4} $ & $^{+2.5}_{-2.5}$\\
$ $ & $0.15$ & $2.04\cdot 10^{-1}$ & $ 52 $ & $^{+16}_{-12}$ & $ ^{+4.6}_{-4.5} $ & $^{+0.4}_{-0.5} $ & $^{+1.3}_{-0.5} $ & $^{-1.2}_{+1.6} $ & $^{+0.3}_{-1.0} $ & $^{+0.1}_{-0.1} $ & $^{+2.4}_{-2.4} $ & $^{+2.5}_{-2.5}$\\
$ $ & $0.22$ & $1.03\cdot 10^{-1}$ & $ 28 $ & $^{+23}_{-15}$ & $ ^{+4.2}_{-4.6} $ & $^{+0.1}_{-0.5} $ & $^{+0.9}_{-1.6} $ & $^{-0.1}_{-0.6} $ & $^{+0.3}_{-0.8} $ & $^{+0.0}_{-0.0} $ & $^{+2.4}_{-2.4} $ & $^{+2.5}_{-2.5}$\\
$ $ & $0.31$ & $3.74\cdot 10^{-2}$ & $ 16 $ & $^{+32}_{-20}$ & $ ^{+5.1}_{-5.3} $ & $^{+1.2}_{-1.5} $ & $^{+1.0}_{-1.9} $ & $^{-2.0}_{+2.5} $ & $^{+0.1}_{-1.1} $ & $^{+0.3}_{-0.3} $ & $^{+2.4}_{-2.4} $ & $^{+2.5}_{-2.5}$\\
$1951$ & $0.07$ & $3.91\cdot 10^{-1}$ & $ 60 $ & $^{+15}_{-11}$ & $ ^{+4.1}_{-4.3} $ & $^{+0.3}_{-0.7} $ & $^{+1.0}_{-0.8} $ & $^{-1.1}_{+0.5} $ & $^{+0.2}_{-0.9} $ & $^{+0.0}_{-0.0} $ & $^{+2.4}_{-2.4} $ & $^{+2.2}_{-2.2}$\\
$ $ & $0.11$ & $1.86\cdot 10^{-1}$ & $ 39 $ & $^{+19}_{-14}$ & $ ^{+4.9}_{-4.4} $ & $^{+0.7}_{-0.6} $ & $^{+2.4}_{-1.3} $ & $^{+1.2}_{-0.7} $ & $^{+0.7}_{-1.0} $ & $^{+0.1}_{-0.1} $ & $^{+2.4}_{-2.4} $ & $^{+2.2}_{-2.2}$\\
$ $ & $0.17$ & $1.22\cdot 10^{-1}$ & $ 37 $ & $^{+19}_{-14}$ & $ ^{+4.1}_{-4.9} $ & $^{+0.2}_{-2.7} $ & $^{+0.3}_{+0.2} $ & $^{-1.2}_{+0.5} $ & $^{+0.6}_{+0.7} $ & $^{+0.1}_{-0.1} $ & $^{+2.4}_{-2.4} $ & $^{+2.2}_{-2.2}$\\
$ $ & $0.24$ & $7.22\cdot 10^{-2}$ & $ 26 $ & $^{+24}_{-16}$ & $ ^{+5.0}_{-4.5} $ & $^{+0.9}_{-0.8} $ & $^{+2.6}_{-0.4} $ & $^{-1.9}_{+1.1} $ & $^{+0.2}_{-0.6} $ & $^{+0.1}_{-0.1} $ & $^{+2.4}_{-2.4} $ & $^{+2.2}_{-2.2}$\\
$ $ & $0.34$ & $3.48\cdot 10^{-2}$ & $ 18 $ & $^{+30}_{-19}$ & $ ^{+4.8}_{-4.7} $ & $^{+1.8}_{-1.8} $ & $^{-0.3}_{-0.2} $ & $^{-1.1}_{+1.9} $ & $^{-0.6}_{-1.0} $ & $^{+0.4}_{-0.4} $ & $^{+2.4}_{-2.4} $ & $^{+2.2}_{-2.2}$\\
$2273$ & $0.07$ & $3.25\cdot 10^{-1}$ & $ 62 $ & $^{+14}_{-11}$ & $ ^{+4.5}_{-4.7} $ & $^{+0.5}_{-1.2} $ & $^{+1.6}_{-2.0} $ & $^{-0.9}_{+0.2} $ & $^{+1.2}_{-0.5} $ & $^{+0.0}_{+0.0} $ & $^{+2.4}_{-2.4} $ & $^{+2.2}_{-2.2}$\\
$ $ & $0.12$ & $1.19\cdot 10^{-1}$ & $ 31 $ & $^{+21}_{-15}$ & $ ^{+4.2}_{-4.8} $ & $^{+0.6}_{-1.0} $ & $^{+0.7}_{-2.4} $ & $^{+0.5}_{-0.6} $ & $^{+0.9}_{-0.5} $ & $^{+0.1}_{-0.1} $ & $^{+2.4}_{-2.4} $ & $^{+2.2}_{-2.2}$\\
$ $ & $0.18$ & $7.59\cdot 10^{-2}$ & $ 29 $ & $^{+22}_{-15}$ & $ ^{+4.2}_{-4.5} $ & $^{+0.3}_{-0.9} $ & $^{+0.6}_{-1.1} $ & $^{-1.6}_{+1.3} $ & $^{+0.5}_{-0.6} $ & $^{+0.1}_{-0.1} $ & $^{+2.4}_{-2.4} $ & $^{+2.2}_{-2.2}$\\
$ $ & $0.26$ & $5.21\cdot 10^{-2}$ & $ 23 $ & $^{+25}_{-17}$ & $ ^{+4.6}_{-4.5} $ & $^{+0.6}_{-0.7} $ & $^{+0.7}_{-1.9} $ & $^{-0.5}_{+1.1} $ & $^{+1.8}_{-0.2} $ & $^{+0.1}_{-0.1} $ & $^{+2.4}_{-2.4} $ & $^{+2.2}_{-2.2}$\\
$ $ & $0.37$ & $1.60\cdot 10^{-2}$ & $ 10 $ & $^{+43}_{-23}$ & $ ^{+6.0}_{-4.4} $ & $^{+2.5}_{-0.8} $ & $^{+2.0}_{-0.8} $ & $^{-1.6}_{+2.5} $ & $^{+1.7}_{-0.1} $ & $^{+0.5}_{-0.5} $ & $^{+2.4}_{-2.4} $ & $^{+2.2}_{-2.2}$\\
$2644$ & $0.09$ & $1.70\cdot 10^{-1}$ & $ 41 $ & $^{+18}_{-13}$ & $ ^{+4.7}_{-4.6} $ & $^{+0.6}_{-1.8} $ & $^{+2.1}_{-1.3} $ & $^{-0.9}_{+0.8} $ & $^{+1.0}_{-0.8} $ & $^{+0.1}_{-0.1} $ & $^{+2.4}_{-2.4} $ & $^{+2.1}_{-2.1}$\\
$ $ & $0.14$ & $7.79\cdot 10^{-2}$ & $ 29 $ & $^{+22}_{-15}$ & $ ^{+4.2}_{-4.5} $ & $^{+0.2}_{-0.6} $ & $^{+1.4}_{-1.7} $ & $^{+0.3}_{+0.3} $ & $^{+0.7}_{-1.2} $ & $^{+0.1}_{-0.1} $ & $^{+2.4}_{-2.4} $ & $^{+2.1}_{-2.1}$\\
$ $ & $0.21$ & $2.68\cdot 10^{-2}$ & $ 12 $ & $^{+38}_{-21}$ & $ ^{+4.1}_{-4.5} $ & $^{+0.4}_{-0.2} $ & $^{+0.9}_{-0.4} $ & $^{-2.1}_{+0.7} $ & $^{+0.0}_{+0.0} $ & $^{+0.0}_{-0.0} $ & $^{+2.4}_{-2.4} $ & $^{+2.1}_{-2.1}$\\
$ $ & $0.29$ & $3.26\cdot 10^{-2}$ & $ 19 $ & $^{+29}_{-18}$ & $ ^{+4.9}_{-4.1} $ & $^{+2.0}_{-0.3} $ & $^{+1.0}_{-0.9} $ & $^{+0.3}_{+2.0} $ & $^{-0.5}_{-0.8} $ & $^{+0.2}_{-0.2} $ & $^{+2.4}_{-2.4} $ & $^{+2.1}_{-2.1}$\\
$ $ & $0.40$ & $1.47\cdot 10^{-2}$ & $ 11 $ & $^{+40}_{-22}$ & $ ^{+4.7}_{-6.3} $ & $^{+1.1}_{-2.1} $ & $^{+0.9}_{-1.7} $ & $^{-3.9}_{+2.2} $ & $^{+0.2}_{-1.2} $ & $^{+0.5}_{-0.5} $ & $^{+2.4}_{-2.4} $ & $^{+2.1}_{-2.1}$\\
$3073$ & $0.06$ & $1.16\cdot 10^{-1}$ & $ 20 $ & $^{+28}_{-18}$ & $ ^{+4.9}_{-4.7} $ & $^{+0.4}_{-1.3} $ & $^{+2.3}_{-2.3} $ & $^{+1.3}_{-0.3} $ & $^{+1.5}_{-0.7} $ & $^{+0.1}_{-0.1} $ & $^{+2.4}_{-2.4} $ & $^{+2.0}_{-2.0}$\\
$ $ & $0.10$ & $9.39\cdot 10^{-2}$ & $ 25 $ & $^{+24}_{-16}$ & $ ^{+4.1}_{-5.8} $ & $^{+0.9}_{-3.8} $ & $^{+0.8}_{-1.5} $ & $^{-1.5}_{+0.7} $ & $^{+0.1}_{-0.4} $ & $^{+0.0}_{-0.0} $ & $^{+2.4}_{-2.4} $ & $^{+2.0}_{-2.0}$\\
$ $ & $0.15$ & $4.29\cdot 10^{-2}$ & $ 19 $ & $^{+29}_{-18}$ & $ ^{+4.6}_{-4.2} $ & $^{+0.6}_{-0.4} $ & $^{+2.0}_{-1.6} $ & $^{+0.4}_{+0.1} $ & $^{+1.2}_{-0.4} $ & $^{+0.1}_{-0.1} $ & $^{+2.4}_{-2.4} $ & $^{+2.0}_{-2.0}$\\
$ $ & $0.23$ & $2.53\cdot 10^{-2}$ & $ 14 $ & $^{+35}_{-20}$ & $ ^{+7.7}_{-4.9} $ & $^{+6.5}_{-0.3} $ & $^{+0.6}_{-1.6} $ & $^{-2.5}_{+1.6} $ & $^{+0.3}_{-0.8} $ & $^{+0.0}_{-0.0} $ & $^{+2.4}_{-2.4} $ & $^{+2.0}_{-2.0}$\\
$ $ & $0.32$ & $1.46\cdot 10^{-2}$ & $ 11 $ & $^{+40}_{-22}$ & $ ^{+4.0}_{-4.1} $ & $^{+0.4}_{-0.6} $ & $^{+0.0}_{-0.4} $ & $^{-1.3}_{+0.1} $ & $^{+0.3}_{+0.9} $ & $^{+0.2}_{-0.2} $ & $^{+2.4}_{-2.4} $ & $^{+2.0}_{-2.0}$\\
$ $ & $0.43$ & $9.60\cdot 10^{-4}$ & $ 1 $ & $^{+220}_{-30}$ & $ ^{+6.8}_{-4.5} $ & $^{+1.4}_{-1.7} $ & $^{-0.8}_{+0.0} $ & $^{-1.2}_{+5.4} $ & $^{+0.4}_{-0.4} $ & $^{+0.5}_{-0.5} $ & $^{+2.4}_{-2.4} $ & $^{+2.0}_{-2.0}$\\
$3568$ & $0.07$ & $1.22\cdot 10^{-1}$ & $ 28 $ & $^{+23}_{-15}$ & $ ^{+4.0}_{-4.6} $ & $^{+0.9}_{-1.2} $ & $^{+1.1}_{-2.3} $ & $^{-0.0}_{-0.2} $ & $^{+0.7}_{-0.7} $ & $^{+0.0}_{-0.0} $ & $^{+2.4}_{-2.4} $ & $^{+1.7}_{-1.7}$\\
$ $ & $0.11$ & $3.80\cdot 10^{-2}$ & $ 14 $ & $^{+35}_{-20}$ & $ ^{+4.1}_{-8.2} $ & $^{+0.4}_{-7.1} $ & $^{+0.9}_{-1.4} $ & $^{+0.2}_{+1.3} $ & $^{+0.5}_{-0.8} $ & $^{+0.1}_{-0.1} $ & $^{+2.4}_{-2.4} $ & $^{+1.7}_{-1.7}$\\
$ $ & $0.17$ & $3.06\cdot 10^{-2}$ & $ 16 $ & $^{+32}_{-19}$ & $ ^{+4.1}_{-4.4} $ & $^{+0.2}_{-0.7} $ & $^{+1.7}_{-1.8} $ & $^{-0.9}_{-0.1} $ & $^{+0.7}_{-1.2} $ & $^{+0.1}_{-0.1} $ & $^{+2.4}_{-2.4} $ & $^{+1.7}_{-1.7}$\\
$ $ & $0.25$ & $2.37\cdot 10^{-2}$ & $ 17 $ & $^{+31}_{-19}$ & $ ^{+4.7}_{-4.2} $ & $^{+0.2}_{-0.4} $ & $^{+1.5}_{-1.3} $ & $^{-1.4}_{+2.3} $ & $^{+1.1}_{-0.5} $ & $^{+0.0}_{-0.0} $ & $^{+2.4}_{-2.4} $ & $^{+1.7}_{-1.7}$\\
$ $ & $0.35$ & $1.02\cdot 10^{-2}$ & $ 10 $ & $^{+43}_{-23}$ & $ ^{+4.6}_{-4.6} $ & $^{+1.9}_{-0.9} $ & $^{+0.8}_{-2.0} $ & $^{-0.9}_{+1.7} $ & $^{+0.8}_{-1.4} $ & $^{+0.3}_{-0.3} $ & $^{+2.4}_{-2.4} $ & $^{+1.7}_{-1.7}$\\
$ $ & $0.47$ & $8.33\cdot 10^{-4}$ & $ 1 $ & $^{+220}_{-30}$ & $ ^{+7.5}_{-5.7} $ & $^{+2.9}_{-1.9} $ & $^{+3.7}_{-2.5} $ & $^{-2.8}_{+4.3} $ & $^{+1.2}_{+0.1} $ & $^{+0.6}_{-0.6} $ & $^{+2.4}_{-2.4} $ & $^{+1.7}_{-1.7}$\\
$4145$ & $0.08$ & $5.23\cdot 10^{-2}$ & $ 15 $ & $^{+33}_{-20}$ & $ ^{+7.8}_{-4.0} $ & $^{+6.5}_{-1.2} $ & $^{+2.0}_{-0.6} $ & $^{+1.3}_{-0.2} $ & $^{+0.6}_{-0.9} $ & $^{+0.0}_{-0.0} $ & $^{+2.4}_{-2.4} $ & $^{+1.6}_{-1.6}$\\
$ $ & $0.13$ & $3.39\cdot 10^{-2}$ & $ 16 $ & $^{+32}_{-19}$ & $ ^{+4.1}_{-4.2} $ & $^{+0.3}_{-0.8} $ & $^{+1.3}_{-0.3} $ & $^{-1.8}_{+1.1} $ & $^{+0.3}_{-0.6} $ & $^{+0.1}_{-0.1} $ & $^{+2.4}_{-2.4} $ & $^{+1.6}_{-1.6}$\\
$ $ & $0.19$ & $1.88\cdot 10^{-2}$ & $ 13 $ & $^{+36}_{-21}$ & $ ^{+4.0}_{-4.2} $ & $^{+0.8}_{-0.5} $ & $^{+1.3}_{-1.6} $ & $^{-0.7}_{+0.4} $ & $^{+0.2}_{-0.9} $ & $^{+0.1}_{-0.1} $ & $^{+2.4}_{-2.4} $ & $^{+1.6}_{-1.6}$\\
$ $ & $0.28$ & $1.49\cdot 10^{-2}$ & $ 14 $ & $^{+35}_{-20}$ & $ ^{+4.0}_{-4.2} $ & $^{+0.3}_{-0.6} $ & $^{+0.4}_{-1.1} $ & $^{-1.3}_{+1.4} $ & $^{+0.2}_{-1.0} $ & $^{+0.1}_{-0.1} $ & $^{+2.4}_{-2.4} $ & $^{+1.6}_{-1.6}$\\
$ $ & $0.39$ & $3.33\cdot 10^{-3}$ & $ 4 $ & $^{+79}_{-29}$ & $ ^{+5.6}_{-3.8} $ & $^{+2.5}_{-0.3} $ & $^{+1.8}_{+0.0} $ & $^{-0.7}_{+2.5} $ & $^{+0.6}_{+1.3} $ & $^{+0.4}_{-0.4} $ & $^{+2.4}_{-2.4} $ & $^{+1.6}_{-1.6}$\\
$ $ & $0.51$ & $1.36\cdot 10^{-3}$ & $ 2 $ & $^{+130}_{-32}$ & $ ^{+5.9}_{-8.9} $ & $^{+3.3}_{-2.6} $ & $^{+1.5}_{-0.6} $ & $^{-6.2}_{+2.9} $ & $^{-4.1}_{-1.7} $ & $^{+0.7}_{-0.7} $ & $^{+2.4}_{-2.4} $ & $^{+1.6}_{-1.6}$\\
$4806$ & $0.11$ & $3.28\cdot 10^{-2}$ & $ 14 $ & $^{+35}_{-20}$ & $ ^{+4.5}_{-4.4} $ & $^{+0.2}_{-0.7} $ & $^{+1.5}_{-2.4} $ & $^{+1.1}_{+1.1} $ & $^{+1.8}_{-0.4} $ & $^{+0.0}_{+0.0} $ & $^{+2.4}_{-2.4} $ & $^{+1.4}_{-1.4}$\\
$ $ & $0.16$ & $1.70\cdot 10^{-2}$ & $ 10 $ & $^{+43}_{-23}$ & $ ^{+4.5}_{-4.5} $ & $^{+0.3}_{-0.8} $ & $^{+0.6}_{-1.2} $ & $^{-2.2}_{+2.6} $ & $^{+0.9}_{-0.9} $ & $^{+0.1}_{-0.1} $ & $^{+2.4}_{-2.4} $ & $^{+1.4}_{-1.4}$\\
$ $ & $0.23$ & $1.38\cdot 10^{-2}$ & $ 13 $ & $^{+36}_{-21}$ & $ ^{+4.1}_{-4.1} $ & $^{+0.6}_{-0.3} $ & $^{+1.7}_{+0.5} $ & $^{-1.6}_{-0.6} $ & $^{+0.6}_{-0.9} $ & $^{+0.0}_{-0.0} $ & $^{+2.4}_{-2.4} $ & $^{+1.4}_{-1.4}$\\
$ $ & $0.33$ & $5.78\cdot 10^{-3}$ & $ 7 $ & $^{+54}_{-25}$ & $ ^{+4.1}_{-5.2} $ & $^{+1.1}_{-0.3} $ & $^{+1.3}_{-2.2} $ & $^{-2.4}_{+1.0} $ & $^{+0.4}_{-1.8} $ & $^{+0.2}_{-0.2} $ & $^{+2.4}_{-2.4} $ & $^{+1.4}_{-1.4}$\\
$ $ & $0.44$ & $5.73\cdot 10^{-3}$ & $ 8 $ & $^{+49}_{-24}$ & $ ^{+5.0}_{-5.4} $ & $^{+1.2}_{-2.8} $ & $^{+0.7}_{-0.7} $ & $^{-2.6}_{+3.2} $ & $^{+0.2}_{-1.2} $ & $^{+0.4}_{-0.4} $ & $^{+2.4}_{-2.4} $ & $^{+1.4}_{-1.4}$\\
$ $ & $0.56$ & $1.23\cdot 10^{-3}$ & $ 2 $ & $^{+130}_{-32}$ & $ ^{+7.8}_{-12} $ & $^{+4.3}_{-1.1} $ & $^{-3.2}_{-3.9} $ & $^{-9.8}_{+3.3} $ & $^{+4.3}_{+0.1} $ & $^{+0.4}_{-0.4} $ & $^{+2.4}_{-2.4} $ & $^{+1.4}_{-1.4}$\\
$5561$ & $0.12$ & $1.90\cdot 10^{-2}$ & $ 10 $ & $^{+43}_{-24}$ & $ ^{+4.1}_{-4.5} $ & $^{+0.2}_{-1.3} $ & $^{+1.8}_{-1.2} $ & $^{-1.7}_{-0.5} $ & $^{+1.2}_{-1.4} $ & $^{+0.0}_{-0.0} $ & $^{+2.4}_{-2.4} $ & $^{+1.2}_{-1.2}$\\
$ $ & $0.18$ & $1.03\cdot 10^{-2}$ & $ 8 $ & $^{+49}_{-24}$ & $ ^{+5.8}_{-3.7} $ & $^{+0.3}_{-0.4} $ & $^{+0.8}_{-1.1} $ & $^{+1.0}_{+4.5} $ & $^{+0.1}_{-0.1} $ & $^{+0.1}_{-0.1} $ & $^{+2.4}_{-2.4} $ & $^{+1.2}_{-1.2}$\\
$ $ & $0.26$ & $4.73\cdot 10^{-3}$ & $ 6 $ & $^{+60}_{-26}$ & $ ^{+4.0}_{-4.0} $ & $^{+0.5}_{-0.3} $ & $^{+1.2}_{-1.2} $ & $^{-1.5}_{+1.5} $ & $^{+0.0}_{-0.4} $ & $^{+0.1}_{-0.1} $ & $^{+2.4}_{-2.4} $ & $^{+1.2}_{-1.2}$\\
$ $ & $0.37$ & $6.69\cdot 10^{-4}$ & $ 1 $ & $^{+220}_{-30}$ & $ ^{+4.2}_{-4.3} $ & $^{+1.2}_{-0.6} $ & $^{+0.8}_{-0.4} $ & $^{-2.4}_{+1.3} $ & $^{+0.6}_{+0.8} $ & $^{+0.3}_{-0.3} $ & $^{+2.4}_{-2.4} $ & $^{+1.2}_{-1.2}$\\
$ $ & $0.49$ & $1.70\cdot 10^{-3}$ & $ 3 $ & $^{+96}_{-30}$ & $ ^{+5.6}_{-7.2} $ & $^{+1.7}_{-3.5} $ & $^{-2.1}_{-4.0} $ & $^{-2.4}_{+3.9} $ & $^{+0.9}_{-0.5} $ & $^{+0.6}_{-0.6} $ & $^{+2.4}_{-2.4} $ & $^{+1.2}_{-1.2}$\\
$ $ & $0.61$ & $ < 6.44\cdot 10^{-4}$ & $ 0 $ & $ $ & $ $ & $ $ & $  $ & $  $ & $  $ & $  $ & $  $ & $ $\\
$6966$ & $0.14$ & $1.37\cdot 10^{-2}$ & $ 22 $ & $^{+26}_{-17}$ & $ ^{+4.1}_{-5.2} $ & $^{+0.9}_{-3.1} $ & $^{+1.8}_{-2.1} $ & $^{-0.9}_{+0.1} $ & $^{+0.8}_{-0.7} $ & $^{+0.0}_{-0.0} $ & $^{+2.4}_{-2.4} $ & $^{+1.0}_{-1.0}$\\
$ $ & $0.21$ & $6.68\cdot 10^{-3}$ & $ 14 $ & $^{+35}_{-20}$ & $ ^{+4.1}_{-4.2} $ & $^{+0.4}_{-0.8} $ & $^{+1.5}_{-1.5} $ & $^{-1.8}_{+1.3} $ & $^{+0.7}_{-0.5} $ & $^{+0.0}_{-0.0} $ & $^{+2.4}_{-2.4} $ & $^{+1.0}_{-1.0}$\\
$ $ & $0.30$ & $2.06\cdot 10^{-3}$ & $ 7 $ & $^{+54}_{-25}$ & $ ^{+4.6}_{-4.1} $ & $^{+0.3}_{-0.5} $ & $^{+1.8}_{-0.9} $ & $^{-2.0}_{+2.3} $ & $^{+0.7}_{-0.6} $ & $^{+0.2}_{-0.2} $ & $^{+2.4}_{-2.4} $ & $^{+1.0}_{-1.0}$\\
$ $ & $0.41$ & $1.87\cdot 10^{-3}$ & $ 7 $ & $^{+54}_{-25}$ & $ ^{+4.7}_{-4.4} $ & $^{+0.8}_{-0.3} $ & $^{+1.7}_{-0.8} $ & $^{-2.4}_{+2.5} $ & $^{+0.6}_{-0.8} $ & $^{+0.4}_{-0.4} $ & $^{+2.4}_{-2.4} $ & $^{+1.0}_{-1.0}$\\
$ $ & $0.53$ & $2.14\cdot 10^{-4}$ & $ 1 $ & $^{+220}_{-30}$ & $ ^{+6.1}_{-4.9} $ & $^{+0.3}_{-0.7} $ & $^{+1.6}_{-0.8} $ & $^{-3.2}_{+4.7} $ & $^{+0.6}_{-1.0} $ & $^{+0.5}_{-0.5} $ & $^{+2.4}_{-2.4} $ & $^{+1.0}_{-1.0}$\\
$ $ & $0.69$ & $1.17\cdot 10^{-4}$ & $ 1 $ & $^{+220}_{-30}$ & $ ^{+18}_{-17} $ & $^{+14}_{-15} $ & $^{+4.2}_{+4.5} $ & $^{-7.0}_{+9.5} $ & $^{-0.5}_{-1.4} $ & $^{+0.6}_{-0.6} $ & $^{+2.4}_{-2.4} $ & $^{+1.0}_{-1.0}$\\
$9059$ & $0.13$ & $5.13\cdot 10^{-3}$ & $ 4 $ & $^{+79}_{-29}$ & $ ^{+8.1}_{-23} $ & $^{+0.9}_{-22} $ & $^{+2.8}_{-3.4} $ & $^{+6.6}_{-2.8} $ & $^{+1.7}_{-1.5} $ & $^{+0.0}_{-0.0} $ & $^{+2.4}_{-2.4} $ & $^{+0.7}_{-0.7}$\\
$ $ & $0.19$ & $3.64\cdot 10^{-3}$ & $ 8 $ & $^{+49}_{-24}$ & $ ^{+4.4}_{-5.0} $ & $^{+0.7}_{-2.1} $ & $^{+2.3}_{-2.5} $ & $^{-1.0}_{+1.1} $ & $^{+1.1}_{-1.5} $ & $^{+0.1}_{-0.1} $ & $^{+2.4}_{-2.4} $ & $^{+0.7}_{-0.7}$\\
$ $ & $0.27$ & $3.10\cdot 10^{-3}$ & $ 11 $ & $^{+40}_{-22}$ & $ ^{+4.0}_{-4.3} $ & $^{+0.5}_{-1.0} $ & $^{+1.1}_{-1.6} $ & $^{-1.8}_{+1.8} $ & $^{+0.6}_{-0.7} $ & $^{+0.1}_{-0.1} $ & $^{+2.4}_{-2.4} $ & $^{+0.7}_{-0.7}$\\
$ $ & $0.38$ & $3.84\cdot 10^{-4}$ & $ 2 $ & $^{+130}_{-32}$ & $ ^{+4.5}_{-5.7} $ & $^{+0.8}_{-0.6} $ & $^{+0.8}_{+0.3} $ & $^{-4.5}_{+2.7} $ & $^{+0.5}_{-0.3} $ & $^{+0.4}_{-0.4} $ & $^{+2.4}_{-2.4} $ & $^{+0.7}_{-0.7}$\\
$ $ & $0.51$ & $1.73\cdot 10^{-4}$ & $ 1 $ & $^{+220}_{-30}$ & $ ^{+5.5}_{-4.9} $ & $^{+1.5}_{-1.2} $ & $^{+1.2}_{-1.0} $ & $^{-3.1}_{+3.8} $ & $^{+0.6}_{-0.7} $ & $^{+0.6}_{-0.6} $ & $^{+2.4}_{-2.4} $ & $^{+0.7}_{-0.7}$\\
$ $ & $0.64$ & $ < 1.83\cdot 10^{-4}$ & $ 0 $ & $ $ & $ $ & $ $ & $  $ & $  $ & $  $ & $  $ & $  $ & $ $\\
$ $ & $0.78$ & $9.12\cdot 10^{-5}$ & $ 1 $ & $^{+220}_{-30}$ & $ ^{+9.9}_{-13} $ & $^{+1.8}_{-1.1} $ & $^{+1.0}_{-1.6} $ & $^{-12}_{+8.8} $ & $^{+0.4}_{-0.3} $ & $^{+2.2}_{-2.2} $ & $^{+2.4}_{-2.4} $ & $^{+0.7}_{-0.7}$\\
$ $ & $0.93$ & $ < 1.08\cdot 10^{-5}$ & $ 0 $ & $ $ & $ $ & $ $ & $  $ & $  $ & $  $ & $  $ & $  $ & $ $\\
$15072$ & $0.61$ & $8.57\cdot 10^{-5}$ & $ 3 $ & $^{+96}_{-30}$ & $ ^{+22}_{-25} $ & $^{+18}_{-20} $ & $^{-5.9}_{-9.5} $ & $^{-8.9}_{+11} $ & $^{+0.4}_{-1.1} $ & $^{+0.9}_{-0.9} $ & $^{+2.4}_{-2.4} $ & $^{+0.5}_{-0.5}$\\
$ $ & $0.75$ & $ < 1.97\cdot 10^{-5}$ & $ 0 $ & $ $ & $ $ & $ $ & $  $ & $  $ & $  $ & $  $ & $  $ & $ $\\
$ $ & $0.91$ & $ < 2.04\cdot 10^{-6}$ & $ 0 $ & $ $ & $ $ & $ $ & $  $ & $  $ & $  $ & $  $ & $  $ & $ $\\
\hline
\end{supertabular}
\end{center}

%% file: DESY-06-116-tab-4.tex
\begin{center}
\footnotesize
\renewcommand{\arraystretch}{1.2}
\tablehead{
\hline
{$Q^2$} & 
\multicolumn{1}{c|}{$x_{\rm edge}$} & 
$\int_{x_{\rm edge}}^{1} {d^2\sigma}/{dxdQ^2}$  &
$N$& 
$\delta_{s}$& 
$\delta_t$ & 
$\delta_u$ &
$\delta_1$ &
$\delta_2$ &
$\delta_3$ &
$\delta_4$ &
$\delta_5$ &
$\delta_6$ \\
{($\Gev^2$)} & 
$$ &
{($\rm pb/GeV^{2}$)} &
$$ &
(\%) &
(\%) &
(\%) &
(\%) &
(\%) &
(\%) &
(\%) &
(\%) &
(\%) \\
\hline \hline}
\tablelasttail{
\hline}
\normalsize
\bottomcaption{
    The integral cross section table for 98-99 $e^-p$ NC scattering.
    The first two columns of the table contain the $Q^2$ and $x_{\rm edge}$ values 
    for the bin, the third contains the measured
    cross section $\int_{x_{\rm edge}}^{1} {d^2\sigma}/{dxdQ^2}$ corrected to the electroweak 
    Born level or the upper limit in case of zero observed 
    events, the fourth contains the number of events reconstructed in 
    the bin, $N$, the fifth contains the 
    statistical uncertainty, $\delta_s$, and the sixth contains the 
     total systematic uncertainty, $\delta_t$.
    The right part of the table lists the total uncorrelated
    systematic uncertainty, $\delta_u$, followed by the bin-to-bin correlated
    systematic uncertainties $\delta_1$--\,$\delta_6$ defined in
    the text.
    The upper (lower) numbers refer to the variation of 
    the cross section, whereas the signs of
    the numbers reflect the direction of change in the cross
    sections. Note that the normalization uncertainty, $\delta_7$ is
    not listed.
    } 
\label{tab:9899crosl}
\begin{supertabular}{|c|l|c|c|c|c||c|c|c|c|c|c|c|}
$ 648$ & $0.22$ & $8.74\cdot 10^{-2}$ & $ 14 $ & $^{+35}_{-20}$ & $ ^{+8.7}_{-8.2} $ & $^{+3.7}_{-1.9} $ & $^{-5.9}_{+5.6} $ & $^{+0.0}_{+0.0} $ & $^{-3.0}_{+3.4} $ & $^{+0.3}_{-0.3} $ & $^{+1.6}_{-1.6} $ & $^{+3.4}_{-3.4}$\\
$ 761$ & $0.24$ & $7.12\cdot 10^{-2}$ & $ 54 $ & $^{+16}_{-12}$ & $ ^{+5.5}_{-6.0} $ & $^{+0.5}_{-0.5} $ & $^{-3.7}_{+2.7} $ & $^{+0.0}_{-0.1} $ & $^{-1.9}_{+2.1} $ & $^{+0.2}_{-0.2} $ & $^{+1.6}_{-1.6} $ & $^{+3.3}_{-3.3}$\\
$ 891$ & $0.26$ & $3.44\cdot 10^{-2}$ & $ 56 $ & $^{+15}_{-12}$ & $ ^{+4.5}_{-4.4} $ & $^{+1.3}_{-0.6} $ & $^{+0.3}_{-0.6} $ & $^{+0.0}_{-0.1} $ & $^{+0.6}_{-0.6} $ & $^{+0.2}_{-0.2} $ & $^{+1.6}_{-1.6} $ & $^{+3.2}_{-3.2}$\\
$1045$ & $0.29$ & $2.29\cdot 10^{-2}$ & $ 54 $ & $^{+16}_{-12}$ & $ ^{+4.5}_{-4.5} $ & $^{+1.3}_{-0.9} $ & $^{+1.5}_{-1.5} $ & $^{+0.0}_{-0.2} $ & $^{+0.7}_{-1.0} $ & $^{+0.3}_{-0.3} $ & $^{+1.6}_{-1.6} $ & $^{+2.9}_{-2.9}$\\
$1224$ & $0.31$ & $9.38\cdot 10^{-3}$ & $ 28 $ & $^{+23}_{-15}$ & $ ^{+4.9}_{-5.5} $ & $^{+0.2}_{-0.5} $ & $^{+2.4}_{-3.6} $ & $^{-0.2}_{-0.2} $ & $^{+1.4}_{-1.1} $ & $^{+0.2}_{-0.2} $ & $^{+1.6}_{-1.6} $ & $^{+2.9}_{-2.9}$\\
$1431$ & $0.34$ & $8.51\cdot 10^{-3}$ & $ 29 $ & $^{+22}_{-15}$ & $ ^{+5.4}_{-5.5} $ & $^{+0.2}_{-0.1} $ & $^{+3.6}_{-2.9} $ & $^{+0.3}_{+0.2} $ & $^{+0.6}_{-2.4} $ & $^{+0.2}_{-0.2} $ & $^{+1.6}_{-1.6} $ & $^{+2.8}_{-2.8}$\\
$1672$ & $0.36$ & $6.31\cdot 10^{-3}$ & $ 25 $ & $^{+24}_{-16}$ & $ ^{+4.4}_{-4.5} $ & $^{+0.8}_{-0.3} $ & $^{+1.4}_{-1.9} $ & $^{-0.4}_{-0.5} $ & $^{+1.7}_{-1.6} $ & $^{+0.2}_{-0.2} $ & $^{+1.6}_{-1.6} $ & $^{+2.5}_{-2.5}$\\
$1951$ & $0.39$ & $3.34\cdot 10^{-3}$ & $ 16 $ & $^{+32}_{-19}$ & $ ^{+6.9}_{-4.3} $ & $^{+1.8}_{-1.6} $ & $^{+5.1}_{-1.6} $ & $^{+1.1}_{+1.0} $ & $^{+2.0}_{-0.9} $ & $^{+0.2}_{-0.2} $ & $^{+1.6}_{-1.6} $ & $^{+2.2}_{-2.2}$\\
$2273$ & $0.43$ & $1.61\cdot 10^{-3}$ & $ 8 $ & $^{+49}_{-24}$ & $ ^{+5.6}_{-7.5} $ & $^{+3.1}_{-2.0} $ & $^{+2.9}_{-5.4} $ & $^{-0.5}_{-0.5} $ & $^{+1.0}_{-3.1} $ & $^{+0.2}_{-0.2} $ & $^{+1.6}_{-1.6} $ & $^{+2.2}_{-2.2}$\\
$2644$ & $0.46$ & $8.11\cdot 10^{-4}$ & $ 5 $ & $^{+67}_{-27}$ & $ ^{+6.3}_{-5.6} $ & $^{+2.1}_{-0.2} $ & $^{+3.9}_{-4.0} $ & $^{+1.4}_{+1.4} $ & $^{+1.9}_{-1.8} $ & $^{+0.2}_{-0.2} $ & $^{+1.6}_{-1.6} $ & $^{+2.1}_{-2.1}$\\
$3073$ & $0.50$ & $ < 1.62\cdot 10^{-4}$ &  $ 0 $ & $  $ & $ $ & $ $ & $  $ & $  $ & $  $ & $  $ & $  $ & $ $\\
$3568$ & $0.54$ & $2.43\cdot 10^{-4}$ & $ 2 $ & $^{+130}_{-32}$ & $ ^{+7.6}_{-4.5} $ & $^{+3.3}_{-0.5} $ & $^{+5.3}_{-2.5} $ & $^{+1.1}_{+0.8} $ & $^{+2.5}_{-1.9} $ & $^{+0.2}_{-0.2} $ & $^{+1.6}_{-1.6} $ & $^{+1.7}_{-1.7}$\\
$4145$ & $0.58$ & $2.06\cdot 10^{-4}$ & $ 2 $ & $^{+130}_{-32}$ & $ ^{+6.2}_{-4.6} $ & $^{+0.7}_{-0.7} $ & $^{+3.6}_{-1.7} $ & $^{+0.1}_{+0.1} $ & $^{+3.9}_{-2.8} $ & $^{+0.1}_{-0.1} $ & $^{+1.6}_{-1.6} $ & $^{+1.6}_{-1.6}$\\
$4806$ & $0.63$ & $ < 1.02\cdot 10^{-4}$ & $0$& $  $ & $ $ & $ $ & $  $ & $  $ & $  $ & $  $ & $  $ & $ $\\
$5561$ & $0.68$ & $ < 8.95\cdot 10^{-5}$ & $0$ & $  $ & $ $ & $ $ & $  $ & $  $ & $  $ & $  $ & $  $ & $ $\\
$6966$ & $0.79$ & $ < 1.38\cdot 10^{-5}$ & $0$ & $  $ & $ $ & $ $ & $  $ & $  $ & $  $ & $  $ & $  $ & $ $\\
\hline
\end{supertabular}
\end{center}

%% file: DESY-06-116-tab-5.tex
\begin{center}
\footnotesize
\renewcommand{\arraystretch}{1.2}
\tablehead{
\multicolumn{13}{l}{
{\normalsize}}\\
\hline
{$Q^2$} & 
\multicolumn{1}{c|}{$x$} & 
${d^2\sigma}/{dxdQ^2}$  &
$N$& 
$\delta_s$& 
$\delta_t$& 
$\delta_u$& 
$\delta_1$ &
$\delta_2$ &
$\delta_3$ &
$\delta_4$ &
$\delta_5$ &
$\delta_6$ \\
{($\Gev^2$)} & 
$$ &
{($\rm pb/GeV^{2}$)} &
$$ &
(\%) &
(\%) &
(\%) &
(\%) &
(\%) &
(\%) &
(\%) &
(\%) &
(\%) \\
\hline \hline}
\tablelasttail{
\hline
\multicolumn{11}{r}{} \\}
\normalsize
\bottomcaption{
    The cross section table for 99-00 $e^+p$ NC scattering.
    The first two columns of the table contain the $Q^2$ and $x$ values 
    at which the cross section is quoted, the third contains the measured
    cross section ${d^2\sigma}/{dxdQ^2}$ corrected to the electroweak 
    Born level or the upper limit in case of zero observed 
    events, the fourth contains the number of events reconstructed in 
    the bin, $N$, the fifth contains the 
    statistical uncertainty, $\delta_s$, and the sixth contains the 
    total systematic uncertainty, $\delta_t$. 
    The right part of the table lists the total uncorrelated
    systematic uncertainty, $\delta_u$, followed by the bin-to-bin correlated
    systematic uncertainties $\delta_1$--\,$\delta_6$ defined in
    the text.
    The upper (lower) numbers refer to the variation of 
    the cross section, whereas the signs of
    the numbers reflect the direction of change in the cross
    sections. Note that the normalization uncertainty, $\delta_7$ is
    not listed.
    } 
\label{tab:9900cros}
\begin{supertabular}{|c|l|c|c|c|c||c|c|c|c|c|c|c|}
$ 648$ & $0.08$ & $3.02$ & $ 255 $ & $^{+6.6}_{-5.9}$ & $ ^{+7.8}_{-7.1} $ & $^{+1.3}_{-1.0} $ & $^{+5.8}_{-5.2} $ & $^{-0.9}_{+1.8} $ & $^{+0.2}_{-0.3} $ & $^{+0.3}_{-0.3}$ & $^{+2.4}_{-2.4}$ & $^{+3.3}_{-3.3}$\\
$ $ & $0.13$ & $1.86$ & $ 116 $ & $^{+10}_{-8.4}$ & $ ^{+7.5}_{-7.5} $ & $^{+2.1}_{-1.3} $ & $^{+5.2}_{-5.6} $ & $^{-1.0}_{+0.6} $ & $^{+1.8}_{+0.3} $ & $^{+0.2}_{-0.2}$ & $^{+2.4}_{-2.4}$ & $^{+3.3}_{-3.3}$\\
$ $ & $0.19$ & $1.05$ & $ 87 $ & $^{+11}_{-9.6}$ & $ ^{+11}_{-9.7} $ & $^{+4.7}_{-3.4} $ & $^{+8.6}_{-7.4} $ & $^{-1.2}_{+1.8} $ & $^{+1.0}_{-2.3} $ & $^{+0.1}_{-0.1}$ & $^{+2.4}_{-2.4}$ & $^{+3.3}_{-3.3}$\\
$ 761$ & $0.09$ & $1.85$ & $ 403 $ & $^{+5.2}_{-4.7}$ & $ ^{+5.2}_{-5.5} $ & $^{+0.4}_{-0.9} $ & $^{+2.1}_{-2.6} $ & $^{-0.6}_{+0.7} $ & $^{+0.0}_{+0.1} $ & $^{+0.2}_{-0.2}$ & $^{+2.4}_{-2.4}$ & $^{+3.3}_{-3.3}$\\
$ $ & $0.14$ & $9.89\cdot 10^{-1}$ & $ 216 $ & $^{+7.3}_{-6.3}$ & $ ^{+6.0}_{-5.2} $ & $^{+0.6}_{-1.2} $ & $^{+3.6}_{-2.0} $ & $^{-0.7}_{+0.9} $ & $^{+0.9}_{-0.2} $ & $^{+0.0}_{-0.0}$ & $^{+2.4}_{-2.4}$ & $^{+3.3}_{-3.3}$\\
$ $ & $0.21$ & $6.18\cdot 10^{-1}$ & $ 161 $ & $^{+8.5}_{-7.3}$ & $ ^{+4.9}_{-5.4} $ & $^{+1.0}_{-0.7} $ & $^{+1.9}_{-3.1} $ & $^{-1.4}_{+1.8} $ & $^{+2.0}_{-2.4} $ & $^{+0.2}_{-0.2}$ & $^{+2.4}_{-2.4}$ & $^{+1.0}_{-1.0}$\\
$ 891$ & $0.10$ & $1.23$ & $ 471 $ & $^{+4.8}_{-4.4}$ & $ ^{+4.7}_{-4.9} $ & $^{+0.3}_{-1.1} $ & $^{+0.5}_{-1.0} $ & $^{-0.7}_{+0.2} $ & $^{+0.3}_{-0.2} $ & $^{+0.1}_{-0.1}$ & $^{+2.4}_{-2.4}$ & $^{+3.3}_{-3.3}$\\
$ $ & $0.15$ & $7.65\cdot 10^{-1}$ & $ 301 $ & $^{+6.1}_{-5.4}$ & $ ^{+5.1}_{-4.9} $ & $^{+1.6}_{-0.4} $ & $^{-0.6}_{+0.1} $ & $^{-0.7}_{+1.1} $ & $^{+0.7}_{-1.1} $ & $^{+0.1}_{-0.1}$ & $^{+2.4}_{-2.4}$ & $^{+3.3}_{-3.3}$\\
$ $ & $0.22$ & $3.64\cdot 10^{-1}$ & $ 189 $ & $^{+7.9}_{-6.8}$ & $ ^{+4.9}_{-4.3} $ & $^{+2.2}_{-0.3} $ & $^{+2.4}_{-1.5} $ & $^{-1.7}_{+1.1} $ & $^{+0.9}_{-1.1} $ & $^{+0.2}_{-0.2}$ & $^{+2.4}_{-2.4}$ & $^{+1.0}_{-1.0}$\\
$1045$ & $0.07$ & $1.49$ & $ 532 $ & $^{+4.5}_{-4.2}$ & $ ^{+5.5}_{-5.2} $ & $^{+0.7}_{-0.6} $ & $^{-2.3}_{+2.9} $ & $^{-0.1}_{+0.2} $ & $^{-0.6}_{+0.6} $ & $^{+0.1}_{-0.1}$ & $^{+2.4}_{-2.4}$ & $^{+3.3}_{-3.3}$\\
$ $ & $0.11$ & $7.98\cdot 10^{-1}$ & $ 388 $ & $^{+5.3}_{-4.8}$ & $ ^{+4.9}_{-4.9} $ & $^{+0.9}_{-0.1} $ & $^{-0.2}_{+0.8} $ & $^{-1.4}_{+1.0} $ & $^{+0.0}_{-0.3} $ & $^{+0.1}_{-0.1}$ & $^{+2.4}_{-2.4}$ & $^{+3.3}_{-3.3}$\\
$ $ & $0.17$ & $4.82\cdot 10^{-1}$ & $ 253 $ & $^{+6.7}_{-5.9}$ & $ ^{+4.9}_{-4.8} $ & $^{+0.9}_{-0.7} $ & $^{+0.1}_{+0.5} $ & $^{-0.1}_{+0.1} $ & $^{+1.3}_{-1.1} $ & $^{+0.1}_{-0.1}$ & $^{+2.4}_{-2.4}$ & $^{+3.3}_{-3.3}$\\
$ $ & $0.24$ & $2.21\cdot 10^{-1}$ & $ 173 $ & $^{+8.2}_{-7.1}$ & $ ^{+4.3}_{-4.5} $ & $^{+0.6}_{-1.1} $ & $^{-2.0}_{+1.9} $ & $^{-0.8}_{+1.1} $ & $^{+1.3}_{-1.6} $ & $^{+0.2}_{-0.2}$ & $^{+2.4}_{-2.4}$ & $^{+1.0}_{-1.0}$\\
$1224$ & $0.07$ & $1.03$ & $ 444 $ & $^{+5.0}_{-4.5}$ & $ ^{+4.9}_{-5.0} $ & $^{+0.8}_{-0.2} $ & $^{-1.7}_{+1.4} $ & $^{-0.1}_{+0.1} $ & $^{-0.4}_{+0.4} $ & $^{+0.2}_{-0.2}$ & $^{+2.4}_{-2.4}$ & $^{+3.3}_{-3.3}$\\
$ $ & $0.12$ & $4.81\cdot 10^{-1}$ & $ 294 $ & $^{+6.2}_{-5.5}$ & $ ^{+4.9}_{-5.1} $ & $^{+0.4}_{-0.9} $ & $^{-1.7}_{+1.6} $ & $^{-0.7}_{+0.1} $ & $^{-0.1}_{-0.3} $ & $^{+0.0}_{-0.0}$ & $^{+2.4}_{-2.4}$ & $^{+3.3}_{-3.3}$\\
$ $ & $0.18$ & $2.99\cdot 10^{-1}$ & $ 208 $ & $^{+7.4}_{-6.4}$ & $ ^{+4.8}_{-5.0} $ & $^{+0.5}_{-0.5} $ & $^{-0.8}_{+0.7} $ & $^{-1.0}_{+0.1} $ & $^{+0.9}_{-1.4} $ & $^{+0.1}_{-0.1}$ & $^{+2.4}_{-2.4}$ & $^{+3.3}_{-3.3}$\\
$ $ & $0.26$ & $1.31\cdot 10^{-1}$ & $ 139 $ & $^{+9.2}_{-7.8}$ & $ ^{+5.3}_{-3.8} $ & $^{+0.4}_{-0.6} $ & $^{-0.9}_{+0.9} $ & $^{-0.2}_{+1.9} $ & $^{+3.4}_{-1.2} $ & $^{+0.1}_{-0.1}$ & $^{+2.4}_{-2.4}$ & $^{+1.0}_{-1.0}$\\
\end{supertabular}
\end{center}

\begin{center}
\footnotesize
\renewcommand{\arraystretch}{1.2}
\tablehead{
\multicolumn{13}{l}{
{\normalsize {\bf Table \thetable\ } (continued):}}\\
\hline
{$Q^2$} & 
\multicolumn{1}{c|}{$x$} & 
${d^2\sigma}/{dxdQ^2}$  &
$N$& 
$\delta_s$& 
$\delta_t$& 
$\delta_u$& 
$\delta_1$ &
$\delta_2$ &
$\delta_3$ &
$\delta_4$ &
$\delta_5$ &
$\delta_6$ \\
{($\Gev^2$)} & 
$$ &
{($\rm pb/GeV^{2}$)} &
$$ &
(\%) &
(\%) &
(\%) &
(\%) &
(\%) &
(\%) &
(\%) &
(\%) &
(\%) \\
\hline \hline}
\tabletail{
\hline
}
\tablelasttail{\hline}
\normalsize
\begin{supertabular}{|c|l|c|c|c|c||c|c|c|c|c|c|c|}
$1431$ & $0.09$ & $5.29\cdot 10^{-1}$ & $ 278 $ & $^{+6.4}_{-5.6}$ & $ ^{+5.0}_{-4.7} $ & $^{+0.5}_{-0.3} $ & $^{-0.4}_{+1.7} $ & $^{-0.0}_{-0.4} $ & $^{-0.2}_{+0.2} $ & $^{+0.1}_{-0.1}$ & $^{+2.4}_{-2.4}$ & $^{+3.3}_{-3.3}$\\
$ $ & $0.14$ & $2.74\cdot 10^{-1}$ & $ 212 $ & $^{+7.3}_{-6.4}$ & $ ^{+5.0}_{-4.8} $ & $^{+0.6}_{-0.3} $ & $^{-0.8}_{+0.9} $ & $^{-0.9}_{+1.4} $ & $^{-0.0}_{-0.1} $ & $^{+0.0}_{-0.0}$ & $^{+2.4}_{-2.4}$ & $^{+3.3}_{-3.3}$\\
$ $ & $0.20$ & $1.51\cdot 10^{-1}$ & $ 126 $ & $^{+9.7}_{-8.1}$ & $ ^{+3.8}_{-4.3} $ & $^{+0.8}_{-1.2} $ & $^{-1.4}_{+0.9} $ & $^{-0.6}_{+0.4} $ & $^{+1.0}_{-1.7} $ & $^{+0.2}_{-0.2}$ & $^{+2.4}_{-2.4}$ & $^{+1.0}_{-1.0}$\\
$ $ & $0.29$ & $0.99\cdot 10^{-1}$ & $ 119 $ & $^{+10}_{-8.3}$ & $ ^{+5.2}_{-4.5} $ & $^{+1.9}_{-0.1} $ & $^{-0.2}_{+2.2} $ & $^{-1.8}_{+1.5} $ & $^{+2.2}_{-2.1} $ & $^{+0.0}_{-0.0}$ & $^{+2.4}_{-2.4}$ & $^{+1.0}_{-1.0}$\\
$1672$ & $0.10$ & $3.71\cdot 10^{-1}$ & $ 249 $ & $^{+6.8}_{-6.0}$ & $ ^{+4.9}_{-5.0} $ & $^{+0.6}_{-0.6} $ & $^{-1.6}_{+1.5} $ & $^{-0.1}_{+0.3} $ & $^{-0.2}_{+0.2} $ & $^{+0.1}_{-0.1}$ & $^{+2.4}_{-2.4}$ & $^{+3.3}_{-3.3}$\\
$ $ & $0.15$ & $1.93\cdot 10^{-1}$ & $ 183 $ & $^{+8.0}_{-6.9}$ & $ ^{+4.8}_{-4.9} $ & $^{+0.5}_{-0.7} $ & $^{-1.1}_{+1.1} $ & $^{-0.7}_{+0.7} $ & $^{+0.2}_{-0.1} $ & $^{+0.1}_{-0.1}$ & $^{+2.4}_{-2.4}$ & $^{+3.3}_{-3.3}$\\
$ $ & $0.22$ & $1.05\cdot 10^{-1}$ & $ 113 $ & $^{+10}_{-8.6}$ & $ ^{+3.7}_{-4.1} $ & $^{+0.2}_{-0.8} $ & $^{-1.3}_{+0.8} $ & $^{-0.3}_{+0.1} $ & $^{+1.0}_{-1.4} $ & $^{+0.1}_{-0.1}$ & $^{+2.4}_{-2.4}$ & $^{+1.0}_{-1.0}$\\
$ $ & $0.31$ & $6.10\cdot 10^{-2}$ & $ 96 $ & $^{+11}_{-9.2}$ & $ ^{+5.0}_{-4.2} $ & $^{+1.0}_{-0.3} $ & $^{-0.6}_{+0.6} $ & $^{-2.1}_{+2.9} $ & $^{+1.7}_{-0.7} $ & $^{+0.1}_{-0.1}$ & $^{+2.4}_{-2.4}$ & $^{+1.0}_{-1.0}$\\
$1951$ & $0.07$ & $3.77\cdot 10^{-1}$ & $ 215 $ & $^{+7.3}_{-6.3}$ & $ ^{+4.9}_{-4.9} $ & $^{+0.6}_{-0.4} $ & $^{-1.5}_{+1.1} $ & $^{-0.0}_{+0.6} $ & $^{+0.1}_{-0.1} $ & $^{+0.1}_{-0.1}$ & $^{+2.4}_{-2.4}$ & $^{+3.3}_{-3.3}$\\
$ $ & $0.11$ & $1.89\cdot 10^{-1}$ & $ 149 $ & $^{+8.9}_{-7.5}$ & $ ^{+5.1}_{-5.1} $ & $^{+0.6}_{-0.9} $ & $^{-1.8}_{+2.0} $ & $^{-0.3}_{-0.6} $ & $^{-0.3}_{+0.2} $ & $^{+0.1}_{-0.1}$ & $^{+2.4}_{-2.4}$ & $^{+3.3}_{-3.3}$\\
$ $ & $0.17$ & $1.10\cdot 10^{-1}$ & $ 126 $ & $^{+9.7}_{-8.1}$ & $ ^{+5.2}_{-4.9} $ & $^{+1.2}_{-0.4} $ & $^{-1.0}_{+1.0} $ & $^{-1.2}_{+1.6} $ & $^{+0.5}_{-0.4} $ & $^{+0.1}_{-0.1}$ & $^{+2.4}_{-2.4}$ & $^{+3.3}_{-3.3}$\\
$ $ & $0.24$ & $7.26\cdot 10^{-2}$ & $ 102 $ & $^{+11}_{-9.0}$ & $ ^{+4.2}_{-3.7} $ & $^{+1.1}_{-0.3} $ & $^{-0.8}_{+1.9} $ & $^{-0.5}_{+0.2} $ & $^{+0.9}_{-0.9} $ & $^{+0.1}_{-0.1}$ & $^{+2.4}_{-2.4}$ & $^{+1.0}_{-1.0}$\\
$ $ & $0.34$ & $3.34\cdot 10^{-2}$ & $ 69 $ & $^{+14}_{-11}$ & $ ^{+4.7}_{-4.2} $ & $^{+0.2}_{-1.0} $ & $^{-1.2}_{+0.8} $ & $^{-1.7}_{+2.6} $ & $^{+1.7}_{-0.9} $ & $^{+0.2}_{-0.2}$ & $^{+2.4}_{-2.4}$ & $^{+1.0}_{-1.0}$\\
$2273$ & $0.07$ & $2.41\cdot 10^{-1}$ & $ 179 $ & $^{+8.1}_{-7.0}$ & $ ^{+5.1}_{-5.0} $ & $^{+0.9}_{-0.4} $ & $^{-1.4}_{+1.7} $ & $^{-1.0}_{+0.8} $ & $^{-0.0}_{-0.0} $ & $^{+0.2}_{-0.2}$ & $^{+2.4}_{-2.4}$ & $^{+3.3}_{-3.3}$\\
$ $ & $0.12$ & $1.55\cdot 10^{-1}$ & $ 150 $ & $^{+8.8}_{-7.5}$ & $ ^{+5.1}_{-5.1} $ & $^{+0.6}_{-1.0} $ & $^{-1.7}_{+2.0} $ & $^{-0.4}_{-0.2} $ & $^{-0.5}_{+0.4} $ & $^{+0.0}_{-0.0}$ & $^{+2.4}_{-2.4}$ & $^{+3.3}_{-3.3}$\\
$ $ & $0.18$ & $8.25\cdot 10^{-2}$ & $ 114 $ & $^{+10}_{-8.5}$ & $ ^{+5.1}_{-4.9} $ & $^{+0.2}_{-0.6} $ & $^{-0.6}_{+1.3} $ & $^{-1.4}_{+1.5} $ & $^{+0.2}_{-0.3} $ & $^{+0.1}_{-0.1}$ & $^{+2.4}_{-2.4}$ & $^{+3.3}_{-3.3}$\\
$ $ & $0.26$ & $3.77\cdot 10^{-2}$ & $ 68 $ & $^{+14}_{-11}$ & $ ^{+4.3}_{-3.7} $ & $^{+1.0}_{-0.2} $ & $^{-0.7}_{+1.8} $ & $^{+0.7}_{+0.4} $ & $^{+1.0}_{-1.1} $ & $^{+0.1}_{-0.1}$ & $^{+2.4}_{-2.4}$ & $^{+1.0}_{-1.0}$\\
$ $ & $0.37$ & $1.71\cdot 10^{-2}$ & $ 40 $ & $^{+19}_{-13}$ & $ ^{+5.1}_{-4.9} $ & $^{+0.8}_{-0.3} $ & $^{-1.4}_{+2.3} $ & $^{-2.8}_{+2.2} $ & $^{+1.9}_{-1.4} $ & $^{+0.3}_{-0.3}$ & $^{+2.4}_{-2.4}$ & $^{+1.0}_{-1.0}$\\
$2644$ & $0.09$ & $1.48\cdot 10^{-1}$ & $ 135 $ & $^{+9.4}_{-7.9}$ & $ ^{+5.2}_{-5.1} $ & $^{+0.5}_{-0.5} $ & $^{-1.3}_{+1.8} $ & $^{-1.5}_{+1.4} $ & $^{-0.3}_{+0.3} $ & $^{+0.1}_{-0.1}$ & $^{+2.4}_{-2.4}$ & $^{+3.3}_{-3.3}$\\
$ $ & $0.14$ & $7.74\cdot 10^{-2}$ & $ 107 $ & $^{+11}_{-8.8}$ & $ ^{+4.9}_{-5.1} $ & $^{+0.5}_{-0.6} $ & $^{-2.1}_{+1.6} $ & $^{-0.1}_{+0.1} $ & $^{+0.1}_{-0.0} $ & $^{+0.1}_{-0.1}$ & $^{+2.4}_{-2.4}$ & $^{+3.3}_{-3.3}$\\
$ $ & $0.21$ & $4.65\cdot 10^{-2}$ & $ 77 $ & $^{+13}_{-10}$ & $ ^{+4.3}_{-4.0} $ & $^{+1.5}_{-0.6} $ & $^{-2.0}_{+0.8} $ & $^{-0.4}_{+1.9} $ & $^{+0.2}_{-0.2} $ & $^{+0.1}_{-0.1}$ & $^{+2.4}_{-2.4}$ & $^{+1.0}_{-1.0}$\\
$ $ & $0.29$ & $2.58\cdot 10^{-2}$ & $ 57 $ & $^{+15}_{-11}$ & $ ^{+3.7}_{-4.6} $ & $^{+0.0}_{-1.8} $ & $^{-1.6}_{+0.1} $ & $^{-1.1}_{+0.7} $ & $^{+1.2}_{-1.4} $ & $^{+0.0}_{-0.0}$ & $^{+2.4}_{-2.4}$ & $^{+1.0}_{-1.0}$\\
$ $ & $0.40$ & $1.11\cdot 10^{-2}$ & $ 33 $ & $^{+21}_{-15}$ & $ ^{+5.1}_{-5.6} $ & $^{+0.6}_{-0.5} $ & $^{-0.0}_{+0.0} $ & $^{-3.3}_{+2.4} $ & $^{+1.2}_{-1.3} $ & $^{+0.4}_{-0.4}$ & $^{+2.4}_{-2.4}$ & $^{+2.7}_{-2.7}$\\
$3073$ & $0.06$ & $1.24\cdot 10^{-1}$ & $ 83 $ & $^{+12}_{-9.8}$ & $ ^{+5.1}_{-5.3} $ & $^{+0.3}_{-0.6} $ & $^{-2.4}_{+1.9} $ & $^{+0.8}_{-0.0} $ & $^{+0.4}_{-0.3} $ & $^{+0.1}_{-0.1}$ & $^{+2.4}_{-2.4}$ & $^{+3.3}_{-3.3}$\\
$ $ & $0.10$ & $8.34\cdot 10^{-2}$ & $ 87 $ & $^{+11}_{-9.7}$ & $ ^{+5.2}_{-4.8} $ & $^{+1.2}_{-0.8} $ & $^{-0.9}_{+1.7} $ & $^{-0.4}_{+0.8} $ & $^{-0.4}_{+0.2} $ & $^{+0.1}_{-0.1}$ & $^{+2.4}_{-2.4}$ & $^{+3.3}_{-3.3}$\\
$ $ & $0.15$ & $5.79\cdot 10^{-2}$ & $ 99 $ & $^{+11}_{-9.0}$ & $ ^{+4.9}_{-4.7} $ & $^{+0.5}_{-0.3} $ & $^{-0.7}_{+1.3} $ & $^{-0.5}_{-0.4} $ & $^{-0.2}_{+0.1} $ & $^{+0.1}_{-0.1}$ & $^{+2.4}_{-2.4}$ & $^{+3.3}_{-3.3}$\\
$ $ & $0.23$ & $2.52\cdot 10^{-2}$ & $ 53 $ & $^{+16}_{-12}$ & $ ^{+3.9}_{-3.8} $ & $^{+0.8}_{-0.2} $ & $^{-0.6}_{+1.1} $ & $^{-1.4}_{+0.9} $ & $^{+0.6}_{-0.4} $ & $^{+0.1}_{-0.1}$ & $^{+2.4}_{-2.4}$ & $^{+1.0}_{-1.0}$\\
$ $ & $0.32$ & $1.79\cdot 10^{-2}$ & $ 49 $ & $^{+16}_{-12}$ & $ ^{+4.7}_{-4.4} $ & $^{+0.2}_{-0.9} $ & $^{-2.2}_{+1.7} $ & $^{-1.0}_{+2.6} $ & $^{+0.7}_{-0.7} $ & $^{+0.1}_{-0.1}$ & $^{+2.4}_{-2.4}$ & $^{+1.0}_{-1.0}$\\
$ $ & $0.43$ & $5.58\cdot 10^{-3}$ & $ 21 $ & $^{+27}_{-17}$ & $ ^{+5.5}_{-5.4} $ & $^{+0.2}_{-0.6} $ & $^{-1.0}_{+1.2} $ & $^{-2.9}_{+2.9} $ & $^{+1.2}_{-1.3} $ & $^{+0.5}_{-0.5}$ & $^{+2.4}_{-2.4}$ & $^{+2.7}_{-2.7}$\\
$3568$ & $0.07$ & $6.73\cdot 10^{-2}$ & $ 59 $ & $^{+15}_{-11}$ & $ ^{+5.2}_{-5.3} $ & $^{+1.4}_{-0.9} $ & $^{-2.2}_{+1.7} $ & $^{+0.6}_{-0.7} $ & $^{+0.3}_{-0.3} $ & $^{+0.1}_{-0.1}$ & $^{+2.4}_{-2.4}$ & $^{+3.3}_{-3.3}$\\
$ $ & $0.11$ & $4.68\cdot 10^{-2}$ & $ 64 $ & $^{+14}_{-11}$ & $ ^{+4.9}_{-5.4} $ & $^{+0.3}_{-1.2} $ & $^{-2.3}_{+1.5} $ & $^{-0.8}_{+0.6} $ & $^{-0.1}_{+0.2} $ & $^{+0.0}_{-0.0}$ & $^{+2.4}_{-2.4}$ & $^{+3.3}_{-3.3}$\\
$ $ & $0.17$ & $3.15\cdot 10^{-2}$ & $ 62 $ & $^{+14}_{-11}$ & $ ^{+4.8}_{-5.3} $ & $^{+0.7}_{-0.2} $ & $^{-2.4}_{+1.2} $ & $^{+0.1}_{-0.7} $ & $^{-0.2}_{+0.1} $ & $^{+0.1}_{-0.1}$ & $^{+2.4}_{-2.4}$ & $^{+3.3}_{-3.3}$\\
$ $ & $0.25$ & $2.06\cdot 10^{-2}$ & $ 56 $ & $^{+15}_{-12}$ & $ ^{+4.1}_{-4.4} $ & $^{+0.3}_{-0.6} $ & $^{-1.6}_{+1.0} $ & $^{-2.1}_{+1.9} $ & $^{+0.5}_{-0.2} $ & $^{+0.1}_{-0.1}$ & $^{+2.4}_{-2.4}$ & $^{+1.0}_{-1.0}$\\
$ $ & $0.35$ & $8.84\cdot 10^{-3}$ & $ 33 $ & $^{+21}_{-15}$ & $ ^{+4.3}_{-3.7} $ & $^{+1.5}_{-0.4} $ & $^{-0.7}_{-0.0} $ & $^{-0.9}_{+1.9} $ & $^{+0.6}_{-0.7} $ & $^{+0.2}_{-0.2}$ & $^{+2.4}_{-2.4}$ & $^{+1.0}_{-1.0}$\\
$ $ & $0.47$ & $4.14\cdot 10^{-3}$ & $ 19 $ & $^{+29}_{-18}$ & $ ^{+6.1}_{-5.4} $ & $^{+0.7}_{-0.4} $ & $^{-0.5}_{+1.3} $ & $^{-3.1}_{+3.6} $ & $^{+1.8}_{-0.6} $ & $^{+0.6}_{-0.6}$ & $^{+2.4}_{-2.4}$ & $^{+2.7}_{-2.7}$\\
$4145$ & $0.08$ & $5.56\cdot 10^{-2}$ & $ 58 $ & $^{+15}_{-11}$ & $ ^{+6.3}_{-5.3} $ & $^{+1.8}_{-1.0} $ & $^{-2.3}_{+3.4} $ & $^{+1.6}_{-0.6} $ & $^{+0.5}_{-0.5} $ & $^{+0.1}_{-0.1}$ & $^{+2.4}_{-2.4}$ & $^{+3.3}_{-3.3}$\\
$ $ & $0.13$ & $3.04\cdot 10^{-2}$ & $ 53 $ & $^{+16}_{-12}$ & $ ^{+5.4}_{-4.9} $ & $^{+0.8}_{-0.4} $ & $^{-1.2}_{+2.5} $ & $^{-0.7}_{+0.4} $ & $^{-0.3}_{+0.5} $ & $^{+0.0}_{-0.0}$ & $^{+2.4}_{-2.4}$ & $^{+3.3}_{-3.3}$\\
$ $ & $0.19$ & $1.74\cdot 10^{-2}$ & $ 47 $ & $^{+17}_{-13}$ & $ ^{+4.8}_{-4.9} $ & $^{+0.3}_{-0.6} $ & $^{-1.3}_{+0.8} $ & $^{-0.5}_{+1.0} $ & $^{+0.1}_{+0.1} $ & $^{+0.2}_{-0.2}$ & $^{+2.4}_{-2.4}$ & $^{+3.3}_{-3.3}$\\
$ $ & $0.28$ & $1.05\cdot 10^{-2}$ & $ 36 $ & $^{+20}_{-14}$ & $ ^{+5.4}_{-4.1} $ & $^{+3.0}_{-0.7} $ & $^{-1.3}_{+2.1} $ & $^{-1.6}_{+1.9} $ & $^{+0.1}_{-0.6} $ & $^{+0.1}_{-0.1}$ & $^{+2.4}_{-2.4}$ & $^{+1.0}_{-1.0}$\\
$ $ & $0.39$ & $5.24\cdot 10^{-3}$ & $ 23 $ & $^{+25}_{-17}$ & $ ^{+4.0}_{-4.4} $ & $^{+0.6}_{-1.3} $ & $^{-1.3}_{+0.0} $ & $^{-1.8}_{+1.7} $ & $^{+0.8}_{-0.9} $ & $^{+0.3}_{-0.3}$ & $^{+2.4}_{-2.4}$ & $^{+1.0}_{-1.0}$\\
$ $ & $0.51$ & $1.21\cdot 10^{-3}$ & $ 7 $ & $^{+54}_{-25}$ & $ ^{+6.1}_{-6.8} $ & $^{+1.2}_{-0.9} $ & $^{-2.4}_{+2.2} $ & $^{-4.5}_{+3.2} $ & $^{+1.4}_{-0.6} $ & $^{+0.5}_{-0.5}$ & $^{+2.4}_{-2.4}$ & $^{+2.7}_{-2.7}$\\
$4806$ & $0.11$ & $2.12\cdot 10^{-2}$ & $ 33 $ & $^{+21}_{-15}$ & $ ^{+4.9}_{-5.3} $ & $^{+0.3}_{-2.0} $ & $^{-1.6}_{+1.1} $ & $^{+0.4}_{+0.7} $ & $^{+0.1}_{-0.4} $ & $^{+0.1}_{-0.1}$ & $^{+2.4}_{-2.4}$ & $^{+3.3}_{-3.3}$\\
$ $ & $0.16$ & $2.12\cdot 10^{-2}$ & $ 47 $ & $^{+17}_{-13}$ & $ ^{+5.0}_{-5.3} $ & $^{+0.7}_{-2.2} $ & $^{-1.1}_{+1.3} $ & $^{-0.7}_{+1.2} $ & $^{-0.4}_{+0.6} $ & $^{+0.1}_{-0.1}$ & $^{+2.4}_{-2.4}$ & $^{+3.3}_{-3.3}$\\
$ $ & $0.23$ & $8.94\cdot 10^{-3}$ & $ 30 $ & $^{+22}_{-15}$ & $ ^{+3.8}_{-3.9} $ & $^{+0.2}_{-0.2} $ & $^{-1.5}_{+1.6} $ & $^{-1.0}_{-0.2} $ & $^{+0.3}_{+0.1} $ & $^{+0.1}_{-0.1}$ & $^{+2.4}_{-2.4}$ & $^{+1.0}_{-1.0}$\\
$ $ & $0.33$ & $5.20\cdot 10^{-3}$ & $ 23 $ & $^{+25}_{-17}$ & $ ^{+4.6}_{-4.1} $ & $^{+1.3}_{-0.3} $ & $^{+0.8}_{+1.2} $ & $^{-2.2}_{+2.3} $ & $^{+0.1}_{-0.6} $ & $^{+0.1}_{-0.1}$ & $^{+2.4}_{-2.4}$ & $^{+1.0}_{-1.0}$\\
$ $ & $0.44$ & $1.88\cdot 10^{-3}$ & $ 10 $ & $^{+43}_{-23}$ & $ ^{+5.2}_{-6.0} $ & $^{+0.5}_{-1.3} $ & $^{-1.2}_{+0.8} $ & $^{-3.6}_{+2.8} $ & $^{+0.6}_{-1.3} $ & $^{+0.4}_{-0.4}$ & $^{+2.4}_{-2.4}$ & $^{+2.7}_{-2.7}$\\
$ $ & $0.56$ & $6.34\cdot 10^{-4}$ & $ 4 $ & $^{+79}_{-29}$ & $ ^{+7.6}_{-5.5} $ & $^{+2.0}_{-1.0} $ & $^{+1.1}_{+1.9} $ & $^{-3.2}_{+5.0} $ & $^{+0.2}_{+2.3} $ & $^{+0.4}_{-0.4}$ & $^{+2.4}_{-2.4}$ & $^{+2.7}_{-2.7}$\\
$5561$ & $0.12$ & $1.44\cdot 10^{-2}$ & $ 30 $ & $^{+22}_{-15}$ & $ ^{+5.0}_{-5.5} $ & $^{+0.5}_{-2.4} $ & $^{-1.3}_{+1.9} $ & $^{-0.8}_{-0.4} $ & $^{+0.3}_{+0.0} $ & $^{+0.0}_{-0.0}$ & $^{+2.4}_{-2.4}$ & $^{+3.3}_{-3.3}$\\
$ $ & $0.18$ & $8.24\cdot 10^{-3}$ & $ 24 $ & $^{+25}_{-17}$ & $ ^{+5.7}_{-5.1} $ & $^{+0.7}_{-0.5} $ & $^{-1.8}_{+2.8} $ & $^{-0.6}_{+1.6} $ & $^{-0.7}_{+0.6} $ & $^{+0.2}_{-0.2}$ & $^{+2.4}_{-2.4}$ & $^{+3.3}_{-3.3}$\\
$ $ & $0.26$ & $5.46\cdot 10^{-3}$ & $ 25 $ & $^{+24}_{-16}$ & $ ^{+4.6}_{-4.4} $ & $^{+0.7}_{-0.6} $ & $^{-2.3}_{+2.8} $ & $^{-1.3}_{+1.1} $ & $^{+0.3}_{-0.1} $ & $^{+0.1}_{-0.1}$ & $^{+2.4}_{-2.4}$ & $^{+1.0}_{-1.0}$\\
$ $ & $0.37$ & $2.41\cdot 10^{-3}$ & $ 13 $ & $^{+36}_{-21}$ & $ ^{+3.8}_{-4.1} $ & $^{+0.3}_{-0.4} $ & $^{-0.5}_{+0.2} $ & $^{-2.1}_{+1.6} $ & $^{+0.1}_{-0.4} $ & $^{+0.2}_{-0.2}$ & $^{+2.4}_{-2.4}$ & $^{+1.0}_{-1.0}$\\
$ $ & $0.49$ & $7.28\cdot 10^{-4}$ & $ 5 $ & $^{+67}_{-27}$ & $ ^{+7.0}_{-6.0} $ & $^{+1.0}_{-0.5} $ & $^{-1.9}_{+3.6} $ & $^{-3.7}_{+3.9} $ & $^{+1.2}_{+0.2} $ & $^{+0.5}_{-0.5}$ & $^{+2.4}_{-2.4}$ & $^{+2.7}_{-2.7}$\\
$ $ & $0.61$ & $ < 1.52\cdot 10^{-4}$ & $ 0 $ & $ $ & $  $ & $  $ & $  $ & $  $ & $  $ & $  $ & $  $ & $  $\\
$6966$ & $0.14$ & $6.39\cdot 10^{-3}$ & $ 39 $ & $^{+19}_{-14}$ & $ ^{+5.7}_{-5.3} $ & $^{+2.1}_{-1.1} $ & $^{-2.4}_{+2.6} $ & $^{+0.3}_{-0.2} $ & $^{+0.2}_{-0.1} $ & $^{+0.0}_{-0.0}$ & $^{+2.4}_{-2.4}$ & $^{+3.3}_{-3.3}$\\
$ $ & $0.21$ & $6.94\cdot 10^{-3}$ & $ 53 $ & $^{+16}_{-12}$ & $ ^{+4.1}_{-4.0} $ & $^{+0.7}_{-0.3} $ & $^{-1.6}_{+1.7} $ & $^{-1.1}_{+1.3} $ & $^{-0.5}_{+0.5} $ & $^{+0.2}_{-0.2}$ & $^{+2.4}_{-2.4}$ & $^{+1.0}_{-1.0}$\\
$ $ & $0.30$ & $1.94\cdot 10^{-3}$ & $ 24 $ & $^{+25}_{-17}$ & $ ^{+4.0}_{-4.1} $ & $^{+0.5}_{-0.1} $ & $^{-1.1}_{+1.2} $ & $^{-1.9}_{+1.5} $ & $^{-0.1}_{+0.0} $ & $^{+0.1}_{-0.1}$ & $^{+2.4}_{-2.4}$ & $^{+1.0}_{-1.0}$\\
$ $ & $0.41$ & $1.10\cdot 10^{-3}$ & $ 16 $ & $^{+32}_{-20}$ & $ ^{+5.1}_{-5.0} $ & $^{+0.5}_{-0.3} $ & $^{-1.0}_{+1.4} $ & $^{-2.2}_{+2.3} $ & $^{+0.4}_{-0.4} $ & $^{+0.2}_{-0.2}$ & $^{+2.4}_{-2.4}$ & $^{+2.7}_{-2.7}$\\
$ $ & $0.53$ & $3.39\cdot 10^{-4}$ & $ 6 $ & $^{+60}_{-26}$ & $ ^{+6.7}_{-6.1} $ & $^{+0.7}_{-0.7} $ & $^{-1.1}_{+1.0} $ & $^{-4.1}_{+4.9} $ & $^{+0.8}_{-0.2} $ & $^{+0.5}_{-0.5}$ & $^{+2.4}_{-2.4}$ & $^{+2.7}_{-2.7}$\\
$ $ & $0.66$ & $4.49\cdot 10^{-5}$ & $ 1 $ & $^{+220}_{-30}$ & $ ^{+9.8}_{-7.8} $ & $^{+1.0}_{-0.9} $ & $^{-0.9}_{+0.8} $ & $^{-6.4}_{+8.6} $ & $^{-2.7}_{+3.1} $ & $^{+0.7}_{-0.7}$ & $^{+2.4}_{-2.4}$ & $^{+0.1}_{-0.1}$\\
$9059$ & $0.13$ & $3.71\cdot 10^{-3}$ & $ 11 $ & $^{+40}_{-22}$ & $ ^{+8.3}_{-9.4} $ & $^{+0.3}_{-5.5} $ & $^{-3.5}_{+4.1} $ & $^{+5.4}_{-4.8} $ & $^{+0.5}_{-0.9} $ & $^{+0.5}_{-0.5}$ & $^{+2.4}_{-2.4}$ & $^{+3.3}_{-3.3}$\\
$ $ & $0.19$ & $3.48\cdot 10^{-3}$ & $ 28 $ & $^{+23}_{-15}$ & $ ^{+5.9}_{-5.7} $ & $^{+1.8}_{-0.6} $ & $^{-3.2}_{+3.2} $ & $^{-0.4}_{+0.4} $ & $^{+0.1}_{+0.3} $ & $^{+0.2}_{-0.2}$ & $^{+2.4}_{-2.4}$ & $^{+3.3}_{-3.3}$\\
$ $ & $0.27$ & $1.73\cdot 10^{-3}$ & $ 22 $ & $^{+26}_{-17}$ & $ ^{+4.1}_{-4.5} $ & $^{+0.7}_{-0.3} $ & $^{-2.8}_{+1.8} $ & $^{-0.7}_{+1.0} $ & $^{-0.4}_{+0.1} $ & $^{+0.2}_{-0.2}$ & $^{+2.4}_{-2.4}$ & $^{+1.0}_{-1.0}$\\
$ $ & $0.38$ & $8.34\cdot 10^{-4}$ & $ 16 $ & $^{+32}_{-19}$ & $ ^{+5.0}_{-4.9} $ & $^{+0.5}_{-0.3} $ & $^{-1.3}_{+1.2} $ & $^{-3.2}_{+3.3} $ & $^{+0.1}_{+0.2} $ & $^{+0.2}_{-0.2}$ & $^{+2.4}_{-2.4}$ & $^{+1.0}_{-1.0}$\\
$ $ & $0.51$ & $1.47\cdot 10^{-4}$ & $ 3 $ & $^{+96}_{-30}$ & $ ^{+6.2}_{-5.5} $ & $^{+0.7}_{-0.1} $ & $^{-0.9}_{+1.7} $ & $^{-3.2}_{+4.0} $ & $^{+0.5}_{-0.7} $ & $^{+0.4}_{-0.4}$ & $^{+2.4}_{-2.4}$ & $^{+2.7}_{-2.7}$\\
$ $ & $0.64$ & $8.16\cdot 10^{-5}$ & $ 2 $ & $^{+130}_{-32}$ & $ ^{+6.5}_{-7.0} $ & $^{+1.2}_{-0.5} $ & $^{-1.8}_{+2.2} $ & $^{-5.8}_{+5.1} $ & $^{-0.4}_{+0.1} $ & $^{+0.0}_{-0.0}$ & $^{+2.4}_{-2.4}$ & $^{+0.1}_{-0.1}$\\
$ $ & $0.78$ & $ < 2.25\cdot 10^{-5}$ & $ 0 $ & $ $ & $  $ & $  $ & $  $ & $  $ & $  $ & $  $ & $  $ & $  $\\
$ $ & $0.93$ & $ < 4.38\cdot 10^{-6}$ & $ 0 $ & $ $ & $  $ & $  $ & $  $ & $  $ & $  $ & $  $ & $  $ & $  $\\
$15072$ & $0.61$ & $3.58\cdot 10^{-5}$ & $ 4 $ & $^{+79}_{-29}$ & $ ^{+6.3}_{-7.8} $ & $^{+2.3}_{-0.9} $ & $^{-2.0}_{+1.5} $ & $^{-6.7}_{+4.5} $ & $^{-0.1}_{+0.3} $ & $^{+0.3}_{-0.3}$ & $^{+2.4}_{-2.4}$ & $^{+0.1}_{-0.1}$\\
$ $ & $0.75$ & $4.98\cdot 10^{-6}$ & $ 1 $ & $^{+220}_{-30}$ & $ ^{+26}_{-6.8} $ & $^{+2.7}_{-0.7} $ & $^{-1.0}_{+2.0} $ & $^{-5.7}_{+25} $ & $^{-0.4}_{+0.2} $ & $^{+1.2}_{-1.2}$ & $^{+2.4}_{-2.4}$ & $^{+0.1}_{-0.1}$\\
$ $ & $0.91$ & $ < 7.42\cdot 10^{-7}$ & $0$ & $ $ & $  $ & $  $ & $  $ & $  $ & $  $ & $  $ & $  $ & $  $\\
\hline
\end{supertabular}
\end{center}

%% file: DESY-06-116-tab-6.tex
\begin{center}
\footnotesize
\renewcommand{\arraystretch}{1.2}
\tablehead{
\hline
{$Q^2$} & 
\multicolumn{1}{c|}{$x_{\rm edge}$} & 
$\int_{x_{\rm edge}}^{1} {d^2\sigma}/{dxdQ^2}$  &
$N$& 
$\delta_{s}$& 
$\delta_t$ & 
$\delta_u$ &
$\delta_1$ &
$\delta_2$ &
$\delta_3$ &
$\delta_4$ &
$\delta_5$ &
$\delta_6$ \\
{($\Gev^2$)} & 
$$ &
{($\rm pb/GeV^{2}$)} &
$$ &
(\%) &
(\%) &
(\%) &
(\%) &
(\%) &
(\%) &
(\%) &
(\%) &
(\%) \\
\hline \hline}
\tablelasttail{
\hline}
\normalsize
\bottomcaption{
    The integral cross section table for 99-00 $e^+p$ NC scattering.
    The first two columns of the table contain the $Q^2$ and $x_{\rm edge}$ values 
    for the bin, the third contains the measured
    cross section $\int_{x_{\rm edge}}^{1} {d^2\sigma}/{dxdQ^2}$ corrected to the electroweak 
    Born level or the upper limit in case of zero observed 
    events, the fourth contains the number of events reconstructed in 
    the bin, $N$, the fifth contains the 
    statistical uncertainty, $\delta_s$, and the sixth contains the 
     total systematic uncertainty, $\delta_t$. 
    The right part of the table lists the total uncorrelated
    systematic uncertainty, $\delta_u$, followed by the bin-to-bin correlated
    systematic uncertainties $\delta_1$--\,$\delta_6$ defined in
    the text.
    The upper (lower) numbers refer to the variation of 
    the cross section, whereas the signs of
    the numbers reflect the direction of change in the cross
    sections. Note that the normalization uncertainty, $\delta_7$ is
    not listed.
    } 
\label{tab:9900crosl}
\begin{supertabular}{|c|l|c|c|c|c||c|c|c|c|c|c|c|}
$ 648$ & $0.22$ & $1.33\cdot 10^{-1}$ & $ 106 $ & $^{+11}_{-8.8}$ & $ ^{+9.8}_{-4.4} $ & $^{+2.4}_{-2.3} $ & $^{+9.0}_{-2.4} $ & $^{+0.2}_{+0.0} $ & $^{-0.6}_{+1.4} $ & $^{+0.1}_{-0.1}$ & $^{+1.6}_{-1.6}$ & $^{+0.1}_{-0.1}$\\
$ 761$ & $0.24$ & $6.73\cdot 10^{-2}$ & $ 212 $ & $^{+7.3}_{-6.4}$ & $ ^{+3.8}_{-5.2} $ & $^{+2.0}_{-3.0} $ & $^{+1.7}_{-3.2} $ & $^{+0.1}_{+0.0} $ & $^{-0.1}_{+0.4} $ & $^{+0.0}_{-0.0}$ & $^{+1.6}_{-1.6}$ & $^{+0.1}_{-0.1}$\\
$ 891$ & $0.26$ & $4.41\cdot 10^{-2}$ & $ 288 $ & $^{+6.2}_{-5.5}$ & $ ^{+3.3}_{-3.3} $ & $^{+1.0}_{-1.0} $ & $^{-1.4}_{+1.6} $ & $^{+0.0}_{+0.0} $ & $^{-0.9}_{+0.3} $ & $^{+0.1}_{-0.1}$ & $^{+1.6}_{-1.6}$ & $^{+0.1}_{-0.1}$\\
$1045$ & $0.29$ & $2.50\cdot 10^{-2}$ & $ 224 $ & $^{+7.1}_{-6.3}$ & $ ^{+4.1}_{-3.3} $ & $^{+1.2}_{-0.7} $ & $^{-1.6}_{+2.7} $ & $^{+0.1}_{+0.0} $ & $^{-0.4}_{+0.3} $ & $^{+0.1}_{-0.1}$ & $^{+1.6}_{-1.6}$ & $^{+0.1}_{-0.1}$\\
$1224$ & $0.31$ & $1.48\cdot 10^{-2}$ & $ 157 $ & $^{+8.6}_{-7.3}$ & $ ^{+3.7}_{-4.2} $ & $^{+1.7}_{-1.0} $ & $^{-3.0}_{+1.5} $ & $^{+0.1}_{+0.0} $ & $^{-0.5}_{+1.1} $ & $^{+0.1}_{-0.1}$ & $^{+1.6}_{-1.6}$ & $^{+0.1}_{-0.1}$\\
$1431$ & $0.34$ & $9.52\cdot 10^{-3}$ & $ 120 $ & $^{+10}_{-8.3}$ & $ ^{+4.0}_{-3.7} $ & $^{+0.3}_{-0.4} $ & $^{-2.3}_{+2.7} $ & $^{+0.0}_{-0.0} $ & $^{-0.8}_{+0.9} $ & $^{+0.1}_{-0.1}$ & $^{+1.6}_{-1.6}$ & $^{+0.1}_{-0.1}$\\
$1672$ & $0.36$ & $3.90\cdot 10^{-3}$ & $ 59 $ & $^{+15}_{-11}$ & $ ^{+4.6}_{-4.5} $ & $^{+0.2}_{-0.7} $ & $^{-3.4}_{+3.3} $ & $^{+0.0}_{+0.0} $ & $^{-0.7}_{+1.4} $ & $^{+0.2}_{-0.2}$ & $^{+1.6}_{-1.6}$ & $^{+0.1}_{-0.1}$\\
$1951$ & $0.39$ & $2.64\cdot 10^{-3}$ & $ 46 $ & $^{+17}_{-13}$ & $ ^{+4.9}_{-4.7} $ & $^{+0.9}_{-1.0} $ & $^{-3.5}_{+3.8} $ & $^{+0.0}_{+0.0} $ & $^{-1.0}_{+1.2} $ & $^{+0.2}_{-0.2}$ & $^{+1.6}_{-1.6}$ & $^{+0.1}_{-0.1}$\\
$2273$ & $0.43$ & $1.52\cdot 10^{-3}$ & $ 29 $ & $^{+22}_{-15}$ & $ ^{+4.8}_{-4.3} $ & $^{+0.6}_{-0.7} $ & $^{-3.0}_{+3.7} $ & $^{+0.0}_{+0.0} $ & $^{-1.2}_{+1.0} $ & $^{+0.1}_{-0.1}$ & $^{+1.6}_{-1.6}$ & $^{+0.1}_{-0.1}$\\
$2644$ & $0.46$ & $9.16\cdot 10^{-4}$ & $ 22 $ & $^{+26}_{-17}$ & $ ^{+4.2}_{-4.8} $ & $^{+1.4}_{-1.4} $ & $^{-3.6}_{+2.4} $ & $^{+0.0}_{+0.0} $ & $^{-0.9}_{+1.5} $ & $^{+0.2}_{-0.2}$ & $^{+1.6}_{-1.6}$ & $^{+0.1}_{-0.1}$\\
$3073$ & $0.50$ & $4.79\cdot 10^{-4}$ & $ 13 $ & $^{+36}_{-21}$ & $ ^{+6.5}_{-7.2} $ & $^{+0.7}_{-0.3} $ & $^{-6.6}_{+5.8} $ & $^{+0.0}_{+0.0} $ & $^{-0.6}_{+0.9} $ & $^{+0.1}_{-0.1}$ & $^{+1.6}_{-1.6}$ & $^{+0.1}_{-0.1}$\\
$3568$ & $0.54$ & $2.27\cdot 10^{-4}$ & $ 7 $ & $^{+54}_{-25}$ & $ ^{+4.8}_{-7.0} $ & $^{+1.1}_{-2.0} $ & $^{-5.8}_{+3.6} $ & $^{+0.0}_{+0.0} $ & $^{-2.0}_{+1.1} $ & $^{+0.1}_{-0.1}$ & $^{+1.6}_{-1.6}$ & $^{+0.1}_{-0.1}$\\
$4145$ & $0.58$ & $7.57\cdot 10^{-5}$ & $ 3 $ & $^{+96}_{-30}$ & $ ^{+6.6}_{-6.4} $ & $^{+1.3}_{-2.2} $ & $^{-4.8}_{+5.4} $ & $^{+0.0}_{+0.0} $ & $^{-2.4}_{+2.3} $ & $^{+0.0}_{-0.0}$ & $^{+1.6}_{-1.6}$ & $^{+0.1}_{-0.1}$\\
$4806$ & $0.63$ & $7.05\cdot 10^{-5}$ & $ 3 $ & $^{+96}_{-30}$ & $ ^{+15}_{-15} $ & $^{+5.4}_{-2.1} $ & $^{-7.1}_{+6.5} $ & $^{+0.0}_{+0.0} $ & $^{-3.6}_{+1.6} $ & $^{+0.2}_{-0.2}$ & $^{+1.6}_{-1.6}$ & $^{+13}_{-13}$\\
$5561$ & $0.68$ & $6.09\cdot 10^{-5}$ & $ 3 $ & $^{+96}_{-30}$ & $ ^{+14}_{-14} $ & $^{+1.2}_{-3.4} $ & $^{-4.4}_{+5.3} $ & $^{+0.0}_{-0.1} $ & $^{-1.6}_{+3.5} $ & $^{+0.3}_{-0.3}$ & $^{+1.6}_{-1.6}$ & $^{+13}_{-13}$\\
$6966$ & $0.79$ & $ < 4.07\cdot 10^{-6}$ & $ 0 $ & $ $ & $  $ & $  $ & $  $ & $  $ & $  $ & $  $ & $  $ & $  $\\
\hline
\end{supertabular}
\end{center}

%% file: DESY-06-116-fig.tex
%%%%%%%%%%%%%%%%%%%% FIGURE%%%%%%%%%%%%%%%%%%%%%%%%%%%%%%
\newpage
\begin{figure}
\begin{center}
\includegraphics[width=12.0cm]{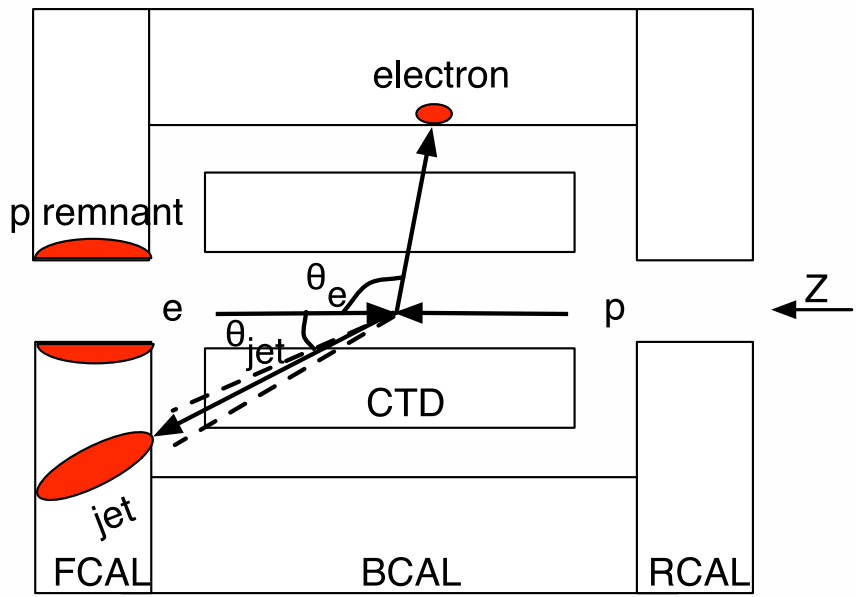} \\
\caption{
A schematic depiction of the ZEUS detector with the main
components used in this analysis labeled. Also shown is
a typical topology for events studied in this analysis.  The
electron is scattered at a large angle and is reconstructed
using the central tracking detector (CTD) and the barrel 
calorimeter (BCAL), while the scattered jet is typically reconstructed 
in the forward calorimeter (FCAL). The jet of particles from the proton 
remnant mostly disappears down the beam pipe.
}
\label{skzeus}
\end{center}
\end{figure}
%%%%%%%%%%%%%%%%%%%%%%%%%%%%%%%%%%%%%%%%%%%%%%%%%%%%%%%%%%%

%%%%%%%%%%%%%%%%%%%% FIGURE%%%%%%%%%%%%%%%%%%%%%%%%%%%%%%
\newpage
\begin{figure}
\begin{center}
\includegraphics[width=16.0cm]{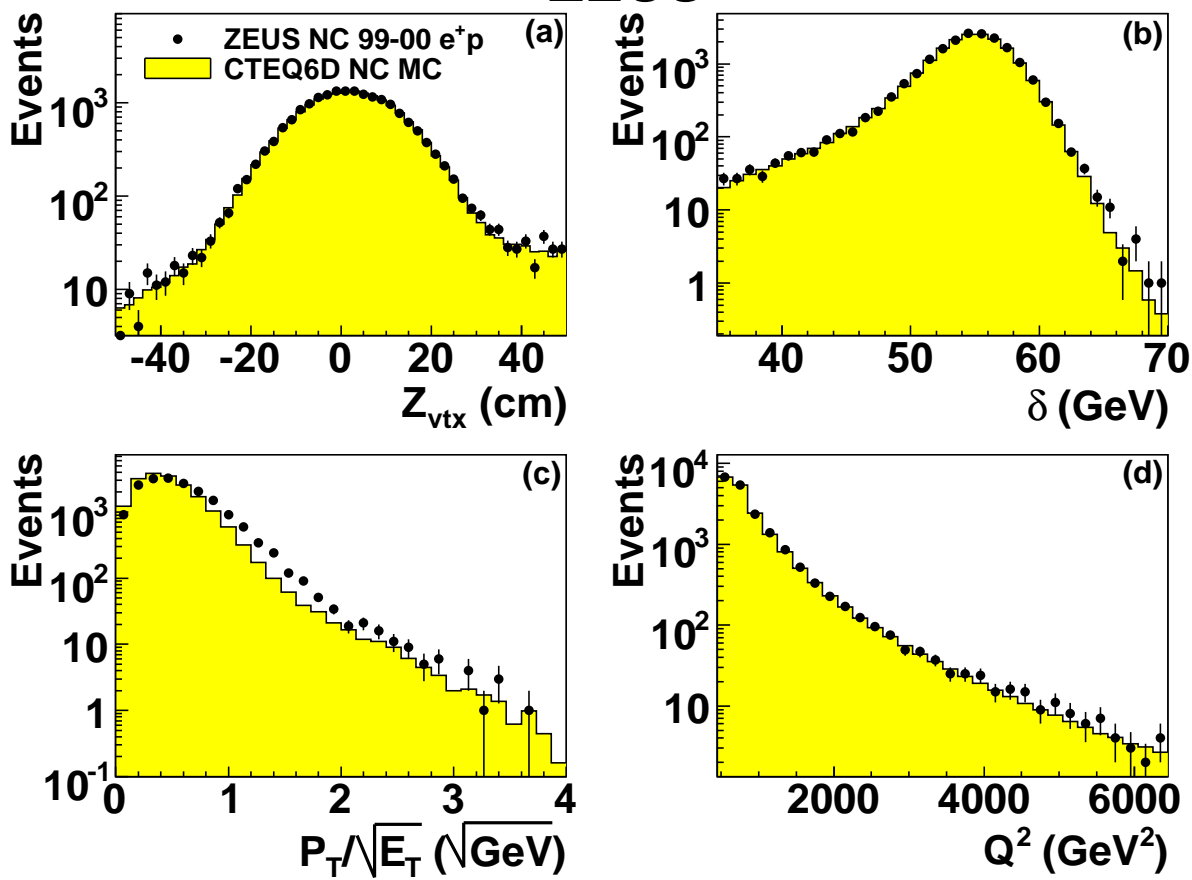}
\caption{
Comparison of NC MC distributions (histograms) with 99-00 $e^{+}p$ 
data (points) 
for: (a) the Z coordinate of the event vertex; (b) $\delta$; 
(c) $P_T/\sqrt{E_T}$ and (d) $Q^2$.  The MC distributions are normalized to 
the measured luminosity.}

\label{kincp}
\end{center}
\end{figure}
%%%%%%%%%%%%%%%%%%%%%%%%%%%%%%%%%%%%%%%%%%%%%%%%%%%%%%%%%%%

%%%%%%%%%%%%%%%%%%%% FIGURE%%%%%%%%%%%%%%%%%%%%%%%%%%%%%%
\newpage
\begin{figure}
\begin{center}
\includegraphics[width=16.0cm]{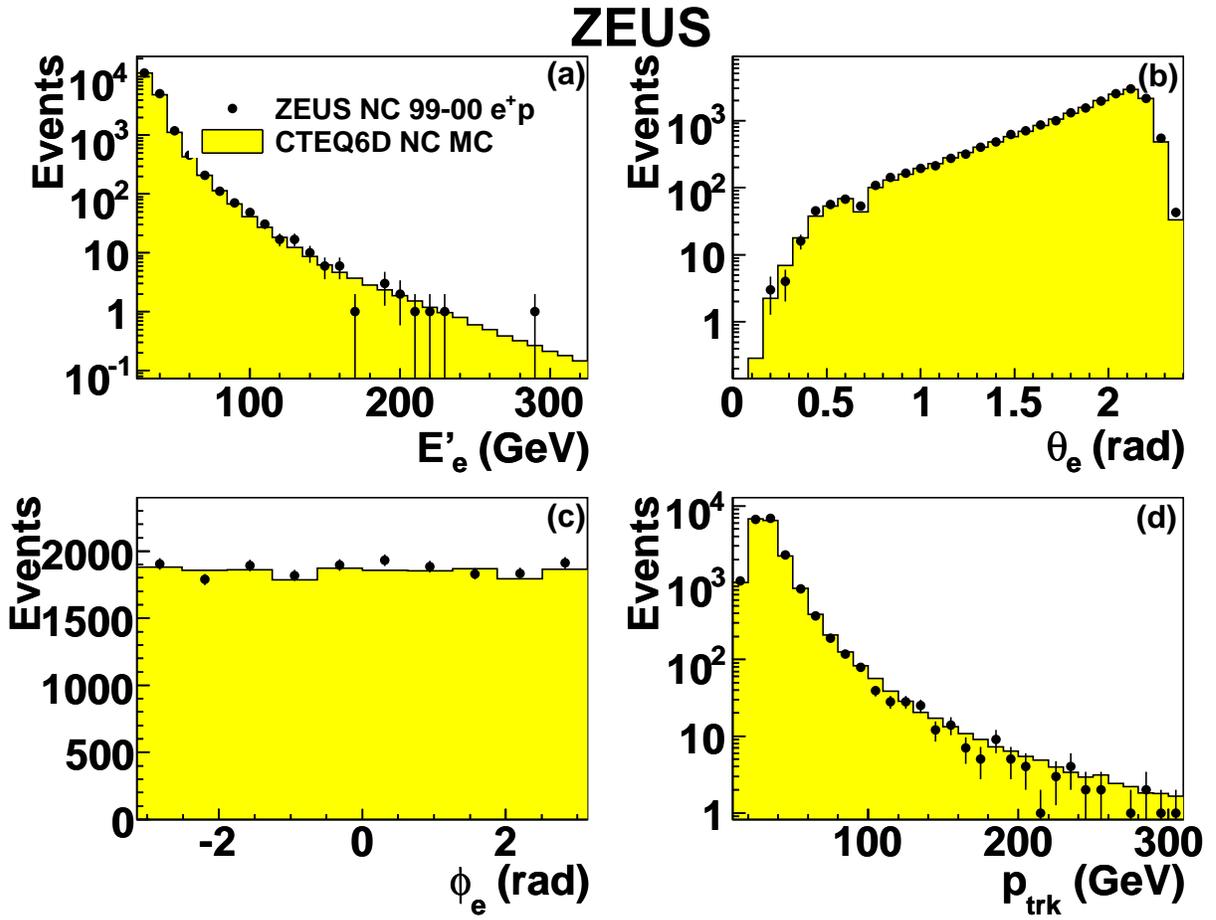}
\caption{
Comparison of NC MC distributions (histograms) with 99-00 $e^{+}p$ 
data (points) for: 
(a) $E'_e$; (b) $\theta_e$; (c) $\phi_e$ and (d) $p_{\rm trk}$, the momentum of the track associated with the scattered electron. The MC distributions are normalized to the measured luminosity.}
\label{ecp}
\end{center}
\end{figure}
%%%%%%%%%%%%%%%%%%%%%%%%%%%%%%%%%%%%%%%%%%%%%%%%%%%%%%%%%%%

%%%%%%%%%%%%%%%%%%%% FIGURE%%%%%%%%%%%%%%%%%%%%%%%%%%%%%%
\newpage
\begin{figure}
\begin{center}
\includegraphics[width=16.0cm]{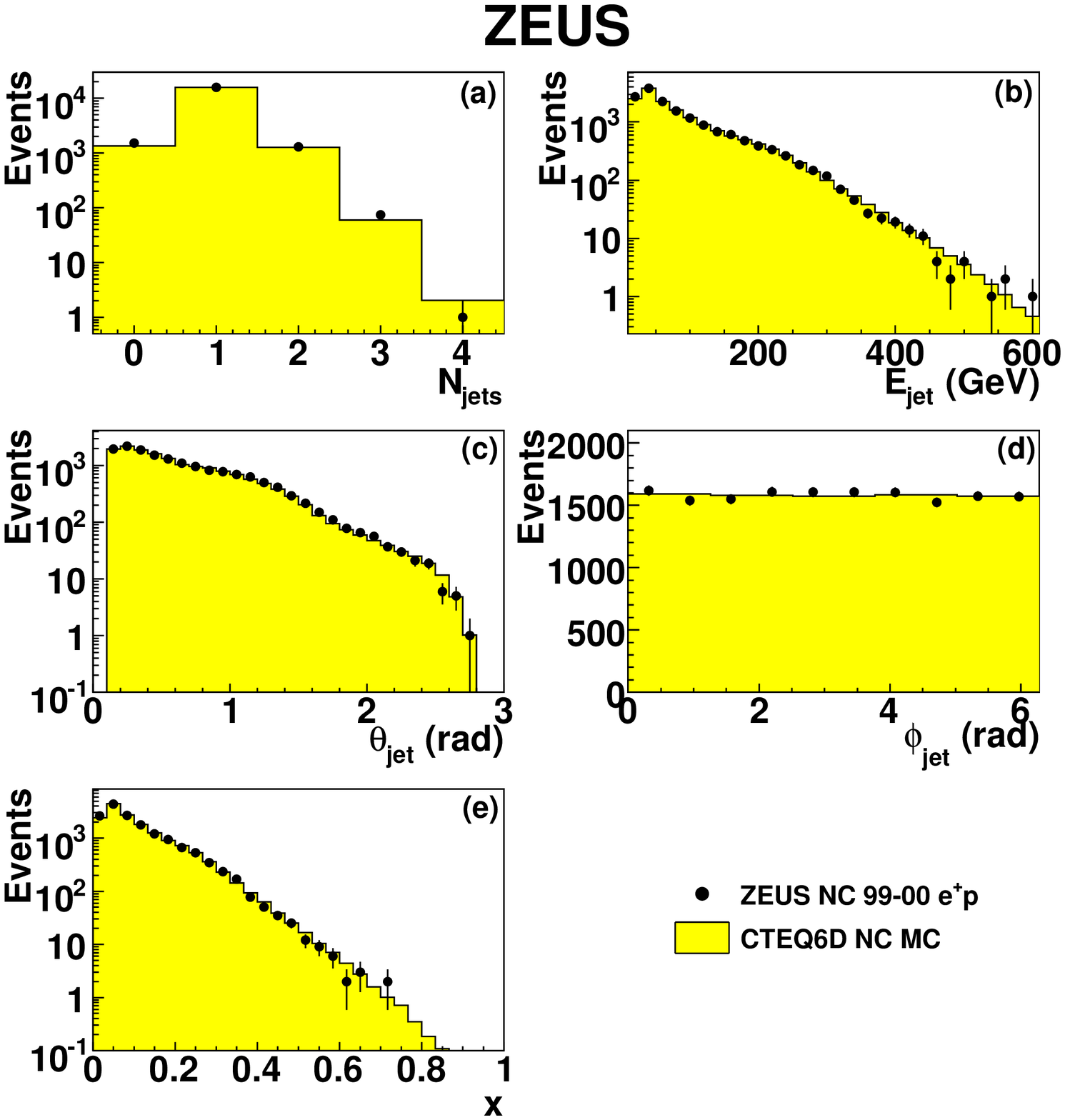}
\caption{
Comparison of NC MC distributions (histograms) with 99-00 $e^{+}p$
 data (points) for: 
(a) the number of reconstructed jets; (b) $E_{jet}$; (c) $\theta_{jet}$; (d) $\phi_{jet}$  and 
(e) x calculated from the jet. The jet distributions are for one jet events.
The MC distributions are normalized to the measured luminosity.}
\label{jcp}
\end{center}
\end{figure}
%%%%%%%%%%%%%%%%%%%%%%%%%%%%%%%%%%%%%%%%%%%%%%%%%%%%%%%%%%%

%%%%%%%%%%%%%%%%%%%% FIGURE%%%%%%%%%%%%%%%%%%%%%%%%%%%%%%
\newpage
\begin{figure}
\begin{center}
\includegraphics[width=16.0cm]{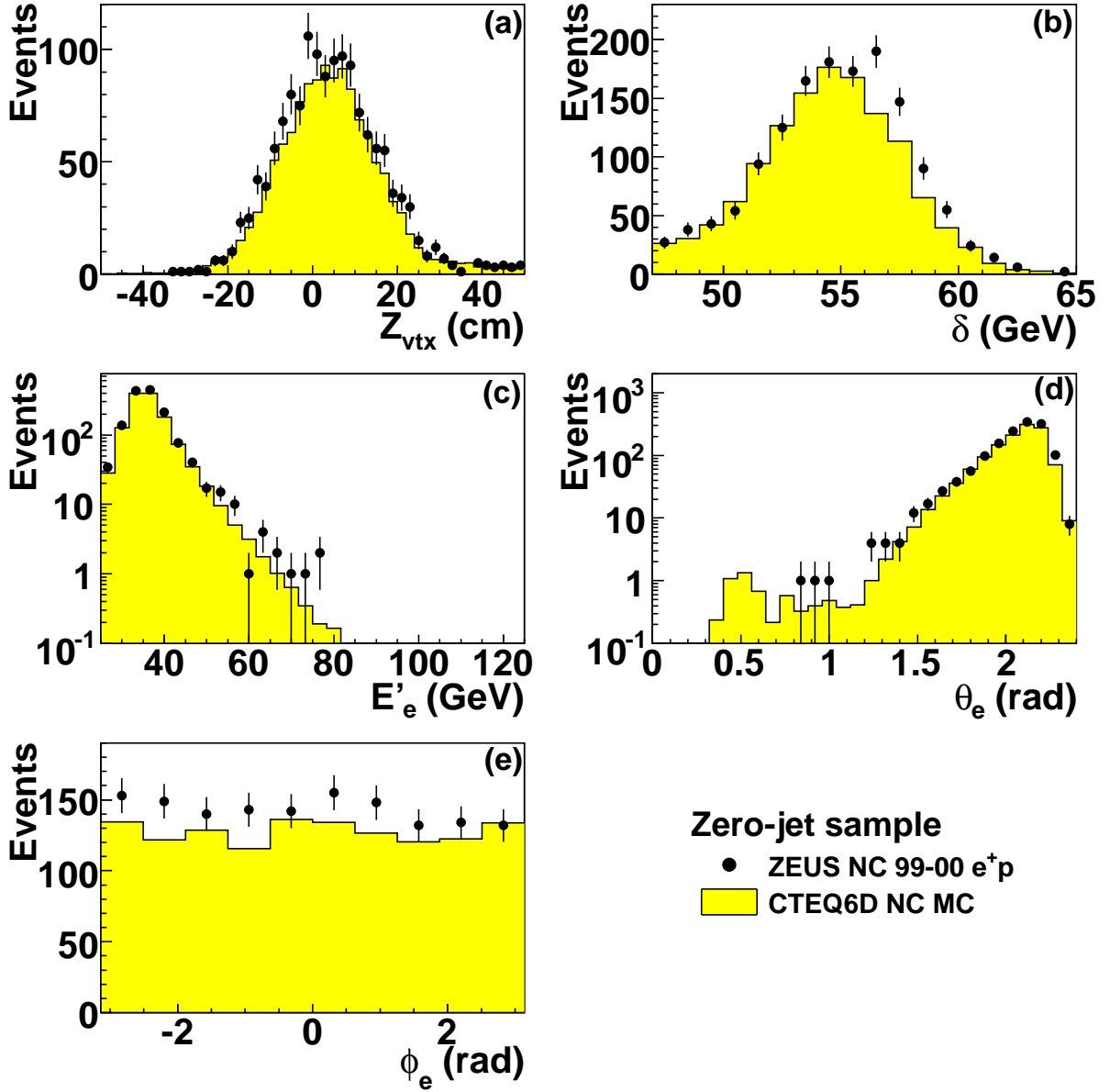}
\caption{
Comparison of NC MC distributions (histograms) with 99-00 $e^{+}p$
 data (points) 
for events with zero jets.  The plots show: 
(a) the Z coordinate of the event vertex; (b) $\delta$; (c) $E'_e$; 
(d) $\theta_e$ and (e) $\phi_e$.
The MC distributions are normalized to the measured luminosity.}
\label{0jcp}
\end{center}
\end{figure}
%%%%%%%%%%%%%%%%%%%%%%%%%%%%%%%%%%%%%%%%%%%%%%%%%%%%%%%%%%%

%%%%%%%%%%%%%%%%%%%% FIGURE%%%%%%%%%%%%%%%%%%%%%%%%%%%%%%
\newpage
\begin{figure}
\begin{center}
$\begin{array}{c}
\includegraphics[width=12.0cm]{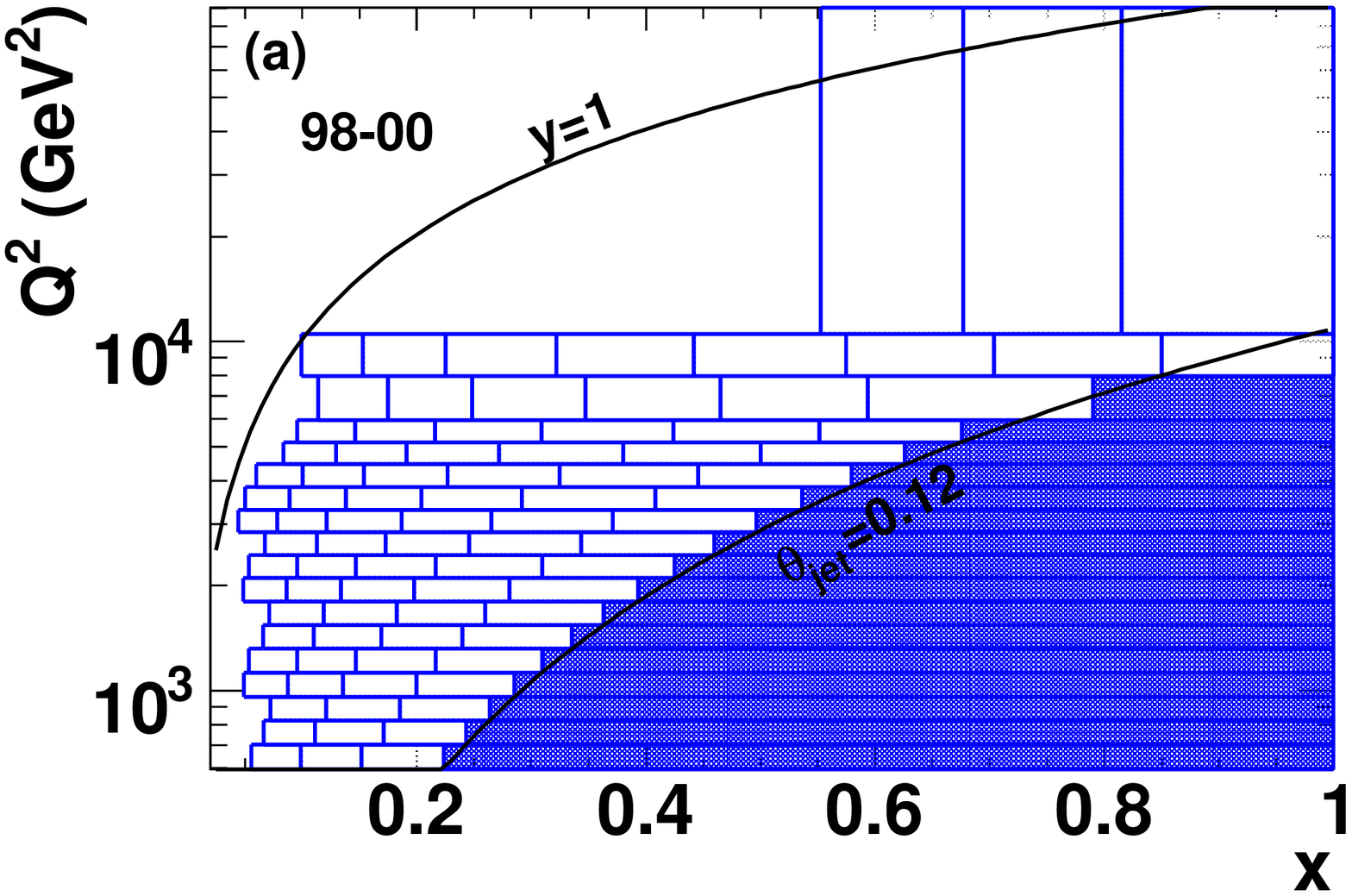} \\
%\bf (a) \\
\includegraphics[width=12.0cm]{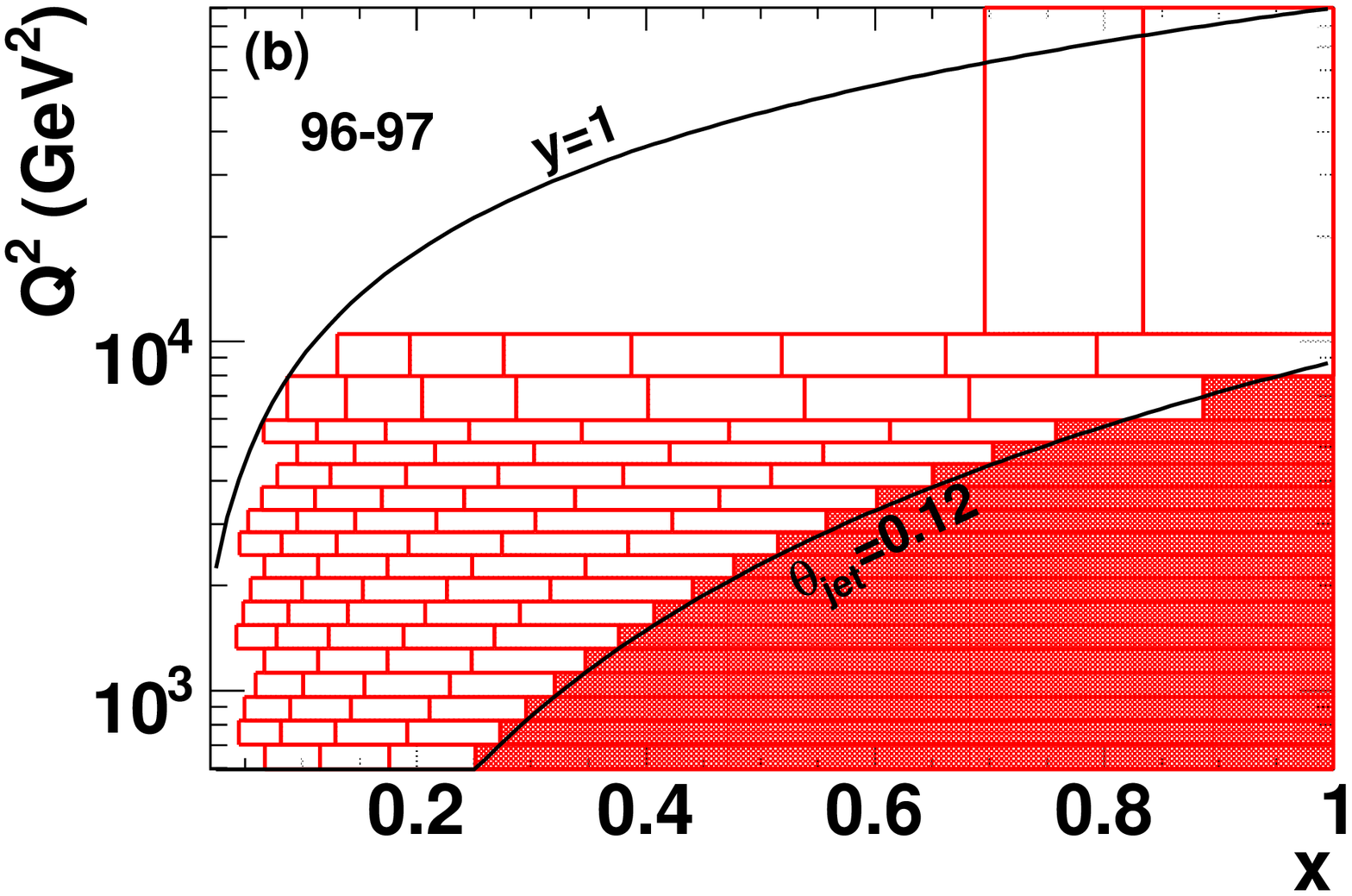} \\
%\bf (b) 
\end{array}$
%\begin{picture} (19.0,21.0)
%\put (0.0,0.0){\epsfig{figure=00bin.eps,width=12.0cm}}
%\put (14.20,20.0){{\bf a)}}
%\end{picture} \\
%\begin{picture} (19.0,21.0)
%\put (0.0,0.0){\epsfig{figure=96bin.eps,width=12.0cm}}
%\put (14.20,20.0){{\bf b)}}
%\end{picture} 

\caption{
Definition of the bins as used in this analysis for: (a) 98-00 data with 
$E_p=920$~$\gev$ and (b) 96-97 data with $E_p=820$~$\gev$.
The shaded bins extending to $x=1$ are for the zero-jet events.
The $y=1$ lines shows the kinematic limit. The $\theta_{jet}=0.12$~rad shows
the selection cut for jets.
}
\label{9600bin}
\end{center}
\end{figure}

%%%%%%%%%%%%%%%%%%%% migration%%%%%%%%%%%%%%%%%%%%%%%%%%%%%%
\newpage
\begin{figure}
\begin{center}
\includegraphics[width=16.0cm]{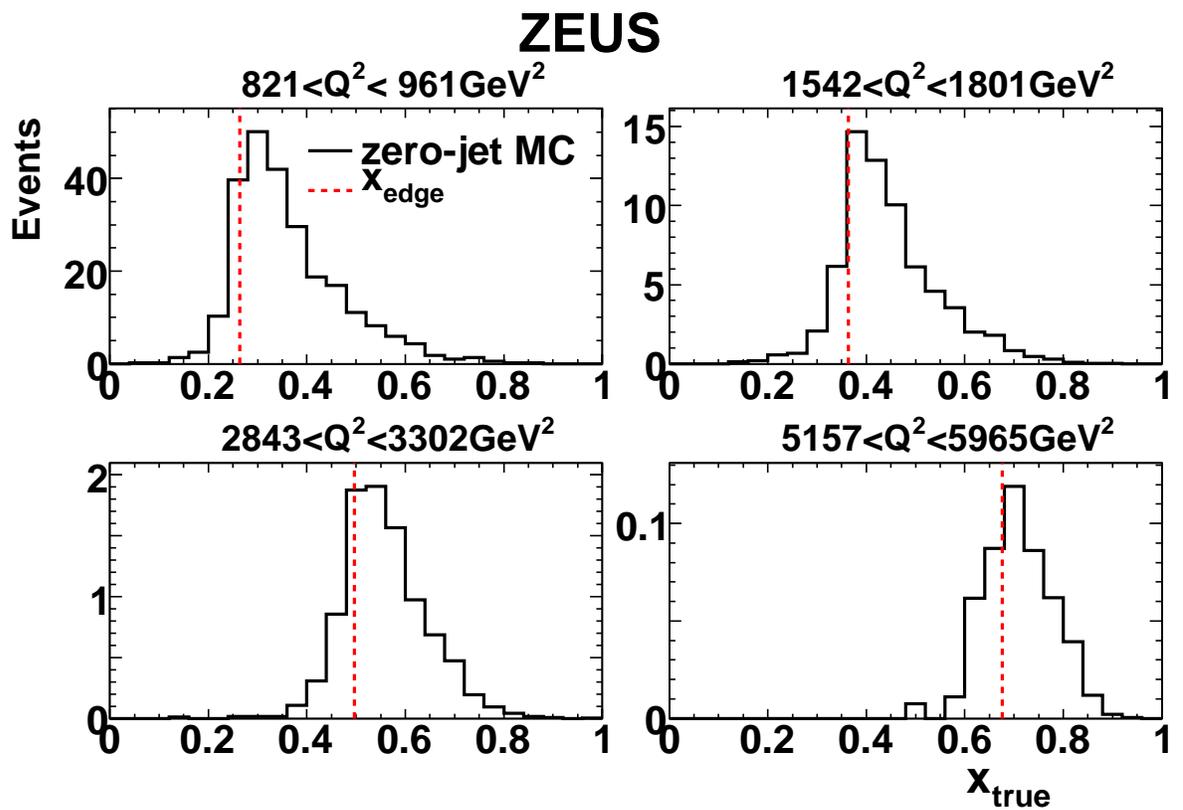}
\caption{
The true x distribution for zero-jet events from 99-00 $e^{+}p$ 
MC simulations in different $Q^2$ bins.
The dashed lines represent the lower edge of the bins, $x_{\rm edge}$.
The MC distributions are normalized to the luminosity
of the data.
}
\label{xmig}
\end{center}
\end{figure}

%%%%%%%%%%%%%%%%%%%% FIGURE%%%%%%%%%%%%%%%%%%%%%%%%%%%%%%
\newpage
\begin{figure}
\begin{center}
\includegraphics[width=16.0cm]{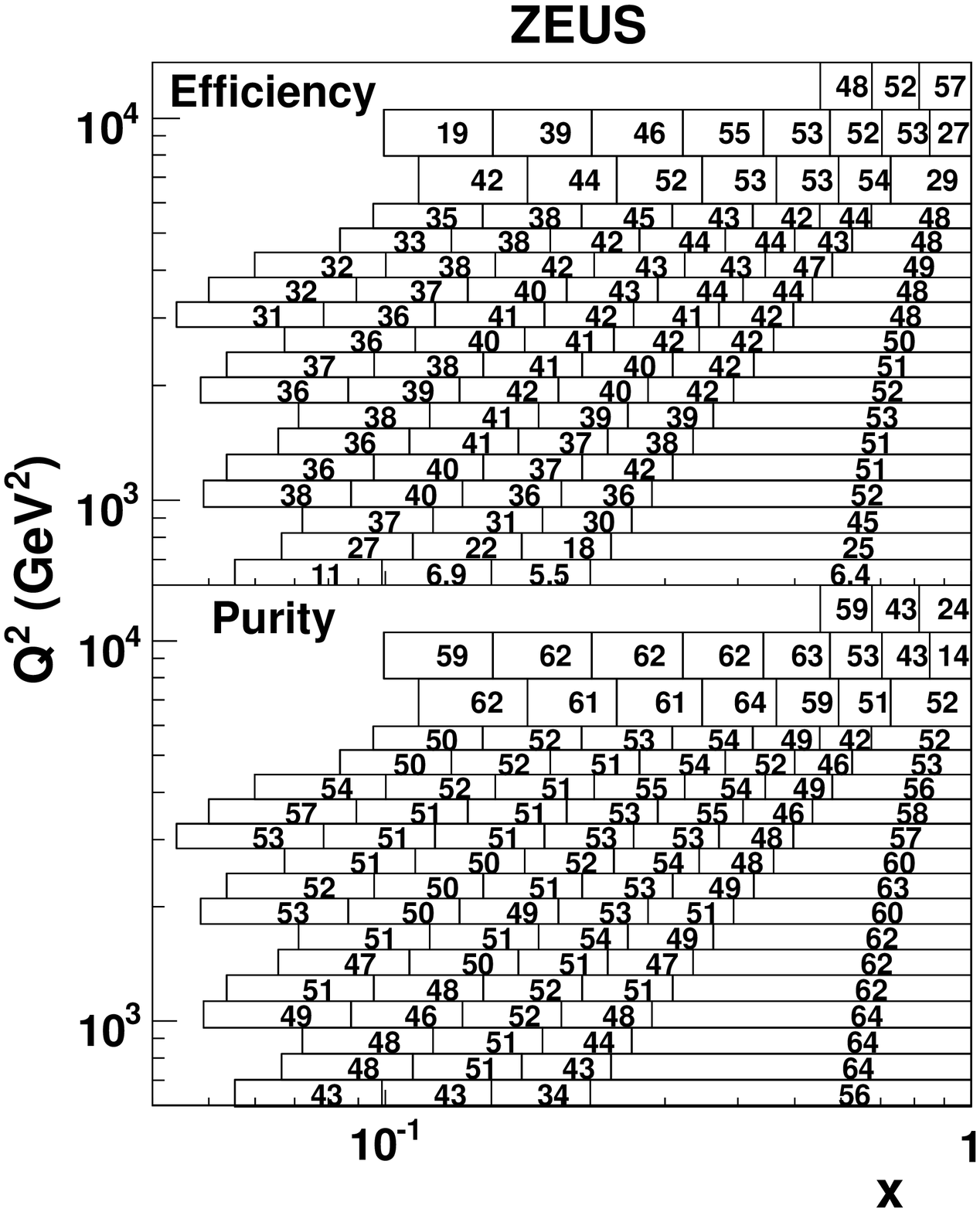}
\caption{
The efficiency and purity in $\%$ for each bin are shown for 99-00 $e^{+}p$
data.
}
\label{ep}
\end{center}
\end{figure}
%%%%%%%%%%%%%%%%%%%%%%%%%%%%%%%%%%%%%%%%%%%%%%%%%%%%%%%%%%%
%%%%%%%%%%%%%%%%%%%% FIGURE%%%%%%%%%%%%%%%%%%%%%%%%%%%%%%
\newpage
\begin{figure}
\begin{center}
\includegraphics[width=16.0cm]{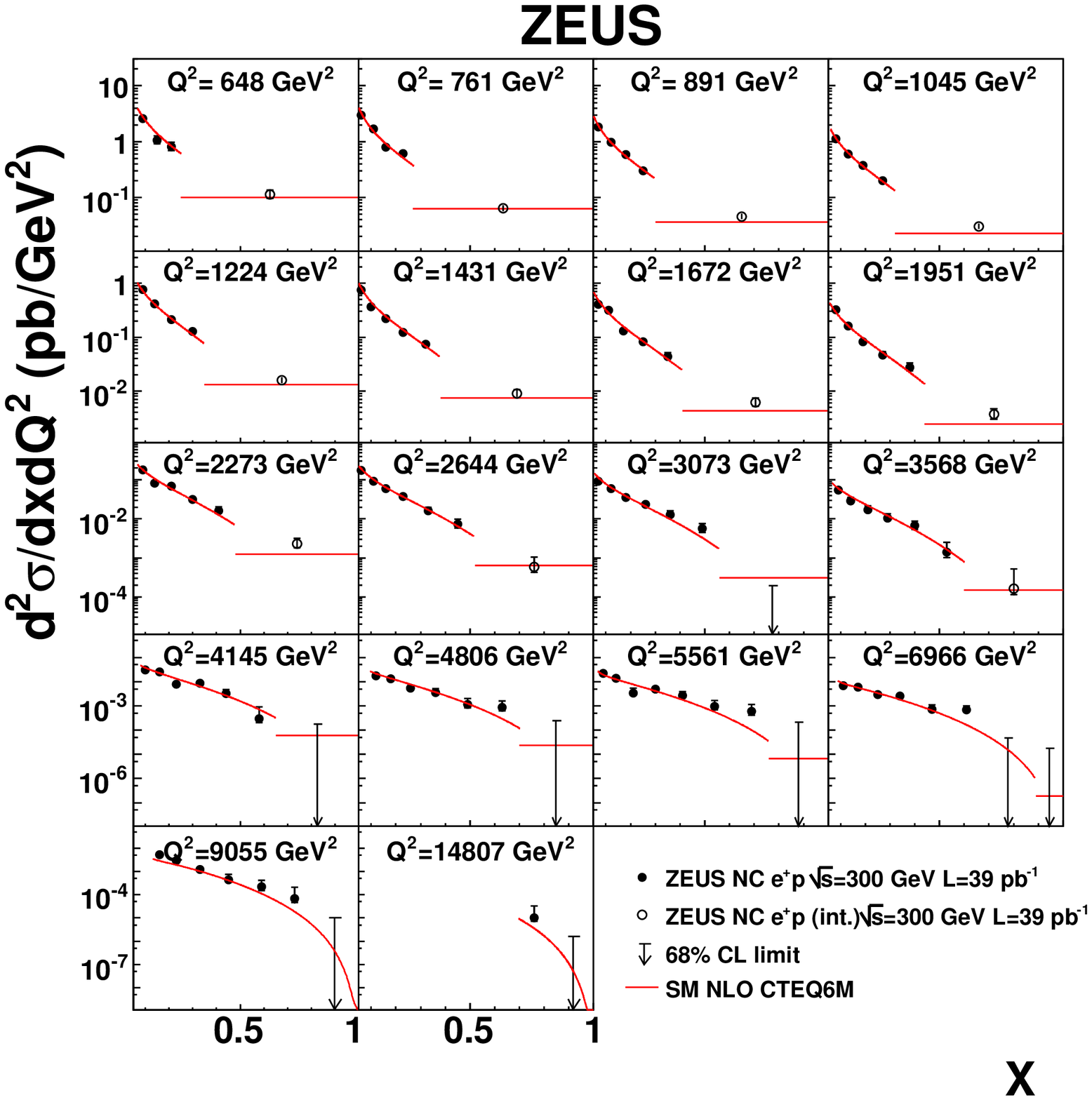}
\caption{
The double-differential cross section for 96-97 $e^+p$ NC scattering 
at $\sqrt{s}=300$~$\gev$
(solid circles) and the integral of the double differential cross section 
(open circles) compared to the Standard Model 
expectations evaluated using CTEQ6M PDFs (lines).The error bars 
show the statistical and systematic uncertainties added in quadrature. 
For bins with zero measured events, a $68\%$ probability limit,
calculated including the uncorrelated systematic uncertainty, is given. 
}
\label{cros96}
\end{center}
\end{figure}
%%%%%%%%%%%%%%%%%%%%%%%%%%%%%%%%%%%%%%%%%%%%%%%%%%%%%%%%%%%
%%%%%%%%%%%%%%%%%%%% FIGURE%%%%%%%%%%%%%%%%%%%%%%%%%%%%%%
\newpage
\begin{figure}
\begin{center}
\includegraphics[width=16.0cm]{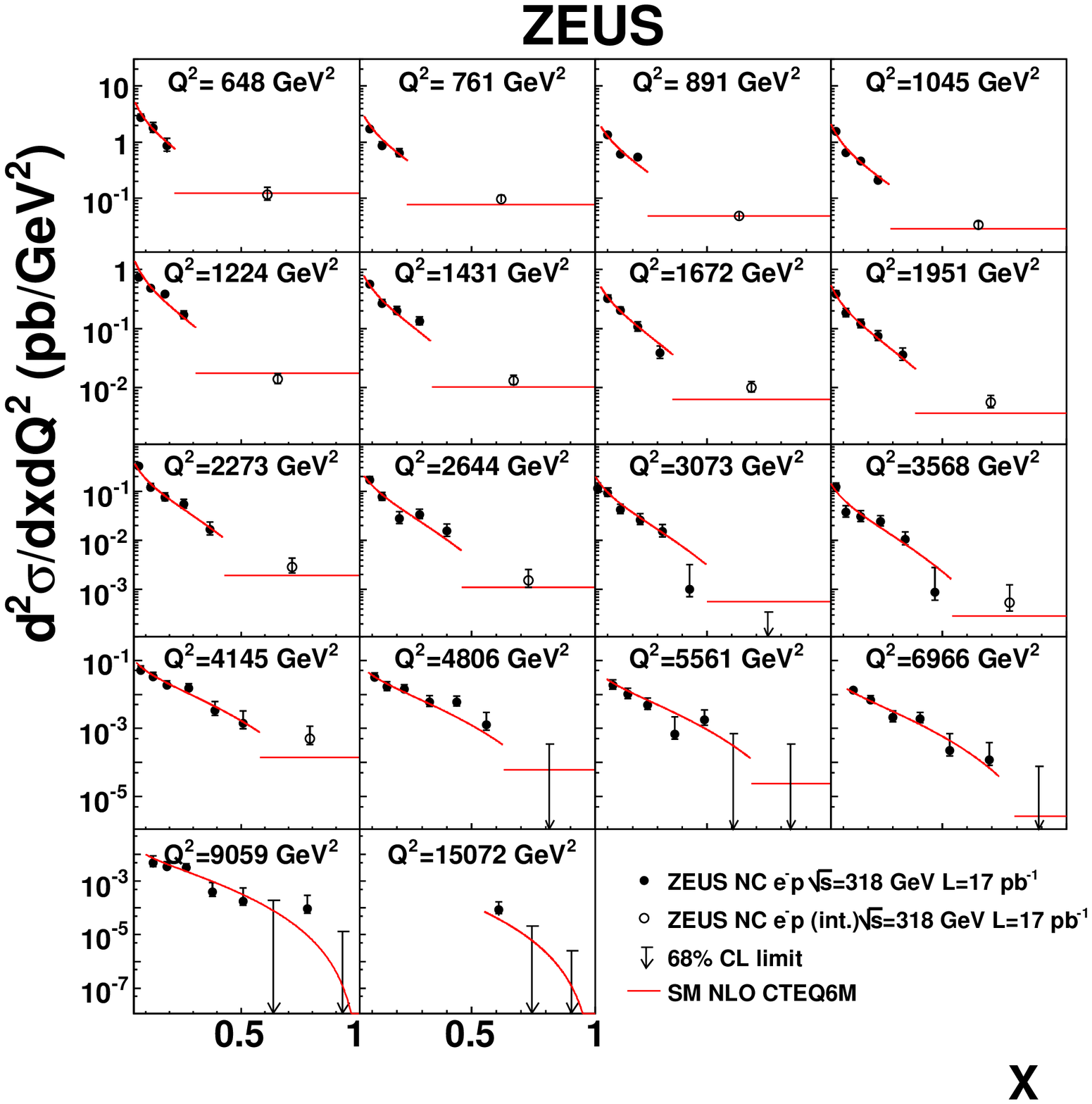}
\caption{
The double-differential cross section for 98-99 $e^-p$ NC scattering 
at $\sqrt{s}=318$~$\gev$
(solid circles) and the integral of the double differential cross section 
(open circles) compared to the Standard Model 
expectations evaluated using CTEQ6M PDFs (lines). The error bars show 
the statistical and systematic uncertainties added in quadrature. 
For bins with zero measured events, a $68\%$ probability limit,
calculated including the uncorrelated systematic uncertainty, is given. 
}
\label{cros98}
\end{center}
\end{figure}
%%%%%%%%%%%%%%%%%%%%%%%%%%%%%%%%%%%%%%%%%%%%%%%%%%%%%%%%%%%

%%%%%%%%%%%%%%%%%%%% FIGURE%%%%%%%%%%%%%%%%%%%%%%%%%%%%%%
\newpage
\begin{figure}
\begin{center}
\includegraphics[width=16.0cm]{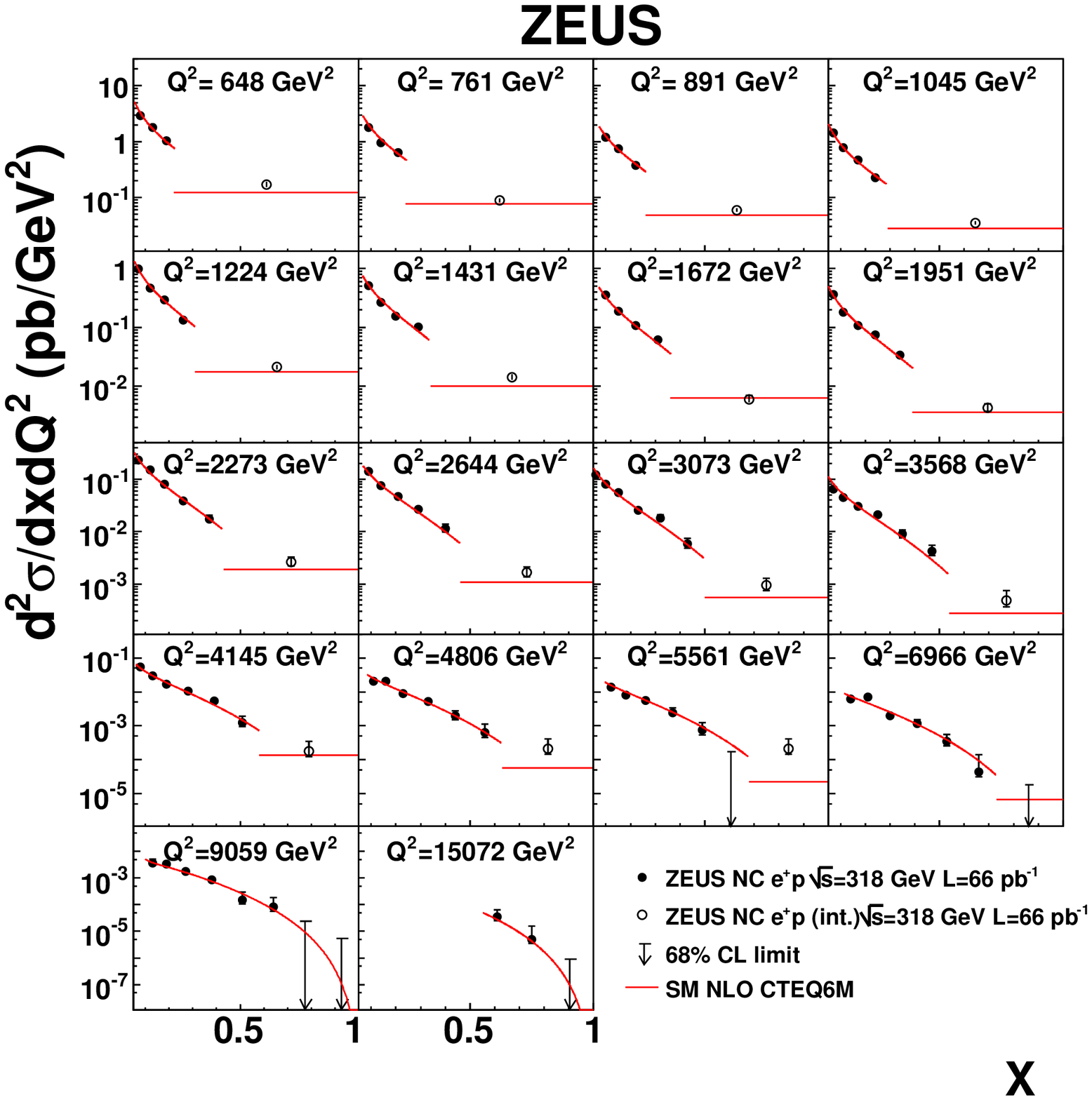}
\caption{
The double-differential cross section for 99-00 $e^+p$ NC scattering  
at $\sqrt{s}=318$~$\gev$
(solid circles) and the integral of the double differential cross section 
(open circles) compared to the Standard Model 
expectations evaluated using CTEQ6M PDFs (lines). The error bars show 
the statistical and systematic uncertainties added in quadrature.
For bins with zero measured events, a $68\%$ probability limit,
calculated including the uncorrelated systematic uncertainty, is given. 
}
\label{cros00}
\end{center}
\end{figure}
%%%%%%%%%%%%%%%%%%%%%%%%%%%%%%%%%%%%%%%%%%%%%%%%%%%%%%%%%%%

%%%%%%%%%%%%%%%%%%%% FIGURE%%%%%%%%%%%%%%%%%%%%%%%%%%%%%%
\newpage
\begin{figure}
\begin{center}
\includegraphics[width=16.0cm]{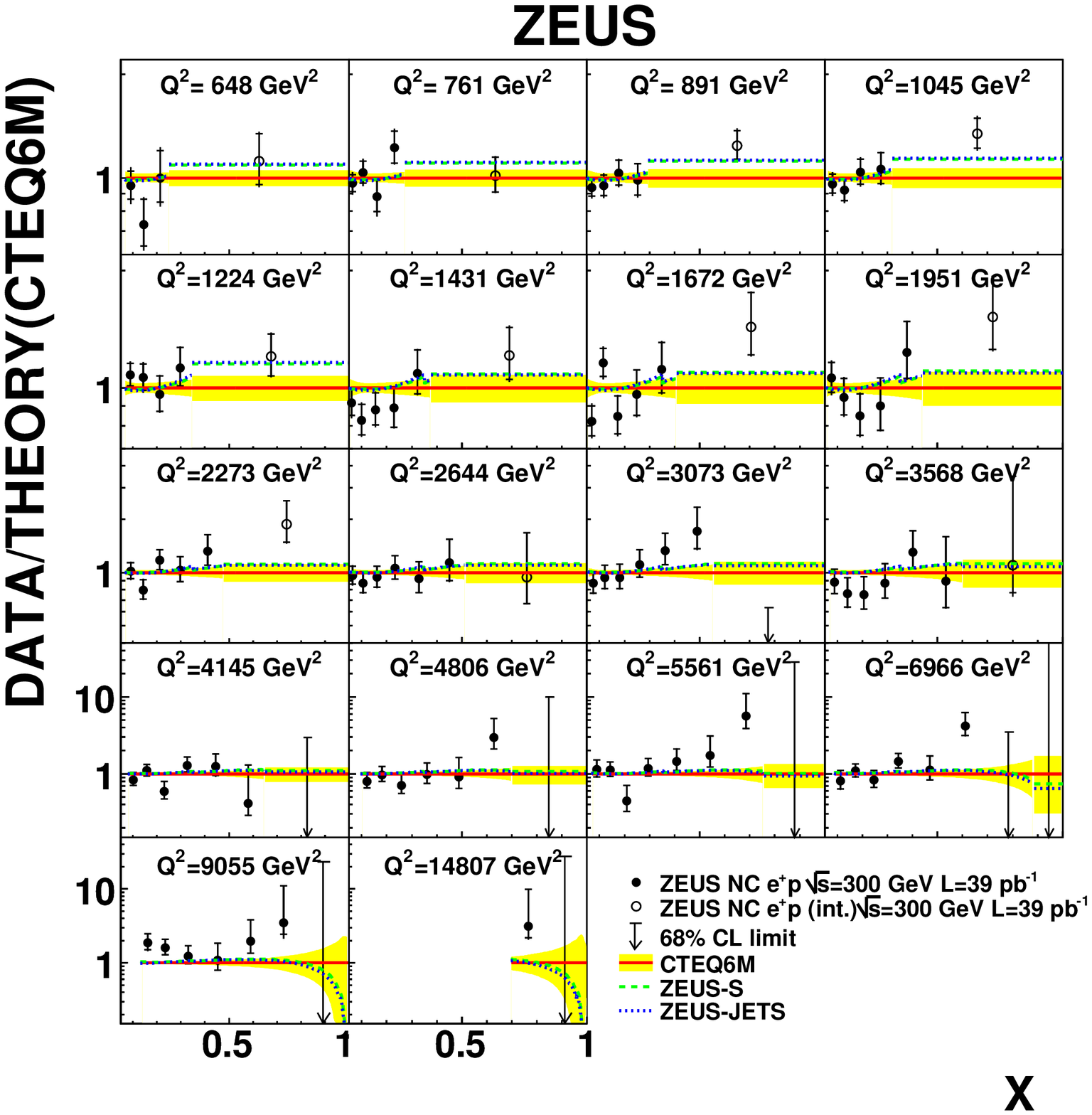}
\caption{
Ratio of the double-differential cross section for 96-97 $e^+p$ NC 
scattering (solid circles) and the integral of the double differential 
cross section (open circles) 
to the Standard Model expectation evaluated using the CTEQ6M PDFs. 
The inner error bars show the statistical uncertainty, while the outer 
ones show the statistical and systematic uncertainties added in quadrature. 
The ratio of the expectations using the ZEUS-S and ZEUS-JET PDFs to 
those using the CTEQ6M predictions are also shown.
For bins with zero measured events, a $68\%$ probability limit,
calculated including the uncorrelated systematic uncertainty, is given. 
}
\label{ratio96}
\end{center}
\end{figure}
%%%%%%%%%%%%%%%%%%%%%%%%%%%%%%%%%%%%%%%%%%%%%%%%%%%%%%%%%%%

%%%%%%%%%%%%%%%%%%%% FIGURE%%%%%%%%%%%%%%%%%%%%%%%%%%%%%%
\newpage
\begin{figure}
\begin{center}
\includegraphics[width=16.0cm]{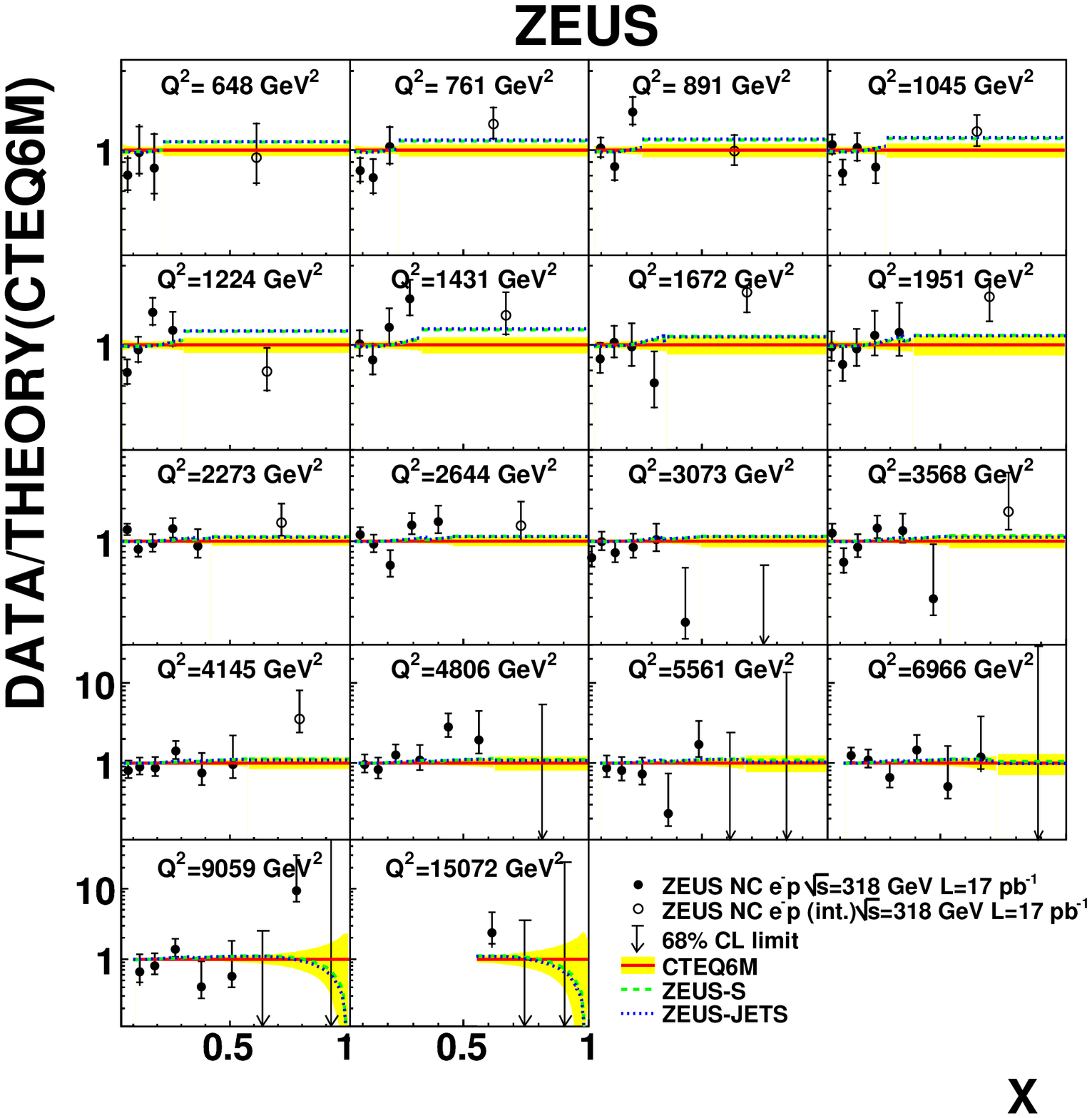}
\caption{
Ratio of the double-differential cross section for 98-99 $e^-p$ NC 
scattering (solid circles) and the integral 
of the double differential cross section (open circles) 
to the Standard Model expectation evaluated using the CTEQ6M PDFs. 
The inner error bars show the statistical uncertainty, while the outer 
ones show the statistical and systematic uncertainties added in quadrature. 
The ratio of the expectations using the ZEUS-S and ZEUS-JET PDFs to 
those using the CTEQ6M predictions are also  shown.
For bins with zero measured events, a $68\%$ probability limit,
calculated including the uncorrelated systematic uncertainty, is given. 
}
\label{ratio98}
\end{center}
\end{figure}
%%%%%%%%%%%%%%%%%%%%%%%%%%%%%%%%%%%%%%%%%%%%%%%%%%%%%%%%%%%

%%%%%%%%%%%%%%%%%%%% FIGURE%%%%%%%%%%%%%%%%%%%%%%%%%%%%%%
\newpage
\begin{figure}
\begin{center}
\includegraphics[width=16.0cm]{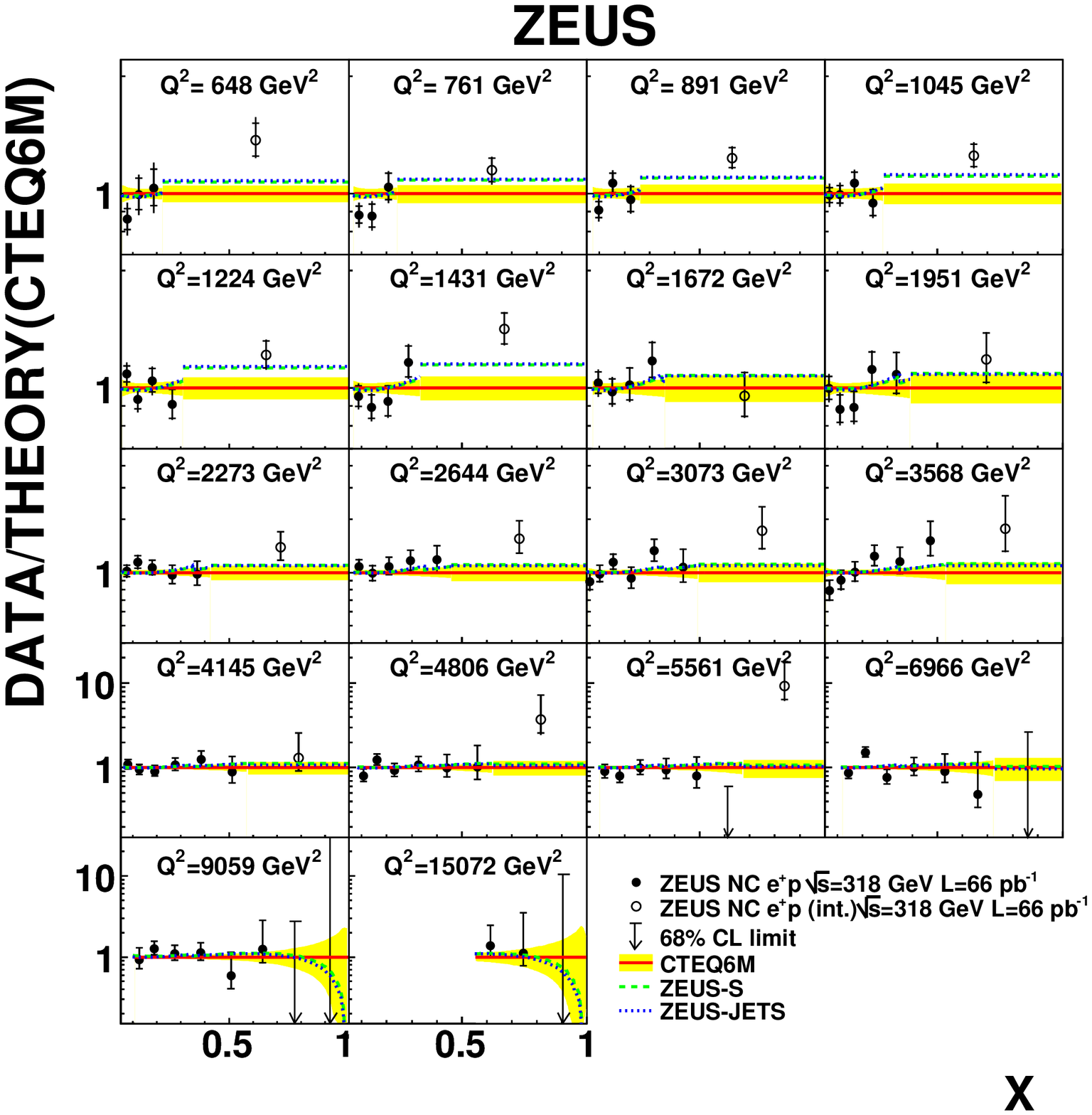}
\caption{
Ratio of the double-differential cross section for 99-00 $e^+p$ NC 
scattering (solid circles) and the integral 
of the double differential cross section (open circles) 
to the Standard Model expectation evaluated using the CTEQ6M PDFs. 
The inner error bars show the statistical uncertainty, while the outer 
ones show the statistical and systematic uncertainties added in quadrature. 
The ratio of the expectations using the ZEUS-S and ZEUS-JET PDFs to 
those using the CTEQ6M predictions are also  shown.
For bins with zero measured events, a $68\%$ probability limit,
calculated including the uncorrelated systematic uncertainty, is given. 
}
\label{ratio00}
\end{center}
\end{figure}
%%%%%%%%%%%%%%%%%%%%%%%%%%%%%%%%%%%%%%%%%%%%%%%%%%%%%%%%%%%